# Algorithmic Meta-Theorems

Stephan Kreutzer

Oxford University Computing Laboratory
`stephan.kreutzer@comlab.ox.ac.uk`

**Abstract.** Algorithmic meta-theorems are general algorithmic results applying to a whole range of problems, rather than just to a single problem alone. They often have a *logical* and a *structural* component, that is they are results of the form: *every computational problem that can be formalised in a given logic $\mathcal{L}$ can be solved efficiently on every class $\mathcal{C}$ of structures satisfying certain conditions.*

This paper gives a survey of algorithmic meta-theorems obtained in recent years and the methods used to prove them. As many meta-theorems use results from graph minor theory, we give a brief introduction to the theory developed by Robertson and Seymour for their proof of the graph minor theorem and state the main algorithmic consequences of this theory as far as they are needed in the theory of algorithmic meta-theorems.

## 1 Introduction

Algorithmic meta-theorems are general algorithmic results applying to a whole range of problems, rather than just to a single problem alone. In this paper we will concentrate on meta-theorems that have a *logical* and a *structural* component, that is on results of the form: *every computational problem that can be formalised in a given logic $\mathcal{L}$ can be solved efficiently on every class $\mathcal{C}$ of structures satisfying certain conditions.*

The first such theorem is Courcelle's well-known result [13] stating that every problem definable in monadic second-order logic can be solved efficiently on any class of graphs of bounded tree-width[1]. Another example is a much more recent result stating that every first-order definable optimisation problem admits a polynomial-time approximation scheme on any class $\mathcal{C}$ of graphs excluding at least one minor (see [22]).

Algorithmic meta-theorems lie somewhere between computational logic and algorithm or complexity theory and in some sense form a bridge between the two areas. In algorithm theory, an active research area is to find efficient solutions to otherwise intractable problems by restricting the class of admissible inputs. For instance, while the dominating set problem is NP-complete in general, it can be solved in polynomial time on any class of graphs of bounded tree-width.

In this line of research, algorithmic meta-theorems provide a simple and easy way to show that a certain problem is tractable on a given class of structures.

---

[1] The definition of tree-width and the other graph parameters and logics mentioned in the introduction will be presented formally in the following sections.

Formalising a problem in MSO yields a formal proof for its tractability on classes of structures of bounded tree-width, avoiding the task of working out the details of a solution using dynamic programming – something that is not always trivial to do but often enough solved by hand-wavy arguments such as "using standard techniques from dynamic programming ...".

Another distinguishing feature of logic based algorithmic meta-theorems is the observation that for a wide range of problems, such as covering or colouring problems, their precise mathematical formulation can often directly be translated into monadic second-order logic. Hence, ideally, instead of having to design an explicit algorithm for solving a problem on bounded tree-width graphs, one can read off tractability results directly from the problem description.

Finally, algorithmic meta-theorems yield tractability results for a whole class of problems providing valuable insight into how far certain algorithmic techniques range. On the other hand, in their negative form of intractability results, they also exhibit some limits to applications of certain algorithmic techniques.

In logic, one of the core tasks is the evaluation of logical formulas in structures – a task underlying problems in a wide variety of areas in computer science from database theory, artificial intelligence to verification and finite model theory.

Among the important logics studied in this context is first-order logic and its various fragments, such as its existential conjunctive fragment known as conjunctive queries in database theory. Whereas first-order model-checking is PSPACE-complete in general, even on input structures with only two elements, it becomes polynomial time for every fixed formula. So what can we possibly gain from restricting the class of admissible structures, if the problem is hard as soon as we have two elements and becomes easy if we fix the formula? Not much, if the distinction is only between taking the formula as full part of the input or keeping it fixed.

A finer analysis of first-order model-checking can be obtained by studying the problem in the framework of parameterized complexity (see [36,46,67]). The idea is to isolate the dependence of the running time on a certain part of the input, called the *parameter*, from the dependence on the rest. We will treat parameterized complexity formally in Section 2.4. The parameterized first-order evaluation problem is the problem, given a structure $A$ and a sentence $\varphi \in \text{FO}$, to decide whether $A \models \varphi$. The parameter is $|\varphi|$, the length of the formula. It is called *fixed parameter tractable* (FPT) if it can be solved in time $f(|\varphi|) \cdot |A|^c$, for some fixed constant $c$ and a computable function $f : \mathbb{N} \to \mathbb{N}$. While first-order model-checking is unlikely to be fixed-parameter tractable in general (unless unexpected results in parameterized complexity happen), Courcelle's theorem shows that even the much more expressive monadic second-order logic becomes FPT on graph classes of bounded tree-width. Hence, algorithmic meta-theorems give us a much better insight into the structure of model-checking problems taking structural information into account.

In this paper we will give an overview of algorithmic meta-theorems obtained so far and present the main methods used in their proofs. As mentioned before, these theorems usually have a logical and a structural component. As for the logic, we will primarily consider first-order and monadic second-order logic (see Section 2). As for the structural component, most meta-theorems have been proved relative to some structure classes based on graph theory, in particular on graph minor theory, such as classes of graphs of bounded tree-width, planar graphs, or $H$-minor free graphs. We will therefore present the relevant parts of graph structure theory needed for the proofs of the theorems presented here.

The paper is organised as follows. In Section 2, we present basic notation used throughout the paper. In Section 2.3 we present the relevant logics and give a brief overview of their model-checking problem. Section 2.4 contains an introduction to parameterized complexity. In Section 3, we introduce the notion of the tree-width of a graph and establish some fundamental properties. We then state and prove theorems by Seese and Courcelle establishing tractability results for monadic second-order logic on graph classes of bounded tree-width. In Section 4 we present an extension of tree-width called clique-width and a more recent, broadly equivalent notion called rank-width. Again we will see that monadic second-order model checking and satisfiability is tractable on graph classes of bounded clique-width. Section 5 contains a brief introduction to the theory of graph minors to the extent needed in later sections of the paper. The results presented in this section are then used in Section 7 to obtain tractability results on graph classes excluding a minor. In Section 7, we also consider the concept of localisation of graph invariants and use it to obtain further tractability results for first-order model checking. But before, in Section 6, we use the results obtained in Section 5 to show limits to MSO-tractability. Finally, we conclude the paper in Section 8.

*Remark.* An excellent survey covering similar topics as this paper has recently been written by Martin Grohe as a contribution to a book celebrating Wolfgang Thomas' 60th birthday [53]. While the two papers share a common core of results, they present the material in different ways and with a different focus.

## 2 Preliminaries

In this section we introduce basic concepts from logic and graph theory and fix the notation used throughout the paper. The reader may safely skip this section and come back to it whenever notation is unclear.

### 2.1 Sets

By $\mathbb{N} := \{0, 1, 2, \ldots\}$ we denote the set of non-negative integers and by $\mathbb{Z}$ the set of integers. For $k \in \mathbb{N}$ we write $[k]$ for the set $[k] := \{0, \ldots, k-1\}$. For a set $M$ and $k \in \mathbb{N}$ we denote by $[M]^k$ and $[M]^{\leq k}$ the set of all subsets of $M$ of size $k$ and size $\leq k$, respectively, and similarly for $[M]^{<k}$.

$[k]$

$[M]^k$, $[M]^{\leq k}$

## 2.2 Graphs

$V(G)$  A *graph* $G$ is a pair consisting of a set $V(G)$ of *vertices* and a set $E(G) \subseteq$
$E(G)$  $[V(G)]^2$ of *edges*. All graphs in this paper are finite, simple, i.e. no multiple
edges, undirected and loop-free. We will sometimes write $G := (V, E)$ for a
graph $G$ with vertex set $V$ and edge set $E$. We denote the class of all (finite)
GRAPH  graphs by GRAPH.

*incident, adjacent*  An edge $e := \{u, v\}$ is *incident* to its end vertices $u$ and $v$ and $u, v$ are *adja-*
$|G|, ||G||$  *cent*. If $G$ is a graph then $|G| := |V(G)|$ is its *order* and $||G|| := \max\{|V(G)|, |E(G)|\}$
its *size*.

For graphs $H, G$ we define the *disjoint union* $G \dot\cup H$ as the graph obtained as
the union of $H$ and an isomorphic copy $G'$ of $G$ such that $V(G') \cap V(H) = \varnothing$.

$H \subseteq G$  *Subgraphs.* A graph $H$ is a *subgraph* of $G$, written as $H \subseteq G$, if $V(H) \subseteq V(G)$
and $E(H) \subseteq E(G) \cap [V(H)]^2$. If $E(H) = E(G) \cap [V(H)]^2$ we call $H$ an *induced*
subgraph.

$G[U]$  Let $G$ be a graph and $U \subseteq V(G)$. The subgraph $G[U]$ *induced by $U$ in $G$* is
the graph with vertex set $U$ and edge set $E(G) \cap [U]^2$.

$G - U$  For a set $U \subseteq V(G)$, we write $G - U$ for the graph induced by $V(G) \setminus U$.
$G - X$  Similarly, if $X \subseteq E(G)$ we write $G - X$ for the graph $(V(G), E(G) \setminus X)$. Finally,
$G - v, \ G - e$  if $U := \{v\} \subseteq V(G)$ or $X := \{e\} \subseteq E(G)$, we simplify notation and write $G - v$
and $G - e$.

*Degree and neighbourhood.* Let $G$ be a graph and $v \in V(G)$. The *neighbour-*
$N_G(v)$  *hood* $N_G(v)$ of $v$ in $G$ is defined as $N_G(v) := \{u \in V(G) : \{u, v\} \in E(G)\}$. The
*distance* $d^G(u, v)$ between two vertices $u, v \in V(G)$ is the length of the shortest
path from $u$ to $v$ or $\infty$ if there is no such path. For every $v \in V(G)$ and $r \in \mathbb{N}$
we define the *r-neighbourhood* of $v$ in $G$ as the set

$$N_r^G(v) := \{w \in V(G) : d^G(v, w) \leq r\}.$$

of vertices of distance at most $r$ from $v$. For a set $W \subseteq V(G)$ we define $N_r^G(W) :=$
$\bigcup_{v \in W} N_r^G(v)$. We omit the index $\cdot^G$ whenever $G$ is clear from the context.

$d_G(v)$  The *degree* of $v$ is defined as $d_G(v) := |N_G(v)|$. We will drop the index $G$
whenever $G$ is clear from the context. Finally, $\Delta(G) := \max\{d(v) : v \in V\}$
$\Delta(G)$  denotes the *maximal degree*, or just *degree*, of $G$ and $\delta(G) := \min\{d(v) : v \in V\}$
$\delta(G)$  the *minimal degree*.

*Paths and walks.* A walk $P$ in $G$ is a sequence $x_1, e_1, \ldots, x_n, e_n, x_{n+1}$ such
that $e_i := \{x_i, x_{i+1}\} \in E(G)$ and $x_i \in V(G)$. The *length* of $P$ is $n$, i.e. the number
of edges. A *path* is a walk without duplicate vertices, i.e. $v_i \neq v_j$ whenever
$i \neq j$. We find it convenient to consider paths as subgraphs and hence use $V(P)$
and $E(P)$ to refer to its set of vertices and edges, resp. An $X - Y$-*path*, for
$X, Y \subseteq V(G)$, is a path with first vertex in $X$ and last vertex in $Y$. If $X := \{s\}$
and $Y := \{t\}$ are singletons, we simply write $s - t$-path.

A graph is *connected* if it is non-empty and between any two vertices $s$ and
$t$ there is an $s - t$-path. A *connected component* of a graph $G$ is a maximal
connected subgraph of $G$.

*Special graphs.* For $n, m \geq 1$ we write $K_n$ for the complete graph on $n$ vertices and $K_{n,m}$ for the complete bipartite graph with one partition of order $n$ and one of order $m$. Furthermore, if $X$ is a set then $K[X]$ denotes the complete graph with vertex set $X$.

For $n, m \geq 1$, the $n \times m$-*grid* $G_{n \times m}$ is the graph with vertex set $\{(i,j) : 1 \leq i \leq n, 1 \leq j \leq m\}$ and edge set $\{((i,j),(i',j')) : |i-i'|+|j-j'| = 1\}$. For $i \geq 1$, the subgraph induced by $\{(i,j) : 1 \leq j \leq m\}$ is called the *ith row* of $G_{n \times m}$ and for $j \geq 1$ the subgraph induced by $\{(i,j) : 1 \leq i \leq n\}$ is called the *jth column*. See Figure 1 for a $3 \times 4$-grid.

$K_n$
$K_{n,m}$
$K[X]$
$G_{n \times m}$

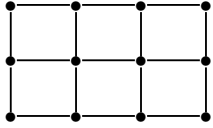

**Fig. 1.** A $3 \times 4$-grid

*Trees.* A *tree* $T$ is a connected acyclic graph. Often we will work with *rooted trees* $T$ with a distinguished vertex $r$, the *root* of $T$. A *leaf* in $T$ is a vertex of degree 1, all other vertices are called *inner vertices*. A tree is *sub-cubic*, if all vertices have degree at most 3. It is *cubic* if every vertex has degree 3 or 1.

A *directed tree* is a rooted tree where all edges are directed away from the root. A *binary tree* is a directed tree where every vertex has at most two outgoing edges. In directed graphs, we view edges as tuples $(u, v)$, where $u$ is the tail and $v$ is the head of the edge, rather than sets $\{u, v\}$.

*Coloured graphs.* Let $\Sigma$ be an alphabet. A $\Sigma$-*labelled tree* is a pair $(T, \lambda)$, where $T$ is a tree and $\lambda : V(T) \to \Sigma$ is a labelling function. Often, $\Sigma$ will be a set $C$ of colours and then we call $C$-labelled trees $C$-*coloured*, or just *coloured*. A $\Sigma$-tree is a $\Sigma$-labelled tree.

### 2.3 Logic

I assume familiarity with basic notions from mathematical logic. See e.g.[38,57] for an introduction to mathematical logic.

A *signature* $\sigma := \{R_1, \ldots, R_k, c_1, \ldots, c_q\}$ is a finite set of relation symbols $R_i$ and constant symbols $c_i$. To each relation symbol $R \in \sigma$ we assign an *arity* $\mathrm{ar}(R)$. A $\sigma$-*structure* $A$ is a tuple $A := \big(V(A), R_1(A), \ldots, R_k(A), c_1(A), \ldots, c_q(A)\big)$ consisting of a set $V(A)$, the *universe*, for each $R_i \in \sigma$ of arity $\mathrm{ar}(R_i) := r$ a set $R_i(A) \subseteq V(A)^r$ and for each $c_i \in \sigma$ a constant $c_i(A) \in V(A)$. We will usually use letters $A, B, \ldots$ for structures. Their universe is denoted as $V(A)$ and for each $R \in \sigma$ we write $R(A)$ for the relation $R$ in the structure $A$ and similarly for constant symbols $c \in \sigma$.

*signature*
$\mathrm{ar}(R)$

Tuples of elements are denoted by $\overline{a} := a_1, \ldots a_k$. We will frequently write $\overline{a}$ without stating its length explicitly, which will then be understood or not

$\overline{a}$

relevant. Abusing notation, we will treat tuples sometimes as sets and write $a \in \overline{a}$, with the obvious meaning, and also $\overline{a} \subseteq \overline{b}$ to denote that every element in $\overline{a}$ also occurs in $\overline{b}$.

$A \cong B$     Two $\sigma$-structures $A, B$ are *isomorphic*, denoted $A \cong B$, if there is a bijection $\pi : V(A) \to V(B)$ such that

- for all relation symbols $R \in \sigma$ of arity $r := \mathrm{ar}(R)$ and all $\overline{a} \in V(A)^r$, $\overline{a} \in R(A)$ if, and only if, $(\pi(a_1), \ldots, \pi(a_r)) \in R(B)$ and
- for all constant symbols $c \in \sigma$, $c(B) = \pi(c(A))$.

Let $\sigma$ be a signature. We assume a countably infinite set of first-order variables $x, y, \ldots$ and second-order variables $X, Y, \ldots$. A $\sigma$-*term* is a first-order variable or a constant symbol $c \in \sigma$. The class of formulas of *first-order logic* over $\sigma$, denoted FO$[\sigma]$, is inductively defined as follows. If $R \in \sigma$ and $\overline{x}$ is a tuple of $\sigma$-terms of length $\mathrm{ar}(R)$, then $R\overline{x} \in$ FO$[\sigma]$ and if $t$ and $s$ are terms then $t = s \in$ FO$[\sigma]$. Further, if $\varphi, \psi \in$ FO$[\sigma]$, then so are $(\varphi \wedge \psi)$, $(\varphi \vee \psi)$ and $\neg\varphi$. Finally, if $\varphi \in$ FO$[\sigma]$ and $x$ is a first-order variable, then $\exists x \varphi \in$ FO$[\sigma]$ and $\forall x \varphi \in$ FO$[\sigma]$.

The class of formulas of *monadic second-order logic* over $\sigma$, denoted MSO$[\sigma]$, is defined by the rules for first-order logic with the following additional rules: if $X$ is a second-order variable and $\varphi \in$ MSO$[\sigma \dot\cup \{X\}]$, then $\exists X \varphi \in$ MSO$[\sigma]$ and $\forall X \varphi \in$ MSO$[\sigma]$. Finally, we define FO $:= \bigcup_{\sigma \text{ signature}}$ FO$[\sigma]$ and likewise for MSO.

First-order variables range over elements of $\sigma$-structures and monadic second-order variables $X$ range over sets of elements. Formulas $\varphi \in$ FO$[\sigma]$ are interpreted in $\sigma$-structures $A$ in the obvious way, where atoms $R\overline{x}$ denote containment in the relation $R(A)$, $=$ denotes equality of elements, $\vee, \wedge, \neg$ denote disjunction, conjunction and negation and $\exists x \varphi$ is true in $A$ if there is an element $a \in V(A)$ such that $\varphi$ is true in $A$ if $x$ is interpreted by $a$. Analogously, $\forall x \varphi$ is true in $A$ if $\varphi$ is true in $A$ for all interpretations of $x$ by elements $a \in V(A)$.

For MSO$[\sigma]$-formulas, $\exists X \varphi$ is true in $A$ if there is a set $U \subseteq V(A)$ such that $\varphi$ is true if $X$ is interpreted by $U$ and analogously for $\forall X \varphi$.

The set of *free* variables of a formula is defined in the usual way. We will write $\varphi(\overline{x})$ to indicate that the variables in $\overline{x}$ occur free in $\varphi$. Formulas without free variables are called *sentences*. If $\varphi$ is a sentence we write $A \models \varphi$ if $\varphi$ is true in $A \models \varphi$    $A$. If $\varphi(\overline{x})$ has free variables $\overline{x}$ and $\overline{a}$ is a tuple of the same length as $\overline{x}$, we write $A \models \varphi(\overline{a})$    $A \models \varphi(\overline{a})$ or $(A, \overline{a}) \models \varphi$ if $\varphi$ is true in $A$ where the free variables $\overline{x}$ are interpreted $(A, \overline{a}) \models \varphi$    by the elements in $\overline{a}$ in the obvious way. We will sometimes consider formulas $\varphi(X)$ with a free second-order variable $X$. The notation extends naturally to free second-order variables.

We will use obvious abbreviations in formulas, such as $\to$ (implication), $x \neq y$ instead of $\neg x = y$ and $\bigvee_{i=1}^{k} \varphi_i$ and $\bigwedge_{i=1}^{k} \varphi_i$ for disjunctions and conjunctions over a range of formulas.

*independent set*    **Example 2.1**   *1. An* independent set, *or* stable set, *in a graph $G$ is a set $X \subseteq V(G)$ such that $\{u, v\} \notin E$ for all $u, v \in X$. The first-order sentence*

$$\varphi_k := \exists x_1 \ldots \exists x_k \bigwedge_{1 \leq i < j \leq k} (x_i \neq x_j \wedge \neg E x_i x_j)$$

is true in a graph $G$ (considered as an $\{E\}$-structure in the obvious way) if, and only if, $G$ contains an independent set of size $k$.

2. A *dominating set* in a graph $G$ is a set $X \subseteq V(G)$ such that for all $v \in V(G)$, either $v \in X$ or there is a $u \in X$ such that $\{v, u\} \in E(G)$. The formula

$$\varphi(X) := \forall x \big( Xx \vee \exists z(Exz \wedge Xz) \big)$$

states that $X$ is a dominating set. Precisely, a set $U \subseteq V(G)$ is a dominating set in $G$ if, and only if, $(G, U) \models \varphi$.

To say that a graph contains a dominating set of size $k$ we can use the formula $\exists x_1 \ldots \exists x_k \forall y \bigvee_{i=1}^{k} \big( y = x_i \vee Ex_i y \big)$. ⊣

Note the difference between the formulas defining an independent set and a dominating set: whereas an independent set of size $k$ can be defined by a formula using existential quantifiers only, i.e. without alternation between existential and universal quantifiers, the formula defining a dominating set of size $k$ contains one alternation of quantifiers. This indicates that the independent set problem might be simpler than the dominating set problem, a realisation that is reflected in the parameterized complexity of the problem as discussed later (see Proposition 2.10).

**Example 2.2**  *1. Consider the following* MSO-*formula*

$$\varphi := \forall X \Big( \big( \exists x Xx \wedge \forall x \forall y (Xx \wedge Exy \rightarrow Xy) \big) \rightarrow \forall x Xx \Big).$$

*The formula says of a graph $G$ that all sets $X \subseteq V(G)$ which are non-empty ($\exists x Xx$) and have the property that whenever $v \in X$ and $\{v, u\} \in E(G)$ then also $u \in X$, already contain the entire vertex set of $G$.*

*Clearly, $G \models \varphi$ if, and only if, $G$ is connected, as the vertex set of any connected component satisfies $\big( \exists x Xx \wedge \forall x \forall y (Xx \wedge Exy \rightarrow Xy) \big)$.*

*2. A* 3-colouring *of a graph $G$ is a function $f : V(G) \rightarrow \{1, 2, 3\}$ such that $f(u) \neq f(v)$ for all $\{u, v\} \in E(G)$. The formula*

$$\varphi := \exists C_1 \exists C_2 \exists C_3 \big( \forall x \bigvee_{i=1}^{3} C_i x \big) \wedge \forall x \forall y \big( Exy \rightarrow \bigwedge_{i=1}^{3} \neg (C_i x \wedge C_i y) \big)$$

*is true in a graph $G$ if, and only if, $G$ is 3-colourable.* ⊣

With any logic $\mathcal{L}$, we can naturally associate the following decision problem, called the *model-checking problem* of $\mathcal{L}$.

> MC($\mathcal{L}$)
> *Input:* Structure $A$ and sentence $\varphi \in \mathcal{L}$.
> *Problem:* Decide $A \models \varphi$.

Much of this paper will be devoted to studying the complexity of model-checking problems on various classes of graphs, primarily in the parameterized setting introduced in the next section.

*satisfiability*      Another natural problem associated with any logic is its *satisfiability problem* defined as the problem to decide for a given sentence $\varphi \in \mathcal{L}$ whether it has a model. We will study this problem relative to a given class $\mathcal{C}$ of structures. This is equivalent to asking whether the $\mathcal{L}$-*theory* of $\mathcal{C}$, i.e. the class of all formulas $\varphi \in \mathcal{L}$ which are true in every structure $A \in \mathcal{C}$, is decidable.

*quantifier rank*    The *quantifier rank* of a formula $\varphi$, denoted $\mathrm{qr}(\varphi)$, is the maximal number of
$\mathrm{qr}(\varphi)$      quantifiers in $\varphi$ nested inside each other. If $\varphi \in$ MSO, we count first- and second-order quantifiers. For instance, the formula in Example 2.2 (1) has quantifier rank 3.

*first-order type*    Let $A$ be a structure and $v_1, \ldots, v_k$ be elements in $V(A)$. For $q \geq 0$, the *first-*
$\mathrm{tp}_q^{\mathrm{FO}}(A, \overline{v})$    *order q-type* $\mathrm{tp}_q^{\mathrm{FO}}(A, \overline{v})$ of $\overline{v}$ is the class of all FO-formulas $\varphi(\overline{x})$ of quantifier-rank
$\mathrm{tp}_q^{\mathrm{MSO}}(A, \overline{v})$    $\leq q$ such that $A \models \varphi(\overline{v})$. *Monadic second-order types* $\mathrm{tp}_q^{\mathrm{MSO}}(A, \overline{v})$ are defined analogously.

By definition, types are infinite. However, it is well known that there are only finitely many FO or MSO-formulas of quantifier rank $\leq q$ which are pairwise not equivalent. Furthermore, we can effectively *normalise* formulas in such a way that equivalent formulas are normalised syntactically to the same formula. Hence, we can represent types by their finite set of normalised formulas.

This has a number of algorithmic applications. For instance, it is decidable if two types are the same and whether a formula $\varphi$ is contained in a type $\Theta$: we simply normalise $\varphi$ to a formula $\psi$ and check whether $\psi \in \Theta$. Note, however, that it is undecidable whether a set of normalised formulas is a type: by definition, types are satisfiable and satisfiability of first-order formulas is undecidable.

The following lemma, which essentially goes back to Feferman and Vaught will be used frequently later on. We refer the reader to [53] or [62] for a proof.

**Lemma 2.3** *Let* tp *be either* $\mathrm{tp}^{\mathrm{MSO}}$ *or* $\mathrm{tp}^{\mathrm{FO}}$ *and let* $H, G$ *be graphs such that* $V(H) \cap V(G) = \{\overline{v}\}$. *Let* $\overline{u} \in V(H)$ *and* $\overline{w} \in V(G)$.

*For all* $q \geq 0$, $\mathrm{tp}_q(G \cup H, \overline{vuw})$ *is uniquely determined by* $\mathrm{tp}_q(G, \overline{vw})$ *and* $\mathrm{tp}_q(H, \overline{uv})$ *and this is effective, i.e. there is an algorithm that computes* $\mathrm{tp}_q(G \cup H, \overline{vuw})$ *given* $\mathrm{tp}_q(G, \overline{vw})$ *and* $\mathrm{tp}_q(H, \overline{uv})$.

Suppose $G = H_1 \cup H_2$ can be decomposed into subgraphs $H_1, H_2$ such that $V(H_1 \cap H_2) = \overline{v}$. The importance of the lemma is that it allows us to infer the truth of a formula in $G$ from the $q$-type of $\overline{v}$ in $H_1$ and $H_2$, where $q := \mathrm{qr}(\varphi)$. Hence, if $G$ is decomposable in this way, we can reduce the question $G \models \varphi$ to the question on smaller graphs $H_1, H_2$. This will be of importance when we study graph-decompositions such as tree-decompositions and similar concepts in Section 3 and 4.

**MSO-Interpretations.**    Let $\mathcal{C}$ be a class of $\sigma$-structures and $\mathcal{D}$ be a class of $\tau$-structures. Suppose we know already that MSO-model-checking is tractable on $\mathcal{C}$ and we want to show that it is tractable on $\mathcal{D}$ also. Here is one way of doing this: find a way to "encode" a given graph $G \in \mathcal{D}$ in a graph $G' \in \mathcal{C}$ and also to "rewrite" the formula $\varphi \in \mathrm{MSO}[\tau]$ into a new formula $\varphi' \in \mathrm{MSO}[\sigma]$ so that $G \models \varphi$ if, and only if, $G' \models \varphi'$. Then tractability of MSO-model checking on $\mathcal{D}$ follows immediately from tractability on $\mathcal{C}$ – provided the encoding is efficient.

MSO-interpretations help us in doing just this: they provide a way to rewrite the formula $\varphi$ speaking about $\mathcal{D}$ to a formula $\varphi'$ speaking about $\mathcal{C}$ and also give us a translation of graphs "in the other direction", namely a way to translate a graph $G' \in \mathcal{C}$ to a graph $G := \Gamma(G') \in \mathcal{D}$ so that $G' \models \varphi'$ if, and only if, $G \models \varphi$. Hence, to reduce the model checking problem for MSO on $\mathcal{D}$ to the problem on $\mathcal{C}$, we have to find an interpretation $\Gamma$ to translate the formulas from $\mathcal{D}$ to $\mathcal{C}$ and an encoding of graphs $G \in \mathcal{D}$ to graphs $G' \in \mathcal{C}$ so that $\Gamma(G') \cong G$. Figure 2 demonstrates the way interpretations are used as reductions.

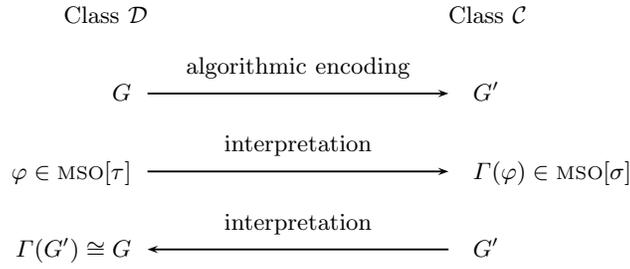

**Fig. 2.** Using interpretations as reductions between problems

We will first define the notion of interpretations formally and then demonstrate the concept by giving an example.

**Definition 2.4** *Let $\sigma := \{E, P_1, \ldots, P_k\}$ and $\tau := \{E\}$ be signatures, where $E$ is a binary relation symbol and the $P_i$ are unary. A (one-dimensional) MSO interpretation from $\sigma$-structures to $\tau$-structures is a triple $\Gamma := (\varphi_{univ}, \varphi_{valid}, \varphi_E)$ of $\mathrm{MSO}[\sigma]$-formulas.*

*For every $\sigma$-structure $T$ with $T \models \varphi_{valid}$ we define a graph (i.e. $\tau$-structure) $G := \Gamma(T)$ as the graph with vertex set $V(G) := \{u \in V(T) : T \models \varphi_{univ}(v)\}$ and edge set*
$$E(G) := \{\{u,v\} \in V(G) : T \models \varphi_E(u,v)\}.$$

*If $\mathcal{C}$ is a class of $\sigma$-structures we define $\Gamma(\mathcal{C}) := \{\Gamma(T) : T \in \mathcal{C}, T \models \varphi_{valid}\}$.*

MSO-*interpretation*

Every interpretation naturally defines a mapping from $\mathrm{MSO}[\tau]$-formulas $\varphi$ to $\mathrm{MSO}[\sigma]$-formulas $\varphi^* := \Gamma(\varphi)$. Here, $\varphi^*$ is obtained from $\varphi$ by recursively replacing

- first-order quantifiers $\exists x \varphi$ and $\forall x \varphi$ by $\exists x(\varphi_{univ}(x) \wedge \varphi^*)$ and $\forall x(\varphi_{univ}(x) \to \varphi^*)$ respectively,
- second-order quantifiers $\exists X \varphi$ and $\forall X \varphi$ by $\exists X\big(\forall y(Xy \to \varphi_{univ}(y)) \wedge \varphi^*\big)$ and $\forall X\big(\forall y(Xy \to \varphi_{univ}(y)) \to \varphi^*\big)$ respectively and
- atoms $E(x,y)$ by $\varphi_E(x,y)$.

The following lemma is easily proved (see [57]).

**Lemma 2.5 (interpretation lemma)** *Let $\Gamma$ be an MSO-interpretation from $\sigma$-structures to $\tau$-structures. Then for all MSO$[\tau]$-formulas and all $\sigma$-structures $G \models \varphi_{valid}$*

$$G \models \Gamma(\varphi) \iff \Gamma(G) \models \varphi.$$

Note that here we are using a restricted form of interpretations. In particular, we only allow one free variable in the formula $\varphi_{univ}(x)$ defining the universe of the resulting graph. A consequence of this is that in any such an interpretation $\Gamma$, we always have $|\Gamma(G)| \leq |G|$. In general interpretations, $\varphi_{univ}(\overline{x})$ can have any number of free variables, so that the universe of the resulting structure consists of tuples of elements and hence can be much (polynomially) larger than the original structure. For our purposes, one-dimensional interpretations are enough and we will therefore not consider more complex forms of interpretations as discussed in e.g. [57].

Initially we studied interpretations to reduce complexity results from one class $\mathcal{C}$ of graphs to another class $\mathcal{D}$. This is done as follows. Let $\Gamma$ be interpretation from $\mathcal{C}$ in $\mathcal{D}$, i.e. $\Gamma$ is a set of formulas speaking about graphs in $\mathcal{C}$ so that for all $G \in \mathcal{C}$, $\Gamma(G) \in \mathcal{D}$.

We first design an algorithm that encodes a given graph $G \in \mathcal{D}$ in a graph $\mathcal{G}' \in \mathcal{C}$ so that $\Gamma(G') \cong G$. Now, given $G \in \mathcal{D}$ and $\varphi \in \text{MSO}$ as input, we translate $G$ to a graph $G' \in \mathcal{C}$ and use the interpretation $\Gamma$ to obtain $\varphi' \in \text{MSO}[\sigma]$ such that $G' \models \varphi'$ if, and only if, $G \models \varphi$. Then we can check – using the model-checking algorithm for $\mathcal{C}$ – whether $G' \models \varphi'$.

**Example 2.6** *Let $\mathcal{C}$ be the class of finite paths and $\mathcal{D}$ be the class of finite cycles. Then $\Gamma(\mathcal{C}) = \mathcal{D}$ for the following interpretation $\Gamma := (\varphi_{univ}, \varphi_{valid}, \varphi_E)$: $\varphi_{univ}(x) = \varphi_{valid} := true$ and*

$$\varphi_E(x,y) := Exy \vee \neg \exists z_1 \exists z_2 \bigl(z_1 \neq z_2 \wedge \bigl((Exz_1 \wedge Exz_2) \vee (Eyz_1 \wedge Eyz_2)\bigr)\bigr)$$

*The formula is true for a pair $x,y$ if there is an edge between $x$ and $y$ or if neither $x$ nor $y$ have two different neighbours. Hence, if $P \in \mathcal{C}$ is a path then $G := \Gamma(P)$ is the cycle obtained from $P$ by connecting the two endpoints.*

*Now, if we know that MSO-model-checking is tractable on $\mathcal{C}$ then we can infer tractability on $\mathcal{D}$ is follows. Given $C \in \mathcal{D}$ and $\varphi \in \text{MSO}$, delete an arbitrary edge from $C$ to obtain a path $P \in \mathcal{C}$ and construct $\varphi' := \Gamma(\varphi)$. Obviously, $\Gamma(P) \cong C$ and hence $P \models \varphi'$, if and only if, $C \models \varphi$.* ⊣

### 2.4 Complexity

We assume familiarity with basic principles of algorithm design and analysis, in particular Big-O notation, as can be found in any standard textbook on algorithms, e.g. [11]. Also, we assume familiarity with basic complexity classes such as PTIME, NP and PSPACE and standard concepts from complexity theory such as polynomial-time reductions as can be found in any text book on complexity theory, e.g. [70]. By reductions we will generally mean polynomial-time many-one reductions, unless explicitly stated otherwise.

The following examples introduce some of the problems we will be considering throughout the paper.

**Example 2.7** 1. Recall from Example 2.1 that an independent set *in a graph G is a set $X \subseteq V(G)$ such that $\{u,v\} \notin E$ for all $u,v \in X$. The* independent set *problem is defined as*

> INDEPENDENT SET
> *Input:* A graph $G$ and $k \in \mathbb{N}$.
> *Problem:* Decide if $G$ contains an independent set of size $k$.

*independent set*

2. *Recall from Example 2.1 that* dominating set *in a graph $G$ is a set $X \subseteq V(G)$ such that for all $v \in V(G)$, either $v \in X$ or there is a $u \in X$ such that $\{v,u\} \in E(G)$. The* dominating set *problem is defined as*

> DOMINATING SET
> *Input:* A graph $G$ and $k \in \mathbb{N}$.
> *Problem:* Decide if $G$ contains a dominating set of size $k$.

*dominating set*

3. *A $k$-*colouring *of a graph $G$ is a function $f : V(G) \to \{1, \ldots, k\}$ such that $f(u) \neq f(v)$ for all $\{u,v\} \in E(G)$. Of particular interest for this paper is the problem to decide if a graph can be coloured by three colours.*

> 3-COLOURING
> *Input:* A graph $G$.
> *Problem:* Decide if $G$ has a 3-colouring.

⊣

It is well known that all three problems in the previous example are NP-complete. Furthermore, we have already seen that the dominating set problem can be reduced to first-order model-checking MC(FO). Hence, the latter is NP-hard as well. However, as the following lemma shows, MC(FO) is (presumably) even much harder than DOMINATING SET.

**Lemma 2.8 (***Vardi* [86]**)** MC(FO) *and* MC(MSO) *are* PSPACE-*complete.*

*Proof (sketch).* It is easily seen that MC(MSO), and hence MC(FO) is in PSPACE: given $A$ and $\varphi \in$ MSO, simply try all possible interpretations for the variables quantified in $\varphi$. This requires only polynomial space.

Hardness of MC(FO) follows easily from the fact that QBF, the problem to decide whether a quantified Boolean formula is satisfiable, is PSPACE-complete. Given a QBF-formula $\varphi := Q_1 X_1 \ldots Q_k X_k \psi$, where $\psi$ is a formula in propositional logic over the variables $X_1 \ldots X_k$ and $Q_i \in \{\exists, \forall\}$, we compute the first-order formula $\varphi' := \exists t \exists f (t \neq f \wedge Q_1 x_1 \ldots Q_k x_k \psi')$, where $\psi'$ is obtained from $\psi$ by replacing each positive literal $X_i$ by $x_i = t$ and each negative literal $\neg X_i$ by $x_i = f$. Here, the variables $t, f$ represent the truth values *true* and *false*. Clearly, for every structure $A$ with at least two elements, $A \models \varphi'$ if, and only if, $\varphi$ is satisfiable. □

An immediate consequence of the proof is that MC(FO) is hard even for very simple structures: they only need to contain at least two elements. An area of computer science where evaluation problems for logical systems have intensively been studied is database theory, where first-order logic is the logical foundation of the query language SQL. A common assumption in database theory is that the size of the query is relatively small compared to the size of the database. Hence, giving the same weight to the database and the query may not truthfully reflect the complexity of query evaluation. It has therefore become standard to distinguish between three ways of measuring the complexity of logical systems:

- *combined complexity*: given a structure $A$ and a formula $\varphi$ as input, what is the complexity of deciding $A \models \varphi$ measured in the size of the structure and the size of the formula?
- *data complexity*: fix a formula $\varphi$. Given a structure $A$ as input, what is the complexity of deciding $A \models \varphi$ measured in the size of the structure only?
- *expression complexity*: fix a structure $A$. Given a formula $\varphi$ as input, what is the complexity of deciding $A \models \varphi$ measured in the size of the formula only?

As seen in Lemma 2.8, the combined complexity of first-order logic is PSPACE-complete. Furthermore, the proof shows that even the expression complexity is PSPACE-complete, as long as we fix a structure with at least two elements. On the other hand, it is easily seen that for a fixed formula $\varphi$, checking whether $A \models \varphi$ can be done in time $|A|^{\mathcal{O}(|\varphi|)}$. Hence, the data complexity of first-order logic is in PTIME.

Besides full first-order logic, various fragments of FO have been studied in database theory and finite model theory. For instance, the combined complexity of the *existential conjunctive fragment* of first-order logic – known as *conjunctive queries* in database theory – is NP-complete. And if we consider the *bounded variable fragment* of first-order logic, the combined complexity is PTIME [87].

Much of this paper is devoted to study model-checking problems for a logic $\mathcal{L}$ on restricted classes $\mathcal{C}$ of structures or graphs, i.e. to study the problem

> MC($\mathcal{L}, \mathcal{C}$)
> *Input:* $A \in \mathcal{C}$ and $\varphi \in \mathcal{L}$.
> *Problem:* Decide $A \models \varphi$.

In Example 2.2, we have already seen that 3-colourability is definable by a fixed sentence $\varphi \in$ MSO. As the problem is NP-complete, this shows that the data-complexity of MSO is NP-hard. In fact, it is complete for the polynomial time hierarchy. There are, however, interesting classes of graphs on which the data-complexity of MSO is PTIME. One example is the class of trees, another are classes of graphs of bounded tree-width.

For first-order logic there is not much to classify in terms of input classes $\mathcal{C}$, as the combined complexity is PSPACE-complete as soon as we have at least one structure of size $\geq 2$ in $\mathcal{C}$ and the data complexity is always PTIME. Hence, the classification into expression and data complexity is too coarse for an interesting theory. However, polynomial time data complexity is somewhat unsatisfactory, as it does not tell us much about the degree of the polynomials. All it says is that for every fixed formula $\varphi$, deciding $A \models \varphi$ is in polynomial time. But the running time of the algorithms depends exponentially on $|\varphi|$ – and this is unacceptably high even for moderate formulas. Hence, the distinction between data and expression complexity is only of limited value for classifying tractable and intractable instances of the model checking problem.

A framework that allows for a much finer classification of model-checking problems is *parameterized complexity*, see [36,46,67]. A *parameterized problem* is a pair $(P, \chi)$, where $P$ is a decision problem and $\chi$ is a polynomial time computable function that associates with every instance $w$ of $P$ a positive integer, called the *parameter*. Throughout this paper, we are mainly interested in parameterized model-checking problems. For a given logic $\mathcal{L}$ and a class $\mathcal{C}$ of structures we define[2]

> $\mathrm{MC}(\mathcal{L}, \mathcal{C})$
>     *Input:* Given $A \in \mathcal{C}$ and $\varphi \in \mathcal{L}$.
> *Parameter:* $|\varphi|$.
>     *Problem:* Decide $A \models \varphi$.

A parameterized problem is *fixed-parameter tractable*, or in the complexity class FPT, if there is an algorithm that correctly decides whether an instance $w$ is in $P$ in time     FPT

$$f(\chi(w)) \cdot |w|^{\mathcal{O}(1)},$$

for some computable function $f : \mathbb{N} \to \mathbb{N}$. An algorithm with such a running time is called an *fpt algorithm*. Sometimes we want to make the exponent of the     *fpt algorithm* polynomial explicit and speak of *linear fpt algorithm*, if the algorithm achieves a running time of $f(\chi(w)) \cdot |w|$, and similarly for quadratic and cubic fpt algorithms. We will sometimes relax the definition of parameterized problems slightly by considering problems $(P, \chi)$ where the function $\chi$ is no longer polynomial time computable, but is itself fixed-parameter tractable. For instance, this will be the case for problems where the parameter is the tree-width of a graph (see Section 3.1), a graph parameter that is computable by a linear fpt-algorithm but not in polynomial time (unless PTIME =NP). Everything we need from parameterized complexity theory in this paper generalises to this parametrization also. See [46, Chapter 11.4] for a discussion of this issue.

In the parameterized world, FPT plays a similar role to PTIME in classical complexity – a measure of tractability. Hence, much work has gone into classifying problems into those which are fixed-parameter tractable and those which are not, i.e. those that can be solved by algorithms with a running time such as $\mathcal{O}(2^{k^2} n^2)$ and those which require something like $\mathcal{O}(n^k)$, where $k$ is the parameter. Running times of the form $\mathcal{O}(n^k)$ yield the parameterized complexity class

---

[2] We abuse notation here and also refer to the parameterized problem as $\mathrm{MC}(\mathcal{L}, \mathcal{C})$. As we will not consider the classical problem anymore, there is no danger of confusion.

XP  XP, defined as the class of parameterized problems that can be solved in time $\mathcal{O}(n^{f(k)})$, for some computable function $f : \mathbb{N} \to \mathbb{N}$.

In terms of model-checking problems, a model-checking problem $\mathrm{MC}(\mathcal{L}, \mathcal{C})$ is in XP if, and only if, the data complexity of $\mathcal{L}$ on $\mathcal{C}$ is PTIME. Obviously, FPT $\subseteq$ XP and this inclusion is strict, as can be proved using the time hierarchy theorem. If FPT is the parameterized analogue of PTIME then XP can be seen as the analogue of EXPTIME. And again, similar to classical complexity, there are hierarchies of complexity classes in between FPT and XP. For our purpose, the most important class is called W[1], which is the first level of the *W-hierarchy* formed by classes W[$i$], for all $i \geq 1$. We refrain from giving the precise definition of W[1] and the W-hierarchy and refer the reader to the monograph [46]. For our purposes, it suffices to know that FPT, XP and the W[$i$]-classes form the following hierarchy

W[1]
W-hierarchy

$$\mathrm{FPT} \subseteq W[1] \subseteq W[2] \subseteq \cdots \subseteq \mathrm{XP}.$$

In some sense, W[1] plays a similar role in parameterized complexity as NP in classical complexity, in that it is generally believed that FPT $\neq$ W[1] (as far as these beliefs go) and proving that a problem is W[1]-hard establishes that it is unlikely to be fixed-parameter tractable, i.e. efficiently solvable in the parameterized sense. The notion of reductions used here is *fpt-reduction*. Again, we refer to [46].

We close the section by stating the parameterized complexity of some problems considered in this paper.

**Definition 2.9** *1. The p-*DOMINATING SET *problem is the problem, given a graph G and $k \in \mathbb{N}$, to decide whether G contains a dominating set of size k. The parameter is k.*
*2. The p-*INDEPENDENT SET *problem is the problem, given a graph G and $k \in \mathbb{N}$, to decide whether G contains an independent set of size k. The parameter is k.*
*3. The p-*CLIQUE *problem is the problem, given a graph G and $k \in \mathbb{N}$, to decide whether G contains a clique of size k. The parameter is k.*

In the sequel, we will usually drop the prefix $p-$ and simply speak about the DOMINATING SET problem. It will always be clear from the context whether we are referring to the parameterized or the classical problem.

**Lemma 2.10** (*Downey, Fellows* [34,35])  *1. p-*DOMINATING SET *is W[2]-complete (see [34]).*
*2. p-*INDEPENDENT SET *is W[1]-complete (see [35]).*
*3. p-*CLIQUE *is W[1]-complete (see [35]).*

We have already seen that dominating and independent sets of size $k$ can uniformly be formalised in first-order logic. Hence MC(FO) is W[2]-hard as well. In fact, it is complete for the parameterized complexity class AW[∗], which contains all levels of the W-hierarchy and is itself contained in XP. Finally, as 3-colourability is expressible in MSO, MSO model-checking is not in XP unless NP=PTIME.

## 3 Monadic Second-Order Logic on Tree-Like Structures

It is a well-known fact, based on the close relation between monadic second-order logic and finite tree- and word-automata (see e.g. [9,31,83,84,10,46,61]), that model-checking and satisfiability for very expressive logics such as MSO becomes tractable on the class of finite trees. At the core of these results is the observation that the validity of an MSO sentence at the root of a tree can be inferred from the label of the root and the MSO-types realised by its successors. There are various ways in which this idea can be turned into a proof or algorithm: we can use effective versions of Feferman-Vaught style theorems (see e.g.[62]) or we can convert formulas into suitable tree-automata and let them run on the trees. The aim of the following sections is to extend the results for MSO and FO from trees to more general classes of graphs. The aforementioned composition methods will in most cases provide the key to obtaining these stronger results.

In this section we generalise the results for MSO model-checking and satisfiability from trees to graphs that are no longer trees but still tree-like enough so that model-checking and satisfiability testing for such graphs can be reduced to the case of trees.

### 3.1 Tree-Width

The precise notion for "tree-likeness" we use is the concept of tree-width. We first introduce tree-decompositions, establish some closure properties and then comment on algorithmic problems in relation to tree-width.

**Tree-Decompositions**

**Definition 3.1** *A* tree-decomposition *of a graph $G$ is a pair $\mathcal{T} := (T, (B_t)_{t \in V(T)})$ consisting of a tree $T$ and a family $(B_t)_{t \in V(T)}$ of sets $B_t \subseteq V(G)$ such that* — tree-decomposition

1. *for all $v \in V(G)$ the set* — $B^{-1}(v)$

$$B^{-1}(v) := \{t \in V(T) : v \in B_t\}$$

   *is non-empty and connected in $T$ and*
2. *for every edge $e \in E(G)$ there is a $t \in V(T)$ with $e \subseteq B_t$.*

*The* width $\mathrm{w}(\mathcal{T})$ *of $\mathcal{T}$ is $\mathrm{w}(\mathcal{T}) := \{|B_t| - 1 : t \in V(T)\}$ and the* tree-width *of $G$ is defined as the minimal width of any of its tree-decompositions.* — tree-width, $\mathrm{w}(\mathcal{T})$

We refer to the sets $B_t$ of a tree-decomposition as *bags*. For any edge $e := \{s,t\} \in E(T)$ we call $B_s \cap B_t$ the *cut at* or *along* the edge $e$. (The reason for this will become clear later. See Lemma 3.13.) — bags, cut

**Example 3.2** *Consider the graph in Figure 3 a). A tree-decomposition of this graph is shown in Figure 3 b).* ⊣

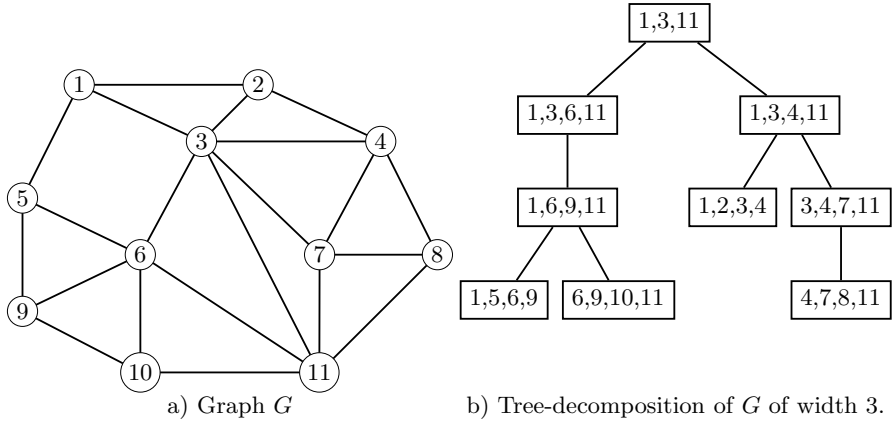

a) Graph $G$  b) Tree-decomposition of $G$ of width 3.

**Fig. 3.** Graph and tree-decomposition from Example 3.2

**Example 3.3** *Trees have tree-width* 1. *Given a tree $T$, the tree-decomposition has a node $t$ for each edge $e \in E(T)$ labelled by $B_t := e$ and suitable edges connecting the nodes.* ⊣

*series-parallel* **Example 3.4** *The class of* series-parallel *graphs $(G, s, t)$ with source $s$ and sink $t$ is inductively defined as follows.*

1. *Every edge $\{s, t\}$ is series-parallel.*
2. *If $(G_1, s_1, t_1)$ and $(G_2, s_2, t_2)$ are series parallel with $V(G_1) \cap V(G_2) = \varnothing$, then so are the following graphs:*
   a) *the graph $(G, s, t)$ obtained from $G_1 \cup G_2$ by identifying $t_1$ and $s_2$ and setting $s = s_1$ and $t = t_2$ (serial composition).*
   b) *the graph $(G, s, t)$ obtained from $G_1 \cup G_2$ by identifying $s_1$ and $s_2$ and also $t_1$ and $t_2$ and setting $s = s_1$ and $t = t_2$ (parallel composition).*

*The class of series-parallel graphs has tree-width* 2. *Following the inductive definition of series-parallel graphs one can easily show that every such graph $(G, s, t)$ has a tree-decomposition of width* 2 *containing a node labelled by $\{s, t\}$. This is trivial for edges. For parallel and serial composition the tree-decompositions of the individual parts can be glued together at the node labelled by the respective source and sink nodes.* ⊣

The final example shows that grids have very high tree-width. Grids play a special role in relation to tree-width. As we will see later, every graph of sufficiently high tree-width contains a large grid minor. Hence, in this sense, grids are the least complex graphs of high tree-width.

**Lemma 3.5** *For all $n > 1$, the $n \times n$-grid $G_{n,n}$ has tree-width $n$.*

In the remainder of this section we will present some basic properties of tree-decompositions and tree-width.

*Closure Properties and Connectivity.* It is easily seen that tree-width is preserved under taking subgraphs. For, if $(T, (B_t)_{t \in V(T)})$ is a tree-decomposition of width $w$ of a graph $G$, then $(T, (B_t \cap V(H))_{t \in V(T)})$ is a tree-decomposition of $H$ of width at most $w$. Further, if $G$ and $H$ are disjoint graphs, we can combine tree-decompositions for $G$ and $H$ to a tree-decomposition of the disjoint union $G \dot\cup H$ by adding one edge connecting the two decompositions.

**Lemma 3.6** *Let $G$ be a graph. If $H \subseteq G$, then $\mathrm{tw}(H) \leq \mathrm{tw}(G)$.*
*Further, if $C_1, \ldots, C_k$ are the components of $G$, then*

$$\mathrm{tw}(G) = \max\{\mathrm{tw}(C_i) : 1 \leq i \leq l\}.$$

To state the next results, we need further notation. Let $G$ be a graph and $(T, (B_t)_{t \in V(T)})$ be a tree-decomposition of $G$.

1. If $H \subseteq G$ we define $B^{-1}(H) := \{t \in V(T) : B_t \cap V(H) \neq \varnothing\}$. $\quad B^{-1}(H)$
2. Conversely, for $U \subseteq T$ we define $B(U) := \bigcup_{t \in V(U)} B_t$. $\quad B(U)$

Occasionally, we will abuse notation and use $B, B^{-1}$ for sets instead of subgraphs. The next lemma is easily proved by induction on $|H|$ using the fact that for each vertex $v \in V(G)$ the set $B^{-1}(v)$ is connected in any tree-decomposition $\mathcal{T}$ of $G$ and that edges $\{u,v\} \in E(G)$ are covered by some bag $B_t$ for $t \in V(T)$. Hence, $B^{-1}(u) \cup B^{-1}(v)$ is connected in $\mathcal{T}$ for all $\{u,v\} \in E(H)$.

**Lemma 3.7** *Let $G$ be a graph and $\mathcal{T} := (T, (B_t)_{t \in V(T)})$ be a tree-decomposition of $G$. If $H \subseteq G$ is connected, then so is $B^{-1}(H)$ in $\mathcal{T}$.*

*Small tree-decompositions.* A priori, by duplicating nodes, tree-decompositions of a graph can be arbitrarily large (in terms of the number of nodes in the underlying tree). However, this is not very useful and we can always avoid this from happening. We will now consider tree-decompositions which are *small* and derive various useful properties from this.

**Definition 3.8** *A tree-decomposition $(T, (B_t)_{t \in V(T)})$ is* small, *if $B_t \not\subseteq B_u$ for all $u, t \in V(T)$ with $t \neq u$.*  $\quad$ small tree-decompositions

The next lemma shows that we can easily convert every tree-decomposition to a small one in linear time.

**Lemma 3.9** *Let $G$ be a graph and $\mathcal{T} := (T, (B_t)_{t \in V(T)})$ a tree-decomposition of $G$. Then there is a small tree-decomposition $\mathcal{T}' := (T', (B'_t)_{t \in V(T')})$ of $G$ of the same width and with $V(T') \subseteq V(T)$ and $B'_t = B_t$ for all $t \in V(T')$.*

*Proof.* Suppose $B_s \subseteq B_t$ for some $s \neq t$. Let $s = t_1 \ldots t_n = t$ be the nodes of the path from $s$ to $t$ in $T$. Then $B_s \subseteq B_{t_2}$, by definition of tree-decompositions. But then, $(T', (B_t)_{t \in V(T')})$ with $V(T') := V(T) \setminus \{s\}$ and

$$E(T') := \begin{pmatrix} E(T) \setminus \{\{v,s\} : \{v,s\} \in E(T)\}\end{pmatrix} \cup \\ \{\{v, t_2\} : \{v,s\} \in E(T) \text{ and } v \neq t_2\}.$$

is a tree-decomposition of $G$ with $V(T)' \subset V(T)$. We repeat this until $T$ is small. $\square$

A consequence of this is the following result, which implies that in measuring the running time of algorithms on graphs whose tree-width is bounded by a constant $k$, it is sufficient to consider the order of the graphs rather than their size.

**Lemma 3.10** *Every (non-empty) graph of tree-width at most $k$ contains a vertex of degree at most $k$.*

*Proof.* Let $G$ be a graph and let $\mathcal{T} := (T, (B_t)_{t \in V(T)})$ be a small tree-decomposition of $G$ of width $k := \mathrm{tw}(G)$. If $|T| = 1$, then $|G| \leq k + 1$ and there is nothing to show. Otherwise let $t$ be a leaf of $T$ and $s$ be its neighbour in $T$. As $\mathcal{T}$ is small, $B_t \not\subseteq B_s$ and hence there is a vertex $v \in B_t \setminus B_s$. By definition of tree-decompositions, $v$ must have all its neighbours in $B_t$ and hence has degree at most $k$. □

**Corollary 3.11** *Every graph $G$ of tree-width $\mathrm{tw}(G) \leq k$ has at most $k \cdot |V(G)|$ edges, i.e., for $k > 0$, $\|G\| \leq k \cdot |G|$.*

*Separators.* We close this section with a characterisation of graphs of small tree-width in terms of separators. This separation property allows for the aforementioned applications of automata theory or Feferman-Vaught style theorems.

**Definition 3.12** *Let $G$ be a graph.*

*(i) Let $X, Y \subseteq V(G)$. A set $S \subseteq V(G)$ separates $X$ and $Y$, or is a* separator *for $X$ and $Y$, if every path containing a vertex of $Y$ and a vertex of $Z$ also contains a vertex of $S$. In other words, $X$ and $Y$ are disconnected in $G - S$.*

*(ii) A* separator *of $G$ is a set $S \subseteq V(G)$, so that $G - S$ has more than one component, i.e. there are sets $X, Y \subseteq V(G)$ such that $S$ separates $X$ and $Y$ and $X \setminus S \neq \emptyset$ and $Y \setminus S \neq \emptyset$.*

**Lemma 3.13** *Let $(T, (B_t)_{t \in V(T)})$ be a small tree-decomposition of a graph $G$.*

*(i) If $e := \{s, t\} \in E(T)$ and $T_1, T_2$ are the components of $T - e$, then $B_t \cap B_s$ separates $B(T_1)$ and $B(T_2)$.*

*(ii) If $t \in V(T)$ is an inner vertex and $T_1, \ldots, T_k$ are the components of $T - t$ then $B_t$ separates $B(T_i)$ and $B(T_j)$, for all $i \neq j$.*

*Proof.* Let $e := \{s, t\} \in E(T)$ and let $T_1, T_2$ be the components of $T - e$. As $\mathcal{T}$ is small, $X := B(T_1) \setminus B(T_2) \neq \emptyset$ and $Y := B(T_2) \setminus B(T_1) \neq \emptyset$. Suppose there was an $X - Y$-path $P$ in $G$ not using any vertex from $B_t \cap B_s$. By Lemma 3.7, $B^{-1}(P)$ is connected and hence there is a path in $T$ from $T_1$ to $T_2$ not using the edge $e$ (as $V(P) \cap B_t \cap B_s = \emptyset$), in contradiction to $T$ being a tree.

Part *(ii)* can be proved analogously. □

Recall from the preliminaries that for an edge $e := \{s, t\} \in E(T)$ we refer to the set $B_s \cap B_t$ as the *cut* at the edge $e$. The previous lemma gives justification to this terminology, as the cut at an edge separates the graph. A simple consequence of this lemma is the following observation, that will be useful later on.

**Corollary 3.14** *Let $G$ be a graph and $\mathcal{T} := (T, (B_t)_{t \in V(T)})$ be a tree-decomposition of $G$. If $X \subseteq V(G)$ is the vertex set of a complete subgraph of $G$, then there is a $t \in V(T)$ such that $X \subseteq B_t$.*

*Proof.* By Lemma 3.9, there is a small tree-decomposition $\mathcal{T}' := (T', (B'_t)_{t \in V(T')})$ such that $V(T') \subseteq V(T)$ and $B'_t = B_t$ for all $t \in V(T')$. Hence, w.l.o.g. we may assume that $\mathcal{T}$ is small.

By Lemma 3.13, every cut at an edge $e \in E(T)$ is a separator of the graph $G$. Hence, as $G[X]$ is complete, if $e \in E(T)$ and $T_1, T_2$ are the two components of $T - e$, then either $X \subseteq B(T_1)$ or $X \subseteq B(T_2)$ but not both. We orient every edge $e \in E(T)$ so that it points towards the component of $T - e$ containing all of $X$. As $T$ is acyclic, there is a node $t \in V(T)$ with no outgoing edge. By construction, $X \subseteq B_t$. □

**Corollary 3.15** $\operatorname{tw}(K_k) = k - 1$ *for all $k \geq 1$.*

**Algorithms and Complexity** The notion of tree-width has been introduced by Robertson and Seymour as part of their proof of the graph minor theorem. Even before that, the notion of *partial $k$-trees*, broadly equivalent to tree-width, had been studied in the algorithms community. The relevance of tree-width for algorithm design stems from the fact that the tree-structure inherent in tree-decompositions can be used to design bottom-up algorithms on graphs of small tree-width to solve problems efficiently which in general are NP-hard. A key step in designing these algorithms is to compute a tree-decomposition of the input graph. Unfortunately, Arnborg, Corneil, and Proskurowski showed that deciding the tree-width of a graph is NP-complete itself.

**Theorem 3.16** (*Arnborg, Corneil, Proskurowski* [3]) *The following problem is* NP-*complete.*

> TREE-WIDTH
> *Input:* Graph $G$, $k \in \mathbb{N}$.
> *Problem:* $\operatorname{tw}(G) = k$?

However, the problem becomes tractable if the tree-width is not a part of the input, i.e. if we are given a constant upper bound on the tree-width of graphs we are dealing with.

A class $\mathcal{C}$ of graphs has *bounded tree-width*, if there is a $k \in \mathbb{N}$ such that $\operatorname{tw}(G) \leq k$ for all $G \in \mathcal{C}$. In [6] Bodlaender proved that for any class of graphs of bounded tree-width tree-decompositions of minimal width can be computed in linear time.   *bounded tree-width*

**Theorem 3.17** (*Bodlaender* [6]) *There is an algorithm which, given a graph $G$ as input, constructs a tree-decomposition of $G$ of width $k := \operatorname{tw}(G)$ in time*

$$2^{\mathcal{O}(k^3)} \cdot |G|.$$

The algorithm by Bodlaender is primarily of theoretical interest. We will see later that many NP-complete problems can be solved efficiently on graph classes of bounded tree-width. For these algorithms to work in linear time, it is essential to compute tree-decompositions in linear time as well. From a practical point of view, however, the cubic dependence on the tree-width in the exponent and the complexity of the algorithm itself poses a serious problem. But there are other simpler algorithms with quadratic or cubic running time in the order of the graph but only linear exponential dependence on the tree-width which are practically feasible for small values of $k$.

### 3.2 Tree-Width and Structures

So far we have only considered graphs and their tree-decompositions. We will do so for most of the remainder, but at least want to comment on tree-decompositions of general structures. We first present the general definition of tree-decompositions of structures and then give an alternative characterisation in terms of the Gaifman- or comparability graph.

**Definition 3.18** *Let $\sigma$ be a signature. A* tree-decomposition *of a $\sigma$-structure $A$ is a pair $\mathcal{T} := (T, (B_t)_{t \in V(T)})$, where $T$ is a tree and $B_t \subseteq V(A)$ for all $t \in V(T)$, so that*

(i) *for all $a \in V(A)$ the set $B^{-1} := \{t \in V(T) : a \in B_t\}$ is non-empty and connected in $T$ and*

(ii) *for every $R \in \sigma$ and all $(a_1, \ldots, a_{\mathrm{ar}(R)}) \in R(A)^{\mathrm{ar}(R)}$ there is a $t \in V(T)$ such that $\{a_1, \ldots, a_{\mathrm{ar}(R)}\} \subseteq B_t$.*

*The width $\mathrm{w}(\mathcal{T})$ is defined as $\max\{|B_t| - 1 : t \in V(T)\}$ and the* tree-width *of $A$ is the minimal width of any of its tree-decompositions.*

The idea is the same as for graphs. We want the tree-decomposition to contain all elements of the structure and at the same time we want each tuple in a relation to be covered by a bag of the decomposition. It is easily seen that the tree-decompositions of a structure coincide with the tree-decompositions of its Gaifman graph, defined as follows.

$\mathcal{G}(A)$ **Definition 3.19 (Gaifman-graph)** *Let $\sigma$ be a signature. The* Gaifman-graph *$\mathcal{G}(A)$ of a $\sigma$-structure $A$ is defined as the graph $\mathcal{G}(A)$ with vertex set $V(A)$ and an edge between $a, b \in V(A)$ if, and only if, there is an $R \in \sigma$ and $\overline{a} \in R(A)$ with $a, b \in \overline{a}$.*

The following observation is easily seen.

**Proposition 3.20** *A structure has the same tree-decompositions as its Gaifman-graph.*

So far we have treated the notion of graphs informally as mathematical structures. As a preparation to the next section, we consider two different ways of modelling graphs by logical structures. The obvious way is to model a graph $G$ as a structure $A$ over the signature $\sigma_{\text{Graph}} := \{E\}$, where $V(A) := V(G)$ and $E(A) := \{(a,b) \in V(A) \times V(A) : \{a,b\} \in E(G)\}$. We write $A(G)$ for this encoding of a graph as a structure and refer to it as the *standard encoding*.

$\sigma_{\text{Graph}}$
$A(G)$

Alternatively, we can model the *incidence graph* of a graph $G$ defined as the graph $G_{\text{Inc}}$ with vertex set $V(G) \cup E(G)$ and edges $E(G_{\text{Inc}}) := \{(v,e) : v \in V(G), e \in E(G), v \in e\}$. The incidence graph gives rise to the following encoding of a graph as a structure, which we refer to as the *incidence encoding*.

*incidence graph*

**Definition 3.21** *Let $G := (V, E)$ be a graph. Let $\sigma_{\text{inc}} := \{P_V, P_E, I\}$, where $P_V, P_E$ are unary predicates and $I$ is a binary predicate. The* incidence structure $A_I(G)$ *is defined as the $\sigma_{\text{inc}}$-structure $A := A_I(G)$ where $V(A) := V \cup E$, $P_E(A) := E$, $P_V(A) := V$ and*

$$I(A) := \{(v,e) : v \in V, e \in E, v \in e\}.$$

The proof of the following lemma is straightforward but may be a good exercise.

**Theorem 3.22** $\text{tw}(G) = \text{tw}(A_I(G))$ *for all graphs $G$.*

It may seem to be a mere technicality how we encode a graph as a structure. However, the precise encoding has a significant impact on the expressive power of logics on graphs. For instance, the following $\text{MSO}[\sigma_{\text{inc}}]$-formula defines that a graph contains a Hamilton-cycle using the incidence encoding, a property that is not definable in MSO on the standard encoding (see e.g. [37, Corollary 6.3.5]).

$$\exists U \subseteq P_E \forall v \text{``}v \text{ has degree 2 in } G[U]\text{''} \wedge \varphi_{\text{conn}}(U),$$

where $\varphi_{\text{conn}}$ is a formula saying that the subgraph $G[U]$ induced by $U$ is connected. Clearly, it is MSO-definable that a vertex $v$ is incident to exactly two edges in $U$, i.e. has degree 2 in $G[U]$. The formula says that there is a set $U$ of edges so that $G[U]$ is connected and that every vertex in $G[U]$ has degree 2. But this means that $U$ is a simple cycle $\mathcal{P}$ in $G$. Further, as all vertices of $G$ occur in $\mathcal{P}$, this cycle must be Hamiltonian.

Hence, MSO is more expressive over incidence graphs than over the standard encoding of graphs. It is clear that MSO interpreted over incidence graphs is the same as considering the extension of MSO by quantification over sets of edges (rather than just sets of vertices) on the standard encoding. This logic is sometimes referred to as $\text{MSO}_2$ in the literature. A more general framework are *guarded logics*, that allow quantification only over tuples that occur together in some relation in the structure. On graphs, *guarded second-order logic* (GSO) is just $\text{MSO}_2$. As we will not be dealing with general structures in the rest of this survey, we refrain from introducing guarded logics formally and refer to [2,51] and references therein instead.

$\text{MSO}_2$

GSO

### 3.3 Coding tree-decompositions in trees

The aim of the following sections is to show that model-checking and satisfiability testing for monadic second-order logic becomes tractable when restricted to graph classes of small tree-width. The proof of these results relies on a reduction from graph classes of bounded tree-width to classes of finite labelled trees. As a first step towards this we show how graphs of tree-width bounded by some constant $k$ can be encoded in $\Sigma_k$-labelled finite trees for a suitable alphabet $\Sigma_k$ depending on $k$. We will also show that the class of graphs of tree-width $k$, for some $k \in \mathbb{N}$, is MSO-interpretable in the class of $\Sigma_k$-labelled trees.

A tree-decomposition $(T, (B_t)_{t \in V(T)})$ of a graph $G$ is already a tree and we will take $T$ as the underlying tree of the encoding. Thus, all we have to do is to define the labelling. Note that we cannot simply take the bags $B_t$ as labels, as we need to work with a finite alphabet and there is no a priori bound on the number of vertices in the bags. Hence we have to encode the vertices in the bags using a finite number of labels. To simplify the presentation we will be using tree-decompositions of a special form.

leaf-decomposition **Definition 3.23** *A* leaf-decomposition *of a graph $G$ is a tree-decomposition $\mathcal{T} := (T, (B_t)_{t \in V(T)})$ of $G$ such that all leaves of $V(T)$ contain exactly one vertex and every $v \in V(G)$ is contained in exactly one leaf of $T$.*

In other words, in leaf-decompositions there is a bijection $\rho$ between the set of leaves of the decomposition and the set of vertices of the graph and the bag $B_t$ of a leaf $t$ contains exactly its image $\rho(t)$. It is easily seen that any tree-decomposition can be converted into a leaf-decomposition of the same width.

**Lemma 3.24** *For every tree-decomposition $\mathcal{T}$ of a graph $G$ there is a leaf-decomposition $\mathcal{T}'$ of $G$ of the same width and this can be computed in linear time, given $\mathcal{T}$.*

To define the alphabet $\Sigma_k$, we will work with a slightly different form of tree-decompositions where the bags are no longer sets but ordered tuples of vertices. It will also be useful to require that all these tuples have the same length and that the tree underlying a tree-decomposition is a binary directed tree.[3]

**Definition 3.25** *An* ordered tree-decomposition *of width $k$ of a graph $G$ is a pair $(T, (\bar{b}_t)_{t \in V(T)})$, where $T$ is a directed binary tree and $\bar{b}_t \in V(G)^k$, so that $(T, (B_t)_{t \in V(T)})$ is a tree-decomposition of $G$, with $B_t := \{b_0, \ldots, b_k\}$ for $\bar{b}_t := b_0, \ldots, b_k$.*

An ordered leaf-decomposition is the ordered version of a leaf-decomposition.

**Example 3.26** *Consider again the graph from Example 3.2. The following shows an ordered leaf-decomposition obtained from the tree-decomposition in Example 3.2 by first adding the necessary leaves containing just one vertex and then converting every bag into an ordered tuple of length 4.*

---

[3] Note that, strictly speaking, to apply the results on MSO on finite trees we have to work with trees where an ordering on the children of a node is imposed. Clearly we can change all definitions here to work with such trees. But as this would make the notation even more complicated, we refrain from doing so.

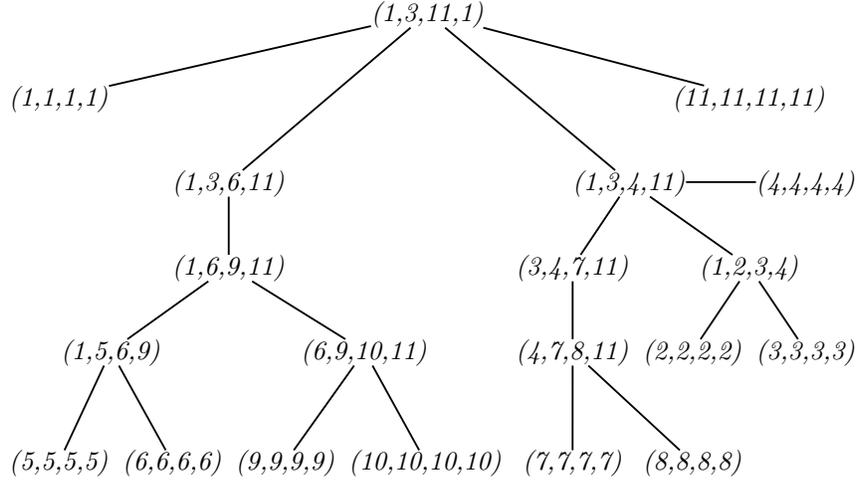

The graph $G$ together with this leaf-decomposition induces the following $\Sigma_3$-labelled tree:

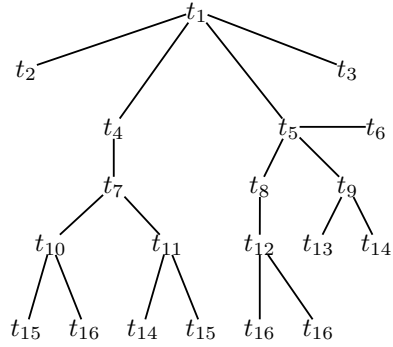

where, for instance, $\lambda(t_4) := \bigl(\mathrm{eq}(t_4), \mathrm{overlap}(t_4), \mathrm{edge}(t_4)\bigr)$, with

- $\mathrm{eq}(t_4) := \varnothing$,
- $\mathrm{overlap}(t_4) := \{(0,0),(0,3),(1,1)\}$, and
- $\mathrm{edge}(t_4) := \{(0,1),(1,2),(1,3),(2,3)\} \cup \{(1,0),(2,1),(3,1),(3,2)\}$.

$\mathrm{eq}(t_4) := \varnothing$, as all positions of $\overline{b}_{t_4}$ correspond to different vertices in $G$. On the other hand, $\mathrm{eq}(t_{15}) := \{(i,j) : i,j \in \{0,\ldots,3\}\}$, as all entries of $\overline{b}_{15}$ refer to the same vertex 5. ⊣

It is easily seen that every tree-decomposition of width $k$ can be converted in linear time to an ordered tree-decomposition of width $k$. Combining this with Bodlaender's algorithm (Theorem 3.17) and Lemma 3.24 above yields the following lemma.

**Lemma 3.27** *There is an algorithm that, given a graph $G$ of tree-width $\leq k$, constructs an ordered leaf-decomposition of $G$ of width $\mathrm{tw}(G)$ in time $2^{\mathcal{O}(k^3)} \cdot |G|$.*

$\lambda(t)$

Now let $G$ be a graph and $\mathcal{L} := (T', (\bar{b}_t)_{t \in V(T')})$ be an ordered leaf-decomposition of $G$ of width $k$. We code $\mathcal{L}$ in a labelled tree $\mathcal{T} := (T, \lambda)$, so that $\mathcal{L}$ and $G$ can be reconstructed from $\mathcal{T}$, and this reconstruction can even be done by MSO formulas.

The tree $T$ underlying $\mathcal{T}$ is the tree $T'$ of $\mathcal{L}$. To define the alphabet and the labels of the nodes let $t \in V(T)$ and let $\bar{b}_t := b_0, \ldots, b_k$. We set

$$\lambda(t) := (\text{eq}(t), \text{overlap}(t), \text{edge}(t))$$

where $\text{eq}(t), \text{overlap}(t), \text{edge}(t)$ are defined as follows:

$\text{eq}(t)$
- $\text{eq}(t) := \{(i, j) : 0 \leq i, j \leq k \text{ and } b_i = b_j\}$.
- If $t$ is the root of $T$, then $\text{overlap}(t) := \varnothing$. Otherwise let $p$ be the predecessor

$\text{overlap}(t)$
of $t$ in $T$ and let $\bar{b}_p := a_0, \ldots, a_k$. We set

$$\text{overlap}(t) := \{(i, j) : 0 \leq i, j \leq k \text{ and } b_i = a_j\}.$$

$\text{edge}(t)$
- Finally, $\text{edge}(t) := \{(i, j) : 0 \leq i, j \leq k \text{ and } \{b_i, b_j\} \in E(G)\}$.

$\Sigma_k$
For every fixed $k$, the labels come from the finite alphabet

$$\Sigma_k := 2^{\{0,\ldots,k\}^2} \times 2^{\{0,\ldots,k\}^2} \times 2^{\{0,\ldots,k\}^2}.$$

$\mathcal{T}(G, \mathcal{L})$
We write $\mathcal{T}(G, \mathcal{L})$ for the labelled tree encoding a leaf-decomposition $\mathcal{L}$ of a graph $G$. Note that the signature depends on the arity $k$ of the ordered leaf-decomposition $\mathcal{L}$, i.e. on the bound on the tree-width of the class of graphs we are working with.

The individual parts of the labelling have the following meaning. Recall that we require all tuples $\bar{b}_t$ to be of the same length $k + 1$ and therefore they may contain duplicate entries. $\text{eq}(t)$ identifies those entries in a tuple relating to the same vertex of the graph $G$. The label $\text{overlap}(t)$ takes care of the same vertex appearing in tuples of neighbouring nodes of the tree. As we are working with directed trees, every node other than the root has a unique predecessor. Hence we can record in the overlap-label of the child which vertices in its bag occur at which positions of its predecessor. Finally, edge encodes the edge relation of $G$. As every edge is covered by a bag of the tree-decomposition, it suffices to record for each node $t \in V(T)$ the edges between elements of its bag $\bar{b}_t$.

The labels $\text{eq}(t), \text{overlap}(t)$ and $\text{edge}(t)$ satisfy some obvious consistency criteria, e.g. $\text{eq}(t)$ is an equivalence relation for every $t$, $\text{eq}(t)$ is consistent with $\text{edge}(t)$ in the sense that if two positions $i, i'$ refer to the same vertex, i.e. $(i, i') \in \text{eq}(t)$ and $(i, j) \in \text{edge}(t)$ then also $(i', j) \in \text{edge}(t)$, and likewise for $\text{eq}(t)$ and $\text{overlap}(t)$. We refrain from giving all necessary details. Note, though, that any $\Sigma_k$-labelled finite tree that satisfies these consistency criteria does encode a graph of tree-width at most $k$. Furthermore, the criteria as outlined above are easily seen to be definable in MSO, in fact even in first-order logic. Again we refrain from giving the exact formula as its definition is long and technical

$\varphi_{cons}$
but absolutely straightforward. Let $\varphi_{cons}$ be the MSO-sentence true in a $\Sigma_k$-labelled tree if, and only if, it satisfies the consistency criteria, i.e. encodes a tree-decomposition of a graph of tree-width at most $k$.

Of course, to talk about formulas defining properties of $\Sigma_k$-labelled trees we first need to agree on how $\Sigma_k$-labelled trees are encoded as structures. For $k \in \mathbb{N}$ we define the signature $\qquad\qquad\sigma_k$

$$\sigma_k := \{E\} \cup \{\mathrm{eq}_{i,j}, \mathrm{edge}_{i,j}, \mathrm{overlap}_{i,j} : 0 \leq i, j \leq k\},$$

where $\mathrm{eq}_{i,j}, \mathrm{overlap}_{i,j}$, and $\mathrm{edge}_{i,j}$ are unary relation symbols. The intended meaning of $\mathrm{eq}_{i,j}$ is that in a $\sigma_k$-structure $A$ an element $t$ is contained in $\mathrm{eq}_{i,j}(A)$ if $(i,j) \in \mathrm{eq}(t)$ in the corresponding tree. Likewise for $\mathrm{overlap}_{i,j}$ and $\mathrm{edge}_{i,j}$. $\sigma_k$-structures, then, encode $\Sigma_k$-labelled trees in the natural way. In the sequel, we will not distinguish notationally between a $\Sigma_k$-labelled tree $\mathcal{T}$ and the corresponding $\sigma_k$-structure $A_{\mathcal{T}}$. In particular, we will write $\mathcal{T} \models \varphi$, for an MSO-formula $\varphi$, instead of $A_{\mathcal{T}} \models \varphi$.

Clearly, the information encoded in the $\Sigma_k$-labelling is sufficient to reconstruct the graph $G$ from a tree $\mathcal{T}(G, \mathcal{L})$, for some ordered leaf-decomposition of $G$ of width $k$. Note that different leaf-decompositions of $G$ may yield non-isomorphic trees. Hence, the encoding of a graph in a $\Sigma_k$-labelled tree is not unique but depends on the decomposition chosen. For our purpose this does not pose any problem, though.

The next step is to define an MSO-interpretation

$$\Gamma := (\varphi_{univ}(x), \varphi_{valid}, \varphi_E(x,y)) \qquad\qquad \Gamma$$

of the class $\mathcal{T}_k$ of graphs of tree-width at most $k$ in the class $\mathcal{T}_{\Sigma_k}$ of $\Sigma_k$-labelled finite trees. To state the interpretation formally, we need to define the three formulas $\varphi_{univ}(x), \varphi_{valid}$, and $\varphi_E(x,y)$. Recall that in a leaf-decomposition $\mathcal{L}$ there is a bijection between the leaves of $\mathcal{T}$ and the vertices of the graph that is being decomposed. Hence, we can take $\varphi_{univ}(x)$ to be the formula

$$\varphi_{univ}(x) := \forall y \neg Exy$$

saying that $x$ is a leaf in $\mathcal{T}$.

Let $G$ be a graph and $\mathcal{L} := (T, (\bar{b}_t)_{t \in V(T)})$ be an ordered leaf-decomposition of $G$ of width $k$. Suppose we are given two leaves $t_u, t_v$ of $\mathcal{L}$ containing $u$ and $v$ respectively and we want to decide whether there is an edge between $u$ and $v$. Clearly, if $e := \{u, v\} \in E(G)$, then $e$ must be covered by some bag, i.e. there are a node $t$ in $\mathcal{L}$ with bag $\bar{b}_t := b_0 \ldots b_k$ and $i \neq j$ such that $b_i = u$ and $b_j = v$ and $(i,j) \in \mathrm{edge}(t)$ in the tree $\mathcal{T} := \mathcal{T}(G, \mathcal{L})$. Further, $u$ occurs in every bag on the path from $t$ to $t_u$ and likewise for $v$. Hence, to define $\varphi_E(x,y)$, where $x, y$ are interpreted by leaves, we have to check whether there is such a node $t$ and paths from $x$ and $y$ to $t$ as before. For this, we need an auxiliary formula which we define next.

Recall that each position $i$ in a bag $\bar{b}_t$ corresponds to a vertex in $G$. Hence, we can associate vertices with pairs $(t,i)$. In general, a vertex can occur at different positions $i$ and different nodes $t \in V(T)$. We can, however, identify any vertex $v$ with the set

$$X_v := \{(t,i) : t \in V(T) \text{ and } v \text{ occurs at position } i \text{ in } \bar{b}_t \}. \qquad X_v$$

We call $X_v$ the *equivalence set* of $v$. If $t \in V(T)$ and $0 \le i \le k$, we define the *equivalence set* of $(t,i)$ as the equivalence set of $b_i$, where $\overline{b}_t := b_0, \ldots, b_k$.

Clearly, this identification of vertices with sets of pairs and the concept of equivalent sets extends to the labelled tree $\mathcal{T} := \mathcal{T}(G, \mathcal{L})$, as $\mathcal{T}$ and $\mathcal{L}$ share the same underlying tree.

To define sets $X_v$ in MSO, we represent $X_v$ by a tuple $\overline{X} := (X_0, \ldots, X_k)$ of sets $X_i \subseteq V(T)$, such that for all $0 \le i \le k$ and all $t \in V(T)$, $t \in X_i$ if, and only if, $(t,i) \in X_v$.

We are going to describe an MSO-formula $\varphi(X_0, \ldots, X_k)$ that is satisfied by a tuple $\overline{X}$ if, and only if, $\overline{X}$ is the equivalence set of a pair $(t,i)$, or equivalently of a vertex $v \in V(G)$. To simplify notation, we will say that a tuple $\overline{X}$ contains a pair $(t,i)$ if $t \in X_i$. Consider the formulas

$$\psi_{\text{eq}}(X_0, \ldots, X_k) := \bigwedge_i \forall t \in X_i \big( \bigwedge_{j \ne i} \text{eq}_{i,j}(t) \to t \in X_j \big)$$

and

$$\psi_{\text{overlap}}(X_0, \ldots, X_k) := \forall s \forall t \bigwedge_{i,j} \Big( E(s,t) \wedge t \in X_i \wedge \text{overlap}_{i,j}(t) \Big) \to s \in X_j.$$

$\psi_{\text{eq}}(\overline{X})$ says of a tuple $\overline{X}$ that $\overline{X}$ is closed under the eq-labels and $\psi_{\text{overlap}}(\overline{X})$ says the same of the overlap-labels. Now let $\psi(\overline{X}) := \psi_{\text{eq}} \wedge \psi_{\text{overlap}}$. $\psi$ is satisfied by a tuple $\overline{X}$ if whenever $\overline{X}$ contains at a pair $(t,i)$, then it contains the complete equivalence set of $(t,i)$. Now, consider the formula

$\varphi_{vertex}$

$$\varphi_{vertex}(X_0, \ldots, X_k) := \psi(\overline{X}) \wedge \overline{X} \ne \varnothing \wedge \forall \overline{X}' \ne \varnothing \big( \overline{X}' \subsetneq \overline{X} \to \neg \psi(\overline{X}') \big)$$

where "$\overline{X} \ne \varnothing$" defines that at least one $X_i$ is non-empty and "$\overline{X}' \subsetneq \overline{X}$" is an abbreviation for a formula saying that $X'_i \subseteq X_i$, for all $i$, and for at least one $i$ the inclusion is strict.

$\varphi_{vertex}(\overline{X})$ is true for a tuple if $\overline{X}$ is non-empty, closed under eq and overlap, but no proper non-empty subset of $\overline{X}$ is. Hence, $\overline{X}$ is the equivalence set of a single vertex $v \in V(G)$. The definition of $\varphi_{vertex}(\overline{X})$ is the main technical part of the MSO-interpretation $\varGamma := (\varphi_{univ}(x), \varphi_{valid}, \varphi_E(x,y))$.

We have already defined $\varphi_{univ}(x) := \forall y \neg Exy$. For $\varphi_{valid}$, recall from above the formula $\varphi_{cons}$ true in a $\Sigma_k$-labelled tree $\mathcal{T}$ if, and only if, $\mathcal{T}$ encodes a tree-decomposition of a graph $G$ of tree-width at most $k$. To define $\varphi_{valid}$ we need a formula that not only requires $\mathcal{T}$ to encode a tree-decomposition of $G$ but a leaf-decomposition.

To force the encoded tree-decomposition to be a leaf-decomposition, we further require the following two conditions.

1. For all leaves $t \in V(T)$ and all $i \ne j$, $(i,j) \in \text{eq}(t)$.
2. For all $t \in V(T)$ and all $0 \le i \le k$ the equivalence set of $(t,i)$ contains exactly one leaf.

Both conditions can easily be defined by MSO-formulas $\varphi_1$ and $\varphi_2$, respectively, where in the definition of $\varphi_2$ we use the formula $\varphi_{vertex}$ defined above.

Hence, the formula

$$\varphi_{valid} := \varphi_{cons} \wedge \varphi_1 \wedge \varphi_2 \qquad\qquad \varphi_{valid}$$

is true in a $\Sigma_k$-labelled tree $\mathcal{T}$ (or the corresponding $\sigma_k$-structure) if, and only if, $\mathcal{T}$ encodes a leaf-decomposition of width $k$.

Finally, we define the formula $\varphi_E(x,y)$ saying that there is an edge between $x$ and $y$ in the graph $G$ encoded by a $\Sigma_k$-labelled tree $\mathcal{T} := (T, \lambda)$. Note that there is an edge in $G$ between $x$ and $y$ if, and only if, there is a node $t \in V(T)$ and $0 \leq i \neq j \leq k$ such that $(i,j) \in \text{edge}(t)$ and $x$ is the unique leaf in the equivalence set of $(t,i)$ and $y$ is the unique leaf in the equivalence set of $(t,j)$. This is formalised by

$$\varphi_E(x,y) := \exists t \bigvee_{i \neq j} \begin{pmatrix} \text{edge}_{i,j}(t) \wedge \exists \overline{X} \exists \overline{Y} \varphi_{vertex}(\overline{X}) \wedge \varphi_{vertex}(\overline{Y}) \wedge \\ X_1(x) \wedge Y_1(y) \wedge X_i(t) \wedge Y_j(t) \end{pmatrix}.$$

This completes the definition of $\Gamma$. Now, the proof of the following lemma is immediate.

**Lemma 3.28** *Let $G$ be a graph of tree-width $\leq k$ and $\mathcal{L}$ be a leaf-decomposition of $G$ of width $k$. Let $\mathcal{T} := \mathcal{T}(G, \mathcal{L})$ be the tree-encoding of $\mathcal{L}$ and $G$. Then $G \cong \Gamma(\mathcal{T})$.*

*Further, by the interpretation lemma, for all MSO-formulas $\varphi$ and all $\Sigma_k$-trees $\mathcal{T} \models \varphi_{valid}$,*

$$\mathcal{T} \models \Gamma(\varphi) \iff \Gamma(\mathcal{T}) \models \varphi.$$

## 3.4 Courcelle's Theorem

In this section and the next we consider computational problems for monadic second-order logic on graph classes of small tree-width. The algorithmic theory of MSO on graph classes of small tree-width has, essentially independently, been developed by Courcelle, Seese and various co-authors. We first consider the model-checking problem for MSO and present Courcelle's theorem. We then state a similar theorem by Arnborg, Lagergreen and Seese concerning the *evaluation problem* of MSO. In the next section, we consider the satisfiability problem and prove Seese's theorem.

**Theorem 3.29 (***Courcelle* [13]**)** *The problem*

> MC(MSO, tw)
>    *Input:* Graph $G$, $\varphi \in$ MSO
>   *Parameter:* $|\varphi| + \text{tw}(G)$
>    *Problem:* $G \models \varphi$?

is fixed parameter tractable and can be solved in time $f(|\varphi|) + 2^{p(\mathrm{tw}(G))} \cdot |G|$, for a polynomial p and a computable function $f : \mathbb{N} \to \mathbb{N}$.

That is, the model-checking problem for a fixed formula $\varphi \in \mathrm{MSO}$ can be solved in linear time on any class of graphs of bounded tree-width.

*Proof.* Let $\mathcal{C}$ be a class of bounded tree-width and let $k$ be an upper bound for the tree-width of $\mathcal{C}$. Let $\varphi \in \mathrm{MSO}$ be given.

On input $G \in \mathcal{C}$ we first compute an ordered leaf-decomposition $\mathcal{L}$ of $G$ of width $k$. From this, we compute the tree $\mathcal{T} := \mathcal{T}(G, \mathcal{L})$. We then check whether $\mathcal{T} \models \Gamma(\varphi)$, where $\Gamma$ is the MSO-interpretation of the previous section.

Correctness of the algorithm follows from Lemma 3.28. The time bounds follow from Lemma 3.24 and the fact that MSO model-checking is in linear time (for a fixed formula) on the class of trees (see e.g. [61, Chapter 7] or [46, Chapter 10]). □

We will see a different proof of this theorem using logical types later when we prove Lemma 7.13. The result immediately implies that parametrized problems such as the independence set or dominating set problem or problems such as 3-colourability and Hamiltonicity are solvable in linear time on classes of graphs of bounded tree-width.

Without proof we state the following extension of Courcelle's theorem which essentially follows from [4]. The proof uses the same methods as described above and the corresponding result for trees.

**Theorem 3.30** (*Arnborg, Lagergreen, Seese* [4]) *The problem*

> *Input:* Graph $G$, $\varphi(X) \in \mathrm{MSO}$, $k \in \mathbb{N}$.
> *Parameter:* $|\varphi| + \mathrm{tw}(G)$.
> *Problem:* Determine whether there is a set $S \subseteq V(G)$ such that $G \models \varphi(S)$ and $|S| \leq k$ and compute one if it exists.

*is fixed-parameter tractable and can be solved by an algorithm with running time* $f(|\varphi|) + 2^{p(\mathrm{tw}(G))} \cdot |G|$, *for a polynomial p and a computable function* $f : \mathbb{N} \to \mathbb{N}$.

Recall that by the results discussed in Section 3.2 the previous results also hold for MSO on incidence graphs, i.e. $\mathrm{MSO}_2$ where quantification over sets of edges is allowed also.

**Corollary 3.31** *The results in Theorem 3.29 and 3.30 extend to* $\mathrm{MSO}_2$.

### 3.5 Seese's Theorem

We close this section with another application of the interpretation defined in Section 3.3. Recall that $\mathrm{MSO}_2$ has set quantification over sets of vertices as well as sets of edges and corresponds to MSO interpreted over the incidence encoding of graphs.

**Theorem 3.32** (*Seese* [79]) *Let $k \in \mathbb{N}$ be fixed. The $\mathrm{MSO}_2$-theory of the class of graphs of tree-width at most $k$ is decidable.*

*Proof.* Let $\Gamma := (\varphi_{univ}, \varphi_{valid}, \varphi_E)$ be the interpretation defined in Section 3.3. On input $\varphi$ we first construct the formula $\varphi^* := \Gamma(\varphi)$. Using the decidability of the MSO-theory of finite labelled trees, we then test whether there is a $\Sigma_k$-labelled tree $\mathcal{T}$ such that $\mathcal{T} \models \varphi_{valid} \wedge \varphi^*$.

If there is such a tree $\mathcal{T}$, then, as $\mathcal{T} \models \varphi_{valid}$, there is a graph $G$ of tree-width at most $k$ encoded by $\mathcal{T}$ which satisfies $\varphi$. Otherwise, $\varphi$ is not satisfiable by any graph of tree-width at most $k$. □

Again without proof, we remark that the following variant of Seese's theorem is also true.

**Theorem 3.33** (*Adler, Grohe, Kreutzer* [1]) *For every $k$ it is decidable whether a given MSO-formula is satisfied by a graph of tree-width exactly $k$.*

We remark that there is a kind of converse to Seese's theorem which we will prove in Section 6 below.

**Theorem 3.34** (*Seese* [79]) *If $\mathcal{C}$ is a class of graphs with a decidable $\mathrm{MSO}_2$-theory, then $\mathcal{C}$ has bounded tree-width.*

The proof of this theorem relies on a result proved by Robertson and Seymour as part of their proof of the graph minor theorem. We will present the graph theory needed for this in Section 5 and a proof of Theorem 3.34 in Section 6.

## 4 From Trees to Cliques

In the previous section we considered graphs that are sufficiently tree-like so that efficient model-checking algorithms for monadic second-order logic can be devised following the tree-structure of the decomposition. On a technical level these results rely on Feferman-Vaught style results allowing to infer the truth of an MSO sentence in a graph from the MSO types of the smaller subgraphs it can be decomposed into. In this section we will see a different property of graphs that also allows for efficient MSO model-checking. It is not based on the idea of decomposing the graph into smaller parts of lower complexity, but instead it is based on the idea of the graphs being *uniform* in some way, i.e. not having too many types of its vertices.

As a first example let us consider the class $\{K_n : n \in \mathbb{N}\}$ of cliques. Obviously, these graphs have as many edges as possible and cannot be decomposed in any meaningful way into parts of lower complexity. However, model-checking for first-order logic or monadic second-order logic is simple, as all vertices look the same. In a way, a clique is no more complex than a set: the edges do not impose any meaningful structure on the graph. This intuition is generalised by the notion of *clique-width* of a graph. It was originally defined in terms of graph grammars by Courcelle, Engelfriet and Rozenberg [17]. Independently, Wanke introduced $k$-NLC graphs, a notion that is equivalent to Courcelle et al.'s definition up to a factor of 2. The term clique-width was introduced in [19]. Clique-decompositions (or $k$-expressions as they are called) are useful for the design of algorithms, as they again provide a tree-structure along which algorithms can work. However, until recently algorithms using clique-decompositions had to be given the decomposition as input, as no fixed-parameter algorithms were known to compute the decomposition.

In 2006, Oum and Seymour [69] introduced the notion of *rank-width* and corresponding *rank-decompositions*, a notion that is broadly equivalent to clique-width in the sense that for every class of graphs, one is bounded if, and only if, the other is bounded. Rank-decompositions can be computed by fpt-algorithms parametrized by the width and from a rank-decomposition a clique-decomposition can be generated. In this way, the requirement of algorithms being given the decomposition as input has been removed. But rank-decompositions are also in many other ways the more elegant notion.

We first recall the definition of clique-width in Section 4.1. In Section 4.2, we then introduce general rank-decompositions of submodular functions, of which the rank-width of a graph is a special case. As a side effect, we also obtain the notion of branch-width, which is another elegant characterisation of tree-width. Model-checking algorithms for MSO on graph classes of bounded rank-width are presented in Section 4.3, where we also consider the satisfiability problem for MSO and a conjecture by Seese.

## 4.1 Clique-Width

*k-expression*    **Definition 4.1 ($k$-expression)** *Let $k \in \mathbb{N}$ be fixed. The set of $k$-expressions is inductively defined as follows:*

  *(i) $\mathbf{i}$ is a $k$-expression for all $i \in [k]$.*
  *(ii) If $i \neq j \in [k]$ and $\varphi$ is a $k$-expression, then so are $\text{edge}_{i-j}(\varphi)$ and $\text{rename}_{i \to j}(\varphi)$.*
  *(iii) If $\varphi_1, \varphi_2$ are $k$-expressions, then so is $(\varphi_1 \oplus \varphi_2)$.*

A $k$-expression $\varphi$ generates a graph $G(\varphi)$ coloured by colours from $[k]$ as
**i**    follows: The $k$-expression $\mathbf{i}$ generates a graph with one vertex coloured by the colour $i$ and no edges.

$\text{edge}_{i-j}$    The expression $\text{edge}_{i-j}$ is used to add edges. If $\varphi$ is a $k$-expression generating the coloured graph $G := G(\varphi)$ then $\text{edge}_{i-j}(\varphi)$ defines the graph $H$ with $V(H) := V(G)$ and

$$E(H) := E(G) \cup \big\{\{u,v\} : u \text{ has colour } i \text{ and } v \text{ has colour } j\big\}.$$

Hence, $\text{edge}_{i-j}(\varphi)$ adds edges between all vertices with colour $i$ and all vertices with colour $j$.

$\text{rename}_{i \to j}(\varphi)$    The operation $\text{rename}_{i \to j}(\varphi)$ recolours the graph. Given the graph $G$ generated by $\varphi$, the $k$-expression $\text{rename}_{i \to j}(\varphi)$ generates the graph obtained from $G$ by giving all vertices which have colour $i$ in $G$ the colour $j$ in $H$. All other vertices keep their colour.

Finally, if $\varphi_1, \varphi_2$ are $k$-expressions generating coloured graphs $G_1, G_2$ respectively, then $(\varphi_1 \oplus \varphi_2)$ defines the disjoint union of $G_1$ and $G_2$.

We illustrate the definition by an example.

**Example 4.2** *Consider again the graph from Example 3.2 depicted in Figure 3. For convenience, the graph is repeated below. We will show how this graph can*

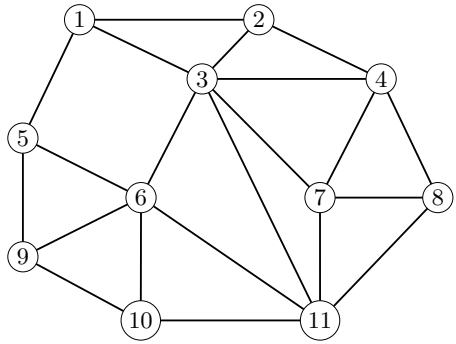

**Fig. 4.** Graph from Example 3.2

*be obtained by a 6-expression.*

*Consider the expression $\varphi_0$ in Figure 5, which generates the graph in Figure 6 a). The labels in the graph represent the colours. Here we use obvious abbreviations such as $\mathrm{edge}_{i-j,s-t}$ to create edges between $i$ and $j$ as well as edges between $s$ and $t$ in one step.*

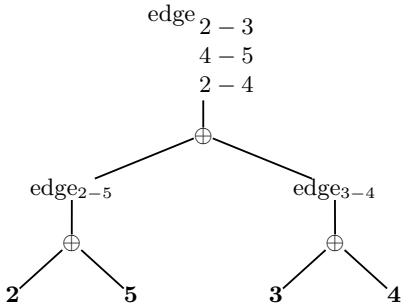

**Fig. 5.** The 6-expression $\varphi_0$ generating the graph in Fig. 6 a)

*The vertices generated so far correspond to the vertices $5, 6, 9, 10$ of the graph in Figure 4. Note that we have already created all edges incident to vertex $9$. Hence, in the construction of the rest of the graph, the vertex $9$ (having colour $2$) does not have to be considered any more. We will use the colour $0$ to mark vertices that will not be considered in further steps of the k-expression. Let $\varphi_1 := \mathrm{rename}_{2\to 0}(\varphi_0)$ be the 6-expression that generates the graph in Figure 6 a), but where the vertex with colour $2$ now has colour $0$.*

*The next step is to generate the vertex $11$ of the graph. This is done by the expression $\varphi_2 := \mathrm{rename}_{5\to 0}\Big(\mathrm{edge}_{1-5,1-4}\big(\mathbf{1}\oplus\varphi_1\big)\Big)$. We proceed by adding the vertices $1$ and $3$ and the appropriate edges. Let*

$$\varphi_3 := \mathrm{rename}_{3\to 0, 4\to 0}\mathrm{edge}_{2-3,4-5,1-5}\Big(\varphi_2 \oplus \big(\mathrm{edge}_{2-5}(\mathbf{2}\oplus\mathbf{5})\big)\Big)$$

*This generates the graph depicted in Figure 6 b). The next step is to add the vertices $7$ and $8$. Let*

$$\varphi_4 := \mathrm{rename}_{1\to 0}\mathrm{edge}_{1-3,1-4,3-5}\big(\varphi_3 \oplus \mathrm{edge}_{3-4}(\mathbf{3}\oplus\mathbf{4})\big)$$

*Finally, we add the vertex $2$ and rename the colour of the vertex $2$ to $0$, i.e. essentially remove the colour, and rename all other colours to $1$.*

$$\varphi_5 := \mathrm{rename}_{2\to 0, 5\to 1, 3\to 1, 4\to 1}\mathrm{edge}_{1-2,1-5}(\mathbf{1}\oplus\varphi_4)$$

*This generates the graph in Figure 6 c).*

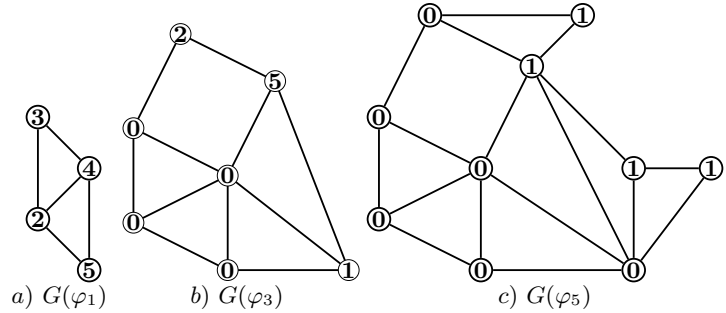

**Fig. 6.** Graphs generated by the 6-expressions in Example 4.2

*Finally, we add the vertex $4$ and edges to all other vertices marked by the colour $1$.*

*The complete expression generating the graph is therefore $\mathrm{edge}_{1-2}(\mathbf{2}\oplus\varphi_5)$.*
⊣

It is easily seen that every finite graph can be generated by a $k$-expression for some $k \in \mathbb{N}$. Just choose a colour for each vertex and add edges accordingly.

**Lemma 4.3** *Every finite graph can be generated[4] by a k-expression for some $k \in \mathbb{N}$.*

Hence, the following concepts are well defined.

**Definition 4.4** *The* clique-width $\mathrm{cw}(G)$ *of a graph $G$ is defined as the least $k \in \mathbb{N}$ such that $G$ can be generated by a k-expression. A class $\mathcal{C}$ of graphs has bounded clique-width if there is a $k \in \mathbb{N}$ such that $\mathrm{cw}(G) \leq k$ for all $G \in \mathcal{C}$.*   clique-width

We give a few more examples.

**Example 4.5** 1. *The class of cliques has clique-width* 2. *(Clique-width* 2, *as the* $\mathrm{edge}_{i,j}$ *operator requires $i \neq j$ to avoid self-loops).*
2. *The class of all trees has clique-width* 3. *By induction on the height of the trees we show that for each tree $T$ there is a 3-expression generating this tree so that the root is coloured by the colour* **1** *and all other nodes are coloured by* **0**. *This is trivial for trees of height* 0. *Suppose $T$ is a tree of height $n+1$ with root $r$ and successors $v_1, \ldots, v_k$ of $r$. For $1 \leq i \leq k$ let $\varphi_i$ be a 3-expression generating the subtree of $T$ rooted at $v_i$. Then $T$ is generated by the expression*

$$\mathrm{rename}_{2 \to 1} \mathrm{rename}_{1 \to 0} \mathrm{edge}_{2-1}(\mathbf{2} \oplus \varphi_1 \oplus \cdots \oplus \varphi_k).$$

3. *It can be shown that the clique-width of the $(n \times n)$-grid is $\Omega(n)$. (This follows, for instance, from Theorem 4.7 below).* ⊣

The next theorem due to Wanke and also Courcelle and Olariu relates clique-width to tree-width.

**Theorem 4.6** ([89,19]) *Every graph of tree-width at most $k$ has clique-width at most $2^{k+1} + 1$.*

As the examples above show, there is no hope to bound the tree-width of a graph in terms of its clique-width. Hence, clique-width is more general than tree-width in the sense that more graph classes have bounded clique-width than bounded tree-width. Gurski and Wanke [55] established the following relation between clique-width and tree-width in terms of complete bipartite subgraphs.

**Theorem 4.7** (*Gurski, Wanke* [55]) *Let $G$ be a graph of clique-width[5] $k$ such that for some $n > 1$ the complete bipartite graph $K_{n,n}$ is not a subgraph of $G$. Then $\mathrm{tw}(G) \leq 3k(n-1) - 1$.*

---

[4] By "generating" we always mean up to isomorphism. That is, a graph $G$ is generated by an expression $\varphi$ if $\varphi$ defines a graph isomorphic to $G$.
[5] In [89] Wanke introduced the notion of $k$-node label controlled graphs ($k$-NLC). They are defined by similar operations as in $k$-expressions and for every graph $G$ we have $\mathrm{cw}(G) \leq \mathrm{nlc}(G) \leq 2 \cdot \mathrm{cw}(G)$, where $\mathrm{nlc}(G)$ denotes the NLC-width. The result in [55] is actually stated and proved in terms of NLC-width.

Another interesting relation between clique-width and tree-width follows from a connection, due to Oum [68], between the branch-width of a graph and the rank-width of its incidence graph which we will present at the end of Section 4.2.

As seen in the previous section, the notion of tree-width is preserved by taking subgraphs, induced subgraphs, minors, and other transformations. Clique-width is less robust. It is easily seen that clique-width is preserved under taking induced subgraphs. But it is not preserved under taking arbitrary subgraphs and hence not preserved under taking minors. For instance, cliques have clique-width 2 but every graph is a subgraph of a clique and we know that there are graphs of arbitrarily high clique-width.

**Proposition 4.8**    (i) *If $G$ is a graph and $H$ is an induced subgraph of $G$, then $\mathrm{cw}(H) \leq \mathrm{cw}(G)$.*
  (ii) *Clique-width is not preserved under taking subgraphs and hence not preserved under taking minors. That is, there are graphs $G$ and $H \subseteq G$ with $\mathrm{cw}(H) > \mathrm{cw}(G)$ and the difference can be arbitrarily large.*

We close this section with a negative result concerning the complexity of deciding clique-width and related measures. Gurski and Wanke showed that deciding the NLC-width of a graph is NP-complete. For clique-width, this was shown by Fellows, Rosamond, Rotics and Szeider.

**Theorem 4.9**   1. *Given a graph $G$ and an integer $k$, the problem to decide whether $G$ has NLC-width at most $k$ is NP-complete (see [56]).*
  2. *Given a graph $G$ and an integer $k$, the problem to decide whether $G$ has clique-width at most $k$ is NP-complete (see [42]).*

However, as we will see in the next section, there are FPT-algorithms, parametrized by the clique-width, to compute an approximate clique-decomposition of a given graph.

Finally, we mention a result by Espelage, Gurski and Wanke [41], that the clique-width of a graph can be computed in linear time on graph classes of bounded tree-width.

### 4.2 Rank-Width

In this section we consider an alternative characterisation of graph classes of bounded clique-width – the *rank-width* of a graph. Rank-width is a special case of abstract *branch-decompositions* of connectivity functions which we present first. Another special case of this abstract notion is the *branch-width of graphs*, a notion that is equivalent up to a small constant factor to tree-width.

*abstract*   **Branch-decompositions of connectivity functions** Let $M$ be a finite non-empty set and $f : 2^M \to \mathbb{R}$ be a function. A *branch-decomposition* of the pair *branch-decomposition*  $(M, f)$ is a pair $(T, \beta)$ consisting of a binary tree $T$ and a bijection $\beta : L(T) \to M$ from the set $L(T)$ of leaves of $T$ to $M$. We inductively define a map $\beta^* : V(T) \to 2^M$ by setting

$$\beta^*(t) := \begin{cases} \{\beta(t)\} & \text{if } t \text{ is a leaf} \\ \beta^*(t_1) \cup \beta^*(t_2) & \text{if } t \text{ is an inner node with successors } t_1 \cup t_2. \end{cases}$$

The *width* of $(T, \beta)$ is defined as $\max\{f(\beta^*(t)) : t \in V(T)\}$ and the *branch-width* of $(M, f)$ is defined as the minimal width of any of its branch-decompositions. If $M$ is empty, we define the branch-width of $M$ to be $f(\varnothing)$. Note that in this case, $(M, f)$ does not have a branch-decomposition, as a tree, being connected, cannot be empty. *width of $(T,\beta)$*  
*abstract branch-width*

Of particular interest are branch-decompositions of connectivity functions $f$ which are integer valued, symmetric and submodular. A function $f : 2^M \to \mathbb{R}$ is *symmetric* if $f(A) = f(M \setminus A)$ for all $A \subseteq M$ and it is *submodular* if $f(A) + f(B) \geq f(A \cap B) + f(A \cup B)$ for all $A, B \subseteq M$. Submodular and symmetric connectivity functions are algorithmically particularly well-behaved. Note that if $f$ is symmetric we can take the tree $T$ of a branch-decomposition of $(M, f)$ to be undirected and cubic (i.e. every vertex has degree 1 or 3). We will occasionally do so, for instance in Figure 7 below. *symmetric*  
*submodular*

In [69], Oum and Seymour showed that optimal branch-decompositions of submodular, symmetric, and integer valued connectivity functions can be approximated up to a factor 3 by an fpt-algorithm. Before we can state the result we need to define how the input to such an algorithm is represented. Let $\mathcal{M}$ be a class of pairs $(M, f)$, where $f : 2^M \to \mathbb{N}$ is symmetric and submodular. $\mathcal{M}$ is a *tractable class of connectivity functions* if there is a representation of the pairs $(M, f) \in \mathcal{M}$ such that, given the representation of a pair $(M, f)$, the underlying set $M$ and the values $f(A)$ can be computed in polynomial time for all $A \subseteq M$. *tractable class*

We are primarily interested in certain connectivity functions naturally associated with graphs and in this case the graph itself will be the representation.

**Theorem 4.10** (*Oum, Seymour* [69]) *Let $\mathcal{M}$ be a tractable class of connectivity functions. Then there is an fpt-algorithm that, on input (the representation of) $(M, f)$ and a parameter $k$, computes a branch-decomposition of $(M, f)$ of width at most $3k$ provided that the branch-width of $(M, f)$ is at most $k$. If the branch-width of $(M, f)$ is greater than $k$, then the algorithm may halt without output or still compute a branch-decomposition of $(M, f)$ of width $\leq 3k$.*

As a first example of abstract branch-decompositions we consider the branch-width of graphs.

**Branch-Width of Graphs** Let $G$ be a graph. The *boundary* $\partial F$ of a set $F \subseteq E(G)$ is defined as the set of vertices incident to an edge in $F$ and also an edge in $E(G) \setminus F$. *boundary, $\partial F$*

We define a function $b_G : 2^{E(G)} \to \mathbb{N}$ by $b_G(F) := |\partial F|$ for all $F \subseteq E(G)$. The function $b_G$ is symmetric and submodular. A *branch-decomposition* of $G$ is a branch-decomposition of $(E(G), b_G)$ and the *branch-width* $\mathrm{bw}(G)$ of $G$ is defined as the branch-width of $(E(G), b_G)$. *branch-decomposition*  
*branch-width*

**Example 4.11** *Figure 7 shows a graph and its branch-decomposition of width 2. For example, $\beta^*(d) = \{\{1,5\},\{3,5\}\}$ and $\partial\beta^*(d) = \{1,3\}$, as the vertex 5 has no edge to a vertex other than $1,3$. Similarly, $\partial\beta^*(b) = \partial\beta^*(e) = \partial\beta^*(e) = \{1,3\}$ and $\partial\beta^*(f) = \{3,4\}$.*

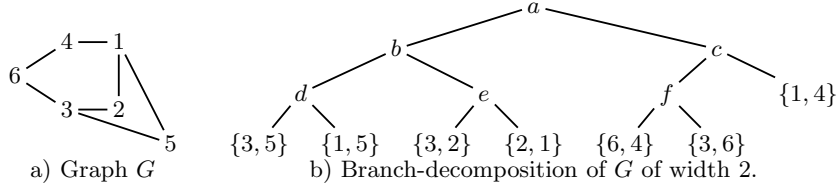

a) Graph $G$  
b) Branch-decomposition of $G$ of width 2.

**Fig. 7.** Branch-decomposition of width 2

⊣

**Example 4.12** (*Robertson, Seymour* [73])  1. For every $n \geq 3$, the $n$-clique $K_n$ has branch-width $\frac{2}{3} \cdot n$.
2. For all $n \geq 2$, the $n \times n$-grid has branch-width $n$.
3. A graph has branch-width 0 if, and only if, it has maximal degree at most 1.
4. Trees and cycles have branch-width at most 2.   ⊣

As the following theorem shows, the branch-width of a graph is equivalent to its tree-width up to a small constant factor.

**Theorem 4.13** (*Robertson, Seymour* [73])  *For all graphs $G$*

$$\mathrm{bw}(G) \quad \leq \quad \mathrm{tw}(G) + 1 \quad \leq \quad \max\{2, \frac{3}{2}\mathrm{bw}(G)\}.$$

*Proof.* To show $\mathrm{bw}(G) \leq \mathrm{tw}(G)+1$, let $\mathcal{T} := (T, (B_t)_{t \in V(T)})$ be a tree-decomposition of $G$ of width $k := \mathrm{tw}(G)$, such that $T$ is a binary tree and every edge of $G$ is covered by exactly one leaf of $T$. Clearly, given a tree-decomposition of $G$ we can easily find one of the same width with this additional property. We define a branch-decomposition $\mathcal{B} := (T', \beta)$ of $G$ as follows: $T' = T$ and for a leaf $t \in L(T)$ of $T$ we set $\beta(t) := e$, where $e$ is the (unique) edge covered by $B_t$. We define $\beta^* : V(T) \to 2^{E(G)}$ as before. It is easily seen that for all $t \in V(T)$, $\partial\beta^*(t) \subseteq B_t$ and hence the width of $\mathcal{B}$ is at most $k+1$.

Conversely, let $\mathcal{B} := (T, \beta)$ be a branch-decomposition of $G$ of width $\mathrm{bw}(G)$. For each $t \in V(T)$ we define $B_t \subseteq V(G)$ as follows. If $t$ is a leaf of $T$ define $B_t := \beta(t)$. Now let $t$ be an inner node with children $t_1, t_2$. For $i = 1, 2$ let $F_i := \beta^*(t_i)$ and let $F_3 := \bigl(E(G) \setminus \beta^*(t)\bigr) = \bigl(E(G) \setminus (F_1 \cup F_2)\bigr)$. We define $B_t := \partial F_1 \cup \partial F_2 \cup \partial F_3$.

By construction, $|F_i| \leq \mathrm{bw}(G)$. We claim that for all $v \in V(G)$, if $v$ occurs in some $\partial F_i$ then it also occurs in $\partial F_j$ for some $j \neq i$. For, if $v \in \partial F_i$ then there must be edges $e \in F_i$ and $e' \in E(G) \setminus F_i$ with $v \in e$ and $v \in e'$. Hence, $e' \in F_j$ for some $j \neq i$ and therefore $v \in \partial F_j$. If follows that $|B_t| \leq \max\{2, \frac{3}{2}\mathrm{bw}(G)\}$.

Now let $\mathcal{T} := (T, (B_t)_{t \in V(T)})$. It is easily verified that $\mathcal{T}$ is indeed a tree-decomposition of $G$.[6] Hence, we obtain a tree-decomposition of $G$ of width $\leq \max\{2, \frac{3}{2}\operatorname{bw}(G)\} - 1$. □

In principle one can use the general algorithm from Theorem 4.10 to compute approximate branch-decompositions of graphs. However, as for the case of tree-width, better algorithms are known.

**Theorem 4.14** (*Bodlaender, Thilikos* [7]) *There is an algorithm that, given a graph $G$ and $k \in \mathbb{N}$, computes a branch-decomposition of $G$ of width at most $k$, if it exists, in time $f(k) \cdot |G|$, for some computable function $f : \mathbb{N} \to \mathbb{N}$.*

**Clique- and Rank-Width** We now turn back to the original goal of giving a different characterisation of clique-width of a graph in terms of its *rank-width*. Recall that the branch-width of a graph is based on a decomposition of its edge set. For rank-width we decompose its vertex set.

Let $G$ be a graph. For $U, W \subseteq V(G)$ we define a $|U| \times |W|$-matrix $M_G(U, W)$ with entries $m_{u,w}$ for $u \in U$ and $w \in W$, where 　　　　$M_G(U,W)$

$$m_{u,w} := \begin{cases} 1 & \text{if } \{u,w\} \in E(G) \\ 0 & \text{otherwise.} \end{cases}$$

Note that $M_G(V(G), V(G))$ is the adjacency matrix of $G$. For all $U, W \subseteq V(G)$ let $\operatorname{rk}(M_G(U, W))$ be its row rank when viewed as a matrix over $\operatorname{GF}(2)$. This 　　$\operatorname{rk}(M_G(U,W))$ induces the following connectivity function $r_G : 2^{V(G)} \to \mathbb{N}$ defined as

$$r_G(U) := \operatorname{rk}(M_G(U, V(G) \setminus U))$$

for $U \subseteq V(G)$. Obviously, $r_G$ is symmetric, as the row and column rank of the matrix coincide. It is left as an exercise to show that it is also submodular.

**Definition 4.15** *A* rank-decomposition *of a graph $G$ is a branch-decomposition of the pair $(V(G), r_G)$. The* rank-width *of $G$, in terms $\operatorname{rw}(G)$, is the minimal width of any of its rank-decompositions.* 　　rank-width, $\operatorname{rw}(G)$

**Example 4.16** *Consider again the graph $G$ from Example 3.2 depicted in Figure 3. The following is a rank-decomposition of $G$ of width 3.*

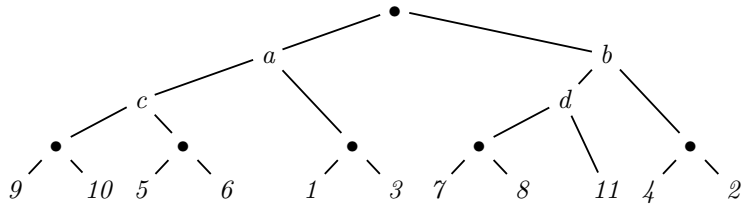

---
[6] At least if $G$ has no isolated vertices. If it does, add a bag for each isolated vertex.

*The relevant matrices determining the width of the decomposition are the matrices $M_a, \ldots, M_d$ at the nodes $a, \ldots, d$.*

$$M_c := M_G\big(\{5,6,9,10\}, \{1,2,3,4,7,8,11\}\big) = \begin{pmatrix} 1 & 0 & 0 & 0 & 0 & 0 & 0 \\ 0 & 0 & 1 & 0 & 0 & 0 & 1 \\ 0 & 0 & 0 & 0 & 0 & 0 & 0 \\ 0 & 0 & 0 & 0 & 0 & 0 & 1 \end{pmatrix}$$

$$M_d := M_G\big(\{7,8,11\}, \{1,2,3,4,5,6,9,10\}\big) = \begin{pmatrix} 0 & 0 & 1 & 1 & 0 & 0 & 0 & 0 \\ 0 & 0 & 0 & 1 & 0 & 0 & 0 & 0 \\ 0 & 0 & 1 & 0 & 0 & 1 & 0 & 1 \end{pmatrix}$$

$$M_a := M_G\big(\{1,3,5,6,9,10\}, \{2,4,7,8,11\}\big) = \begin{pmatrix} 1 & 0 & 0 & 0 & 0 \\ 1 & 1 & 1 & 0 & 1 \\ 0 & 0 & 0 & 0 & 0 \\ 0 & 0 & 0 & 0 & 1 \\ 0 & 0 & 0 & 0 & 0 \\ 0 & 0 & 0 & 0 & 1 \end{pmatrix}$$

$$M_b := M_G\big(\{2,4,7,8,11\}, \{1,3,5,6,9,10\}\big) = \begin{pmatrix} 1 & 1 & 0 & 0 & 0 & 0 \\ 0 & 1 & 0 & 0 & 0 & 0 \\ 0 & 1 & 0 & 0 & 0 & 0 \\ 0 & 1 & 0 & 0 & 0 & 0 \\ 0 & 0 & 0 & 0 & 0 & 0 \\ 0 & 1 & 0 & 1 & 0 & 1 \end{pmatrix}$$

*Obviously, $\mathrm{rk}(M_a) = \mathrm{rk}(M_b) = \mathrm{rk}(M_c) = 3$ and this is the maximal rank occurring in the decomposition. Hence, the decomposition has width 3.* ⊣

It is not too hard to see that the rank-width of a graph can be bounded in terms of its branch-width and hence its tree-width. The following theorem due to Oum gives an exact bound.

**Theorem 4.17** (*Oum* [68]) $\mathrm{rw}(G) \leq \max\{1, \mathrm{bw}(G)\}$ *for all graphs $G$.*

It is easily seen that the rank of width a complete graph is 1 (all entries in all matrices are 1). Hence, there can be an arbitrarily large difference between the rank-width and the branch-width of a graph. On the other hand, Oum [68] proved that if $I(K_n)$ denotes the incidence graph of the $n$-clique $K_n$, then for all $n \geq 3$ with $n = 0, 1 \mod 3$ we have $\mathrm{rw}(I(K_n)) = \mathrm{bw}(I(K_n)) = \lceil \frac{2}{3} \cdot n \rceil$.

Another example of graphs of high tree- and high rank-width are $n \times n$-grids, whose rank-width has been shown by Jelínek [58] to be $n$.

An fpt-algorithm for computing rank-decompositions follows from Theorem 4.10 but more efficient algorithms are known.

**Theorem 4.18** (*Hlineny, Oum* [40]) *There is an algorithm that, given a graph $G$ and $k \in \mathbb{N}$, computes a rank-decomposition of $G$ of width at most $k$, provided $\mathrm{rw}(G) \leq k$, in time $f(k) \cdot |G|^3$, for some computable function $f : \mathbb{N} \to \mathbb{N}$.*

Oum and Seymour [69] established the following connection between rank-width and clique-width:

$$\mathrm{rw}(G) \leq \mathrm{cw}(G) \leq 2^{\mathrm{rw}(G)+1} - 1.$$

In particular, a class of graphs has bounded clique-width if, and only if, it has bounded rank-width (see [69]). Together with Theorem 4.18 this yields a parameterized algorithm for computing approximate clique-decompositions of graphs.

We have already seen that clique-width and tree-width and hence branch-width of graphs can differ arbitrarily and this clearly extends to rank-width. However, Oum [68] established the following relation between the branch-width of a graph and the rank-width of the incidence graph.

$$\mathrm{bw}(G) - 1 \leq \mathrm{rw}(I(G)) \leq \mathrm{bw}(G)$$

### 4.3 Monadic Second-Order Logic and Bounded Clique-Width

In this section we aim at extending Courcelle's and Seese's theorems from tree-width to clique-width. As in Section 3, we will do so by a reduction to MSO model-checking and satisfiability on trees. In particular, we show next that for each $k$ the class of graphs of clique-width $k$ can be interpreted in the class of coloured trees for a suitable set of colours depending on $k$. The idea is simple: the class of graphs of clique-width $k$ is the class of graphs generated by $k$-expressions whose syntax trees will be the class of trees we are looking for. Hence, let

$$\Sigma_k := \{\mathbf{0}, \ldots, \mathbf{k-1}, \oplus, \mathrm{edge}_{i,j}, \mathrm{rename}_{i \to j} : 0 \leq i \neq j < k\}$$

be the symbols used in $k$-expressions and let $\mathcal{T}_{\Sigma_k}$ be the class of all $\Sigma_k$-labelled directed trees. Obviously, not every $\Sigma_k$-labelled tree is the syntax tree of a $k$-expression. However, every $\Sigma_k$-labelled directed tree such that the symbol $\oplus$ occurs precisely at the nodes with two successors, no node has more than two successors and the leaves are precisely the nodes labelled by a symbol from $\{\mathbf{0}, \ldots, \mathbf{k-1}\}$ are syntax trees of $k$-expressions. These conditions are easily expressed by an MSO-sentence $\varphi_{valid}$. Hence, for all $T \in \mathcal{T}_{\Sigma_k}$, $T \models \varphi_{valid}$ if, and only if, $T$ is the syntax tree of a $k$-expression. The formula $\varphi_{valid}$ is one part of an interpretation $\varGamma_k := \big(\varphi_{univ}, \varphi_{valid}, \varphi_E(x,y)\big)$ from $\Sigma_k$-labelled trees to graphs of clique-width at most $k$.

The formula $\varphi_{univ}(x)$ defining the universe of a graph generated by a $k$-expression coded in a tree $T$ is trivial: $\varphi_{univ}(x)$ just defines the set of leaves.

Finally, we have to define the formula $\varphi_E(x,y)$ such that for all $T \in \mathcal{T}_\Sigma$ with $T \models \varphi_{valid}$ and all leaves $u, v \in V(T)$ we have $T \models \varphi_E(u,v)$ if, and only if, there is an edge between $u$ and $v$ in the graph $G$ generated by $T$. Note that such an edge exists if, and only if, there is a common ancestor $t$ of $u$ and $v$ in $T$ labelled by $\mathrm{edge}_{i-j}$, for some $0 \leq i \neq j < k$, so that at the node $t$, one of $u, v$ has colour $i$ and the other the colour $j$. To check this, we only need to look at the unique path from $t$ to $u$ (and $v$ respectively) and keep track of how the colour of $u$ (resp. $v$) changes along this path. This can easily be formalised in MSO by a formula $\varphi_E(x,y)$ as required. Hence, the triple $\varGamma_k := (\varphi_{univ}(x), \varphi_{valid}, \varphi_E(x,y))$ is an interpretation from $\sigma_k$-structures to graphs, where $\sigma_k := \{E\} \cup \varSigma_k$ is the signature of $\varSigma_k$-labelled trees.

The interpretation is the key to tractability results for MSO model-checking and satisfiability. We consider model-checking first and prove the following extension of Courcelle's theorem. It was first proved by Courcelle in terms of certain graph grammars (see [12,14]) and then by Courcelle, Makowski and Rotics for graph classes of bounded clique-width.

**Theorem 4.19** (*Courcelle, Makowski, Rotics* [18]) *Let $\mathcal{C}$ be a class of graphs of bounded clique-width. Then the model-checking problem for* MSO *on $\mathcal{C}$ is fixed-parameter tractable.*

*Proof.* Let $\varphi \in$ MSO be fixed and let $k$ be an upper bound for the clique-width of the graphs in $\mathcal{C}$. Given a graph $G$ we first compute a $k$-expression $\vartheta$ generating $G$. This can be done in polynomial time (see Section 4.2). Let $T$ be the $\varSigma_k$-labelled syntax tree of $\vartheta$. We can now test whether $T \models \varGamma_k(\varphi)$. □

We now consider the satisfiability problem for monadic second-order logic.

**Theorem 4.20** *For every $k$, the* MSO*-theory of the class $CW_k$ of graphs of clique-width at most $k$ is decidable.*

*Proof.* Let $\varphi \in$ MSO$[\{E\}]$ be given. By the interpretation lemma, $\varphi$ is valid in $CW_k$ if, and only if, $\varGamma_k(\varphi) \in$ MSO$[\sigma_k]$ is valid in the class $\{T \in \mathcal{T}_\varSigma : T \models \varphi_{valid}\}$ if, and only if, $\varGamma_k(\varphi) \wedge \varphi_{valid}$ is valid in the class of finite $\varSigma_k$-trees. The latter is well known to be decidable [31,83]. □

Seese conjectured a kind of converse to the theorem, the famous Seese conjecture [79].

**Conjecture 4.21 (Seese's conjecture)** *Every class $\mathcal{C}$ of structures with decidable* MSO$_1$*-theory has bounded clique-width.*

This conjecture can be rephrased in terms of MSO-interpretations using the following result due to Engelfriet and V. van Oostrom and also Courcelle and Engelfriet.

**Lemma 4.22** ([39,16]) *A class of graphs has bounded clique-width if, and only if, it is interpretable in the class of coloured trees for some suitable set of colours.*

Note that these papers use so-called MSO-transductions instead of interpretations. An MSO-transduction is essentially the same as an interpretation except that the formulas are allowed to have free second-order variables, the parameters. A graph is then interpretable in a tree if there is an interpretation of the parameters by sets of tree-nodes satisfying the formulas in the MSO-transduction. Hence, the parameters play exactly the same role as the colours of the trees we use here. As the colours/parameters in our context are the symbols of $k$-expressions, we prefer to have them as labels of the syntax trees rather than as free variables in the interpretation.

Using the previous lemma we can rephrase Seese's conjecture as follows:

**Conjecture 4.23 (Seese's conjecture)** *Every class $\mathcal{C}$ of structures with decidable $\mathrm{MSO}_1$-theory is MSO-interpretable in the class of coloured trees for some set of colours.*

In [20], Courcelle and Oum prove the following weakening of the conjecture. Let $C_2\mathrm{MSO}$ be the extension of MSO by atoms $\mathrm{EVEN}(X)$, where $X$ is a monadic second-order variable, stating that the interpretation of $X$ has even cardinality. Hence, $C_2\mathrm{MSO}$ extends MSO by counting modulo 2.

**Theorem 4.24 (*Courcelle, Oum* [20])** *Every class of graphs with a decidable $C_2\mathrm{MSO}$ theory has bounded clique-width, i.e. is interpretable in a class of coloured trees.*

Note that the theorem is weaker than Seese's conjecture as there are less classes of graphs whose $C_2\mathrm{MSO}$ theory is decidable than there are classes of graphs with a decidable MSO-theory.

### 4.4 MSO Model-Checking Beyond Tree- and Clique-Width

In the previous section we showed that the model-checking problem for monadic second-order logic is fixed-parameter tractable on classes of graphs with bounded tree- or clique-width. There is not much hope for extending these results to other or larger classes of graphs such as planar graphs or graphs of bounded degree. This follows immediately from the following theorem by Garey, Johnson and Stockmeyer and the fact that 3-colourability is MSO-definable.

**Theorem 4.25 (*Garey, Johnson, Stockmeyer* [49])** *3-colourability is $\mathrm{NP}$-complete on the class of planar graphs of degree at most 4.*

However, first-order logic is tractable on many more classes of graphs. For instance, Seese [80] showed that first-order logic admits linear time model-checking (for a fixed formula) on any class of graphs of bounded degree. The same complexity bound was later obtained by Frick and Grohe [47] for planar graphs and classes of graphs of bounded *local tree-width*, a notion that properly extends both planarity and bounded degree (see below).

The most general results in this respect are fixed-parameter algorithms for first-order model-checking on $H$-minor free graphs and an extension thereof, called locally excluded minors. These results make heavy use of concepts and results developed by Robertson and Seymour in their celebrated proof of the graph minor theorem. In the next section, we will therefore give a brief overview of the relevant concepts of the graph minor theory used in the proofs. One such theorem, the excluded grid theorem, will be used later to prove the converse of Seese's theorem mentioned above. This will be the topic of Section 6. We return to first-order model-checking in Section 7.

# 5 Graph Minors

In this section we present relevant terminology and results from graph minor theory used later in the paper. Most of the results were developed in Robertson and Seymour's celebrated proof of the graph minor theorem (Theorem 5.2 below) presented in a series [74] of 23 papers, with additions and improvements by other authors.

## 5.1 Minors and Minor Ideals

*G/e*  
*contraction*

Let $G$ be a graph and $e := \{v, w\} \in E(G)$ be an edge. The graph $G/e$ obtained from $G$ by *contracting* the edge $e$ is the graph obtained from $G$ by removing $e$, identifying its two endpoints, and possibly removing parallel edges. Formally, $G/e$ is defined by

$$V(G/e) := V(G) \setminus \{v, w\} \cup \{x_e\},$$

where $x_e$ is a new vertex, and

$$E(G/e) := \begin{matrix}\left(E(G) \setminus \{\{u, u'\} : \{u, u'\} \cap e \neq \varnothing\}\right) \cup \\ \{\{u, x_e\} : u \in V(G/e) \text{ and} \{u, v\} \in E(G) \text{ or } \{u, w\} \in E(G)\}.\end{matrix}$$

Figure 8 illustrates edge contraction.

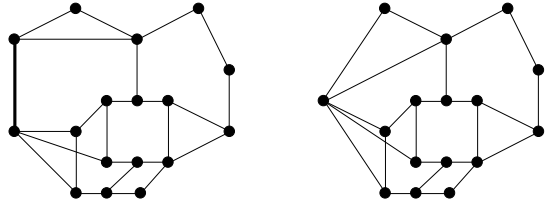

**Fig. 8.** Contracting an edge

*minor*

A graph $H$ is a *minor* of a graph $G$ if $H$ can be obtained from $G$ by deleting vertices and edges and contracting edges. We write $H \preccurlyeq G$ to denote that $H$ is isomorphic to a minor of $G$.

An alternative definition of minors is in terms of minor maps. A *minor map* from $H$ to $G$ is a function $\mu$ that associates with every vertex $v \in V(H)$ a connected subgraph $\mu(v) \subseteq G$ and with every edge $e \in E(H)$ an edge $\mu(e) \in E(G)$ such that

- if $u, v \in V(H)$ and $u \neq v$ then $\mu(v)$ and $\mu(u)$ are vertex disjoint and
- if $e := \{u, v\} \in E(H)$ then $\mu(e) := \{u', v'\}$ for some $u' \in V(\mu(u))$ and $v' \in V(\mu(v))$.

The subgraph $G_\mu \subseteq G$ with

$$V(G_\mu) := \bigcup \{V(\mu(v)) : v \in V(H)\}$$

and

$$E(G_\mu) := \bigcup \{E(\mu(v)) : v \in V(H)\} \cup \{\mu(e) : e \in E(H)\}$$

is called a *model* or *image* of $H$ in $G$. In graph theory literature, the term model is commonly used. We prefer the name image here to avoid confusion with logical models. Figure 9 illustrates an image of $K_5$ in a graph $G$. *model, image*

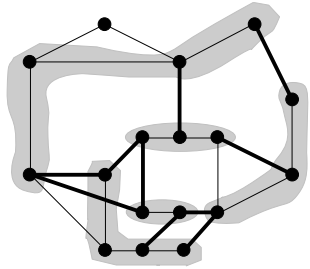

**Fig. 9.** Image of $K_5$ in a graph $G$

It is easily seen that we can always choose an image of $H$ in $G$ so that each vertex is represented by a tree in $G$.

Let $G, H$ be graphs. $G$ is a *subdivision* of $H$ if $H$ can be obtained from $G$ by replacing some edges in $G$ by paths which are pairwise internally vertex disjoint, i.e. $H$ can be constructed from $G$ by repeatedly *subdividing* edges. If a subgraph of $G$ is isomorphic to a subdivision of $H$, then $H$ is called a *topological minor* of $G$. Now suppose $H \preccurlyeq G$ and $H$ has maximal degree 3. Let $\mu$ be a minor map from $H$ into $G$ so that the image of all vertices of $H$ are trees in $G$. Then each of these trees has at most 3 leaves and hence at most one vertex of degree more than two. It follows that every graph $H$ of maximal degree $\leq 3$ that is a minor of $G$ also is a topological minor of $G$. *subdivision*

*topological minor*

**Lemma 5.1** *Let $H, G$ be graphs. If $\Delta(H) \leq 3$ and $H \preccurlyeq G$, then $H$ is a topological minor of $G$.*

If $H \not\preccurlyeq G$, we say that $H$ is a *forbidden* minor of $G$, or that $G$ excludes $H$. For any graph $H$ let $Excl(H) := \{G : H \not\preccurlyeq G\}$ be the class of graphs not containing $H$ as a minor. Analogously, if $\mathcal{H}$ is a set of graphs, then $Excl(\mathcal{H}) := \bigcap \{Excl(H) : H \in \mathcal{H}\}$ is the class of graphs not containing any member of $\mathcal{H}$ as a minor. *excluded minor* *Excl(H)* *Excl($\mathcal{H}$)*

A class $\mathcal{C}$ of graphs is a *minor ideal* if for all $G \in \mathcal{C}$ and $H \preccurlyeq G$ also $H \in \mathcal{C}$. It is *proper* if it is not the class of all graphs. *minor ideal* *proper minor ideal*

A class $\mathcal{C}$ is *characterised* by a class $\mathcal{F}$ of graphs if $\mathcal{C} = Excl(\mathcal{F})$. Note that any minor ideal $\mathcal{C}$ can be characterised by a class of excluded minors, e.g. $\mathcal{C} = \textit{Excl}(\text{GRAPHS} \setminus \mathcal{C})$. As the main result of their fundamental work on graph minors, Robertson and Seymour proved that any minor ideal can in fact be characterised by a *finite* set of forbidden minors.

**Theorem 5.2** (*Robertson, Seymour* [78]) *For every minor ideal $\mathcal{C}$ there is a finite set $\mathcal{F}$ of graphs such that $\mathcal{C} = \textit{Excl}(\mathcal{F})$.*

There are many natural examples of minor ideals.

- Every cycle can be contracted to a triangle. Hence, $\textit{Excl}(K_3)$ is the class of acyclic graphs.
- Kuratowski's theorem [59] (or rather a variant established by Wagner [88]) implies that planar graphs are characterised by excluding $K_{3,3}$ and $K_5$.
- *Series-parallel* graphs and *outerplanar* graphs exclude $K_4$. It can be shown that $\textit{Excl}(K_4)$ is the class of subgraphs of series-parallel graphs and the class of outerplanar graphs is characterised by $\textit{Excl}(K_4, K_{2,3})$. (See e.g. [30, Exercises 7.32 and 4.20].)
- The class of graphs *not* having $k$ vertex disjoint cycles, for any fixed $k \in \mathbb{N}$. For $k \in \mathbb{N}$ let $T_k$ be the graph consisting of $k$ disjoint copies of a triangle. Clearly, every graph containing $k$ vertex disjoint cycles contains $T_k$ as a minor. Conversely, every graph containing $T_k$ as a minor also contains $k$ vertex disjoint cycles. Hence the class $\mathcal{C}_k$ of graphs *not* having $k$ disjoint cycles is characterised by $T_k$.

It is easily seen that for each $k \in \mathbb{N}$ the class $\mathcal{T}_k$ of graphs of tree-width at most $k$ and the class $\mathcal{B}_k$ of graphs of branch-width at most $k$ are minor ideals and so is the class of graphs of genus at most $k$. Finally, let us mention another famous example of a minor ideal: the class of *knotlessly* embeddable graphs.

On the other hand, the class of graphs of clique-width at most $k$ is not minor closed and hence not a minor ideal. Also, the class of graphs of crossing number $k \geq 1$ is *not* minor closed.

Robertson and Seymour also proved that for any fixed graph $H$, testing if a graph $G$ contains $H$ as a minor can be done in cubic time (we will say more about this later in this section). Hence, combining this minor test with Theorem 5.2 implies that every minor-ideal can be decided in cubic time.

**Corollary 5.3** *Every minor ideal can be decided in cubic time.*

The various concepts and results developed in the course of the proof of Theorem 5.2 have sparked of a rich algorithmic theory of graphs based on structural restrictions of instances. We have already hinted at the algorithmic theory of graphs of bounded tree-width. However, the algorithmic applications of the graph minor theory developed by Robertson and Seymour extend far beyond tree-like graphs. In the following two sections we present some of the results and methods with implications for algorithms and model-checking on graphs.

However, the following can only give a glimpse into the deep results underlying the proof of the graph minor theorem – we will not even be able to state the relevant results in full detail let alone attempt to prove them. While we are trying to give an intuitive account of the results and proof methods, we will necessarily have to be brief and the presentation may not always reflect the actual proofs.

## 5.2 Disjoint Paths and the Trinity Lemma

Let us try to prove Theorem 5.2. Clearly, the statement of the theorem is equivalent to the statement that in every infinite class of finite graphs one graph is a minor of another. Let $\mathcal{C} := \{H, G_1, G_2, \dots\}$ be an infinite class of finite graphs. If $H$ is a minor of some $G_i$, then the claim is trivially satisfied by $H$. Hence, the only interesting case is when no $G_i \in \mathcal{C}$ contains $H$ as a minor. For this reason, much of the theory developed by Robertson and Seymour deals with graphs not containing another fixed graph $H$ as a minor. We refer to such graphs as *H-minor free*. Clearly, if $G$ is $H$-minor free, then $G$ also excludes a clique $K_k$ as a minor, for instance taking $k := |V(H)|$. Let us fix $k$ for the rest of the section.  *H-minor free*

The key to studying the structure of $K_k$-minor free graphs is the following theorem, proved by Robertson and Seymour in [72]. Recall from Section 2 that $G_{k \times k}$ denotes the $k \times k$-grid.

**Theorem 5.4 (Excluded Grid Theorem [72])** *There is a computable function $f : \mathbb{N} \to \mathbb{N}$ such that every graph of tree-width at least $f(k)$ contains $G_{k \times k}$ as a minor.*

We refer to [30] for a proof of this theorem. As every planar graph is a minor of a suitably large grid, the theorem implies – is equivalent, in fact – to the following statement.

**Corollary 5.5** *For all $H$, the class $Excl(H)$ of $H$-minor free graphs has bounded tree-width if, and only if, $H$ is planar.*

The function $f$ in the original proof of Theorem 5.4 was huge. In [71], Robertson, Seymour and Thomas significantly improved the bounds on $f$ to $20^{2k^5}$. However, no matching lower bounds have been established and it is conjectured that the actual bound may be as small as polynomial in $k$. For planar graphs $G$ a much better bound can be obtained.

**Theorem 5.6 (***Robertson, Seymour, Thomas* [71]) *Every planar graph with no $k \times k$-grid minor has tree-width $\leq 6k - 5$.*

For branch-width a slightly tighter bound has been established: every planar graph of branch-width at least $4k - 3$ contains a $k \times k$-grid minor (see [71]). Whereas it is still open whether optimal tree-decompositions of planar graphs can be computed in polynomial time, in [82] Seymour and Thomas proved that optimal branch-decompositions of planar graphs can be computed in time $\mathcal{O}(n^4)$. This has later been improved to $\mathcal{O}(n^3)$ by Gu and Tamaki [54]. It should be noted that these algorithms do not contain any large hidden constants and perform reasonably well in practise. Optimal branch-decompositions of planar graphs with up to 50.000 edges have been computed by actual implementations of the algorithms (see e.g. [5]).

To give an application of the grid-theorem on planar graphs, we note that it implies an $2^{\mathcal{O}(\sqrt{k})} \cdot n^c$ algorithm, for some $c \in \mathbb{N}$, for deciding whether a planar graph has a path of length $k$. For this, use an $\mathcal{O}(n^3)$ algorithm for testing whether a given planar graph $G$ has branch-width at most $4\sqrt{k}-3$. If so, then one can compute a suitable branch-decomposition and use dynamic programming to decide whether a path of length $k$ exists. Otherwise, the planar grid theorem tells us that the graph contains a $\sqrt{k} \times \sqrt{k}$ grid as a minor and hence a path of length at least $k$ following the grid structure. A similar algorithmic idea has found numerous applications, for instance on $H$-minor free graphs, in the form of *bidimensionality theory*. See e.g. [25,32,27,33,24,26,28] and references therein.

*wall*    For the rest of this section we will work with a somewhat simpler structure than grids, called *walls*.

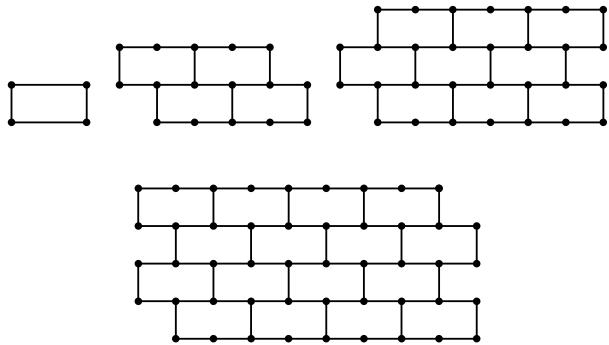

**Fig. 10.** Elementary walls of height 1–4

An *elementary wall* is a graph as displayed in Figure 10. A wall of height $h$ is a subdivision of an elementary wall of height $h$. See Figure 11 for a wall of height 4. The induced cycles of a wall, i.e. the cycles of length 6 in an elementary *brick*    wall or their subdivisions in general walls, are called the *bricks* of the wall. We assign coordinates $(i,j) = (row, col)$ to the bricks of a wall. The brick in the lower left corner is assigned $(1,1)$, its neighbour to the right $(1,2)$, the brick just *central brick*    above it $(2,1)$ and so on. The *central brick* of $H$ is the brick with coordinates *central vertex*    $(\lceil h/2 \rceil, \lceil h/2 \rceil)$. A *central vertex* of a wall is a vertex contained in the central brick but not in its neighbours to the left or right.

*perimeter*    The outermost (non-induced) cycle of a wall $W$ is called its *perimeter*.

Clearly, every large grid contains a large wall as a subgraph and conversely every large wall contains a large grid as a minor. The main advantage of working with walls rather than grids is that if $G$ contains an elementary wall as a minor then, by Lemma 5.1, it contains a wall of the same height as a subgraph.

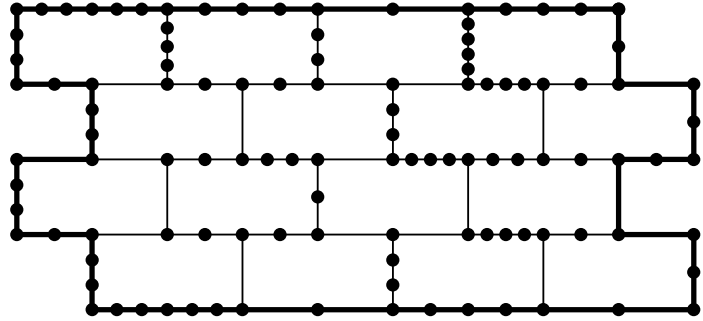

**Fig. 11.** A wall of height 4

Let us come back to the analysis of the structure of graphs. Let $t$ be a bound on the tree-width we want to consider. If $G$ has tree-width at most $t$, then it is sufficiently tree-like and its structure is well understood. So suppose $G$ has large tree-width. By the Excluded Grid Theorem 5.4, we know that $G$ contains a large wall $W$ as a subgraph. We can use $W$ as a drawing board on which we *draw* the rest of the graph $G$. Clearly, as $G$ is not required to be planar, this "drawing" will not necessarily be plane, i.e. edges may cross. In particular, edges or paths may span over different bricks of the wall. This is called a "crossing". More formally, a crossing consists of two pairwise vertex-disjoint paths with endpoints $v_1, v_3$ and $v_2, v_4$ such that $v_1, v_2, v_3, v_4$ occur clockwise in this order on some cycle of the grid. Figure 12 illustrates the concept of crossings.

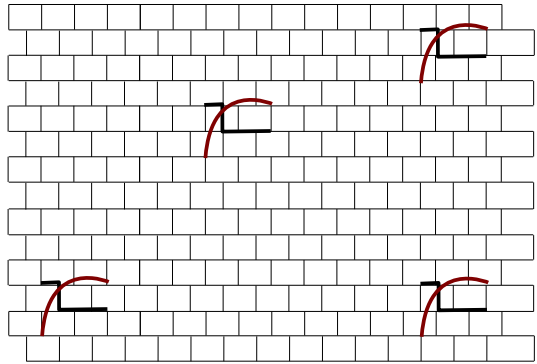

**Fig. 12.** Crossings in a graph

Crossings are important for our purpose. For, if $G$ contains many crossings which, in addition, are sufficiently far apart from each other on the wall used to draw $G$, then we can use the crossings to find a large clique minor of the graph. To see this, take a large clique and draw it "flat" on the wall $W$. Necessarily (unless your clique has less than five vertices) some of the edges in the clique will cross each other. However, if the wall $W$ is large enough and there are sufficiently many crossings far apart from each other, then we can replace the edges of the clique by disjoint paths in $G$ so that edges that cross are replaced by disjoint paths that cross each other using a "crossing" in the drawing of $G$. The following Figure 13 illustrates this with $K_5$ and one crossing. The grey areas are (essentially) the parts that are being contracted for each vertex in the clique.

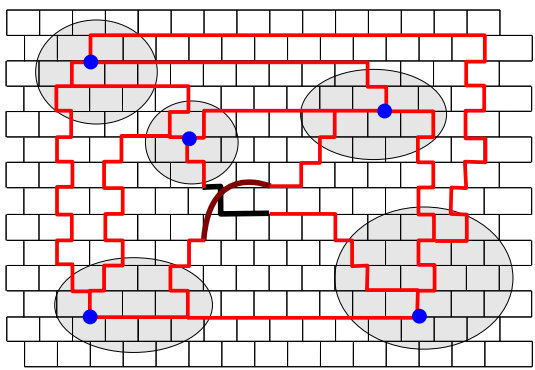

**Fig. 13.** A $K_5$-minor in a wall with one crossing

Hence, if $W$ is large enough and there are many crossings pairwise far apart in $W$, then $G$ contains a large clique minor. So, how does a graph $G$ drawn on a large wall look like if it does not contain a large clique minor?

As explained before, all but a small number of crossings must be grouped together in a bounded number of small parts of the wall. These regions with many crossings are called *vortices*. Further, there can be some vertices which are very well connected to the rest of the graph, i.e. a set $X$ of vertices that have edges to arbitrary vertices in the graph, where edges can be replaced by paths of arbitrary length. The vertices in $X$ are called *apices* (see Figure 14).

*vortex*

*apex*

However, any such well-connected vertex in $X$ can be used as a crossing and hence, if $G$ excludes $K_k$, there are either at most $|X| \leq \binom{k}{2}$ such elements, or their connections to the wall are concentrated on a small part of the wall $W$ (and hence they are part of the vortices) so that the crossings cannot be used to route the edges of a $K_k$-minor. In this case, we will find a subwall of $W$ which is still "large" and is connected only to a subset of $X$ of size $\leq \binom{k}{2}$. Hence, we can continue the discussion with the subwall $W'$ where we do not have vortices and only a bounded number of apices.

Besides the apices, there can be other parts of the graph with direct connections to the interior of the wall,[7] which do not induce any further crossings.

---

[7] There may also be parts of the graph connected to the wall only through its perimeter. These parts are not relevant here but we come back to this in the next section.

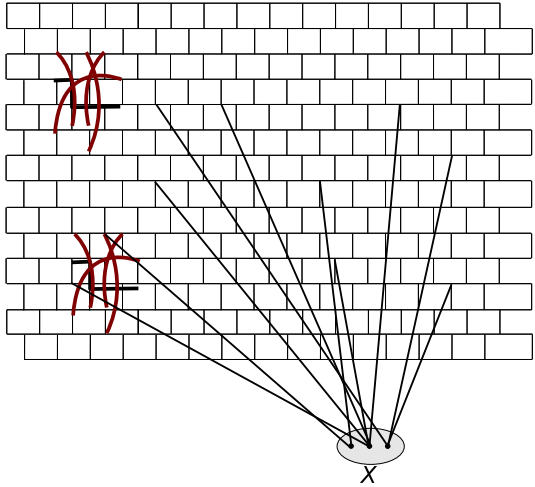

**Fig. 14.** *Vortices* and *apices* in a graph drawn on a wall

We call these *extensions*. Essentially, an extension is a subgraph $D$ of $G$ that is connected to the wall only within a brick and only with at most 3 vertices. This is important as with three vertices the extensions cannot induce further crossings in the wall.

Furthermore, we can assume that the tree-width of any such extension is bounded, as otherwise we could forget about the rest of the graph and do the same analysis within the extension, either producing a large clique minor or a large wall with vortices, apices and extensions. Note, though, that the apices may have connections to the extensions. See Figure 15 for an illustration.

The discussion so far presents the main ideas in the proof of the next lemma, one of the important results in the Graph Minor Series. To state it precisely, we need some further notation.

For a subgraph $D$ of a graph $G$, we let $\partial^G D$ be the set of all vertices of $D$ that are incident with an edge in $E(G) \setminus E(D)$. In the following, let $W$ be a wall of height at least 2 in a graph $G$ and let $P$ be the perimeter of $W$, i.e. the boundary cycle of $W$. Let $K'$ be the unique connected component of $G \setminus P$ that contains $W \setminus P$. The graph $K = K' \cup P$ is called the *compass* of $W$ in $G$. A *layout of $K$ (with respect to the wall $W$ in $G$)* is a family $(C, D_1, \ldots, D_m)$ of connected subgraphs of $K$ such that:

1. $K = C \cup D_1 \cup \ldots \cup D_m$,
2. $W \subseteq C$ and there is no separation $(X, Y)$ of $C$ of order $\leq 3$ with $V(W) \subseteq X$ and $Y \setminus X \neq \varnothing$,
3. $\partial^G D_i \subseteq V(C)$ for all $i \in \{1, \ldots, m\}$,
4. $|\partial^G D_i| \leq 3$ for all $i \in \{1, \ldots, m\}$,
5. $\partial^G D_i \neq \partial^G D_j$ for all $i \neq j \in \{1, \ldots, m\}$.

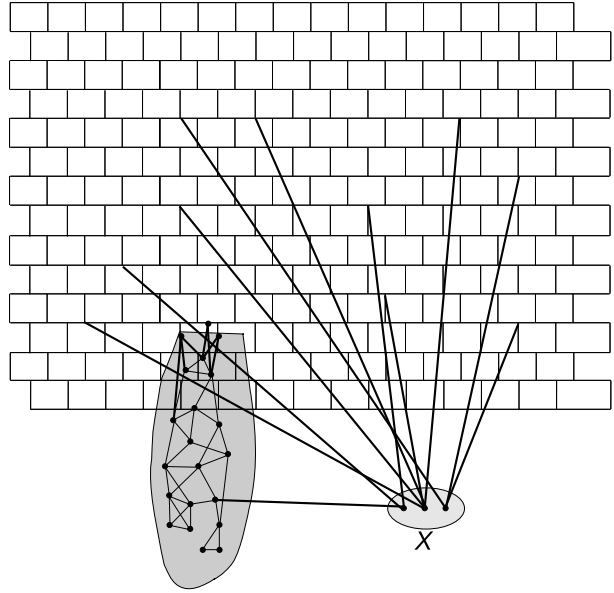

**Fig. 15.** Apices, extensions and connections within the subwall $W'$

We let $\overline{C}$ be the graph obtained from $C$ by adding new vertices $d_1, \ldots, d_m$ and, for $1 \leq i \leq m$, edges between $d_i$ to the vertices in $\partial^G D_i$ and edges between all vertices in $\partial^G D_i$. Hence, for each $i \in \{1, \ldots, m\}$, the vertex $d_i$ together with the (at most 3) vertices in $\partial^G D_i$ form a clique. We call $\overline{C}$ the *core* of the layout and $D_1, \ldots, D_m$ its *extensions*. The layout $(C, D_1, \ldots, D_m)$ is *flat* if its core $\overline{C}$ is planar. Note that this implies that the core has an embedding in the plane that extends the "standard planar embedding" of the wall $W$ (as shown in Figure 10), because the wall $W$ has a unique embedding into the sphere. We call the wall $W$ *flat* (in $G$) if the compass of $W$ has a flat layout.

The following lemma, which we refer to as the trinity lemma, is (essentially) Lemma 9.8 of [76]. Concerning the uniformity, see the remarks at the end of [76] (on page 109).

**Lemma 5.7 (Trinity Lemma [76])** *There are computable functions $f, g : \mathbb{N}^2 \to \mathbb{N}$ and an algorithm A that, given a graph $G$ and non-negative integers $k, h$, computes either*

1. *a tree-decomposition of $G$ of width $f(k, h)$,*
2. *a $K_k$-minor of $G$, or*
3. *a subset $X \subseteq V(G)$ with $|X| < \binom{k}{2}$, a wall $W$ of height $h$ in $G \setminus X$, and a flat layout $(C, D_1, \ldots, D_m)$ of the compass of $W$ in $G \setminus X$ such that the tree-width of each of the extensions $D_1, \ldots, D_m$ is at most $f(k, h)$.*

*Furthermore, the running time of the algorithm is bounded by $g(k, h) \cdot |V(G)|^2$.*

Using the trinity lemma, we can now sketch the proof of the following theorem due to Robertson and Seymour [76].

**Theorem 5.8** (*Robertson, Seymour* [76]) *The following problem is fixed-parameter tractable with a cubic fpt algorithm.*

> $p$-DISJOINT-PATHS
> *Input:* Graph $G$, $s_1, \ldots, s_k, t_1, \ldots, t_k \in V(G)$.
> *Parameter:* k.
> *Problem:* Are there $k$ vertex disjoint paths connecting
> $s_i$ and $t_i$, $1 \leq i \leq k$?

The idea of the algorithm is as follows. Apply the trinity lemma on $G$ for suitable values of $k$ and $h$. If $G$ has tree-width $\leq f(k, h)$, then the disjoint paths problem can be solved by standard techniques using dynamic programming (or by formalising the problem in MSO and using Courcelle's theorem). Otherwise, if $G$ contains a large clique minor (say at least $K_{3k}$), then we can do the following. To simplify the presentation, let us assume that $G$ actually contains the $3k$-clique as a subgraph. If there are $2k$ vertex disjoint paths connecting $\{s_1, \ldots, s_k, t_1, \ldots, t_k\}$ to the clique, then these paths together with the edges of the clique yield the $k$ vertex-disjoint paths connecting $s_i, t_i$ as desired. Otherwise, by Menger's theorem, there is a separator $X \subseteq V(G)$ of size at most $2k$ separating the clique and (part of) the $\{s_i, t_i : 1 \leq i \leq k\}$. But now, the problem can be reduced to a constant number of disjoint paths problems on smaller subgraphs, trying to connect $s_i, t_i$ with all possible combinations of elements in the separator.

If $G$ does not contain the clique as a subgraph but as a minor, then the argument becomes considerably more complicated, but can still be done. Hence, the case where $G$ contains a large enough clique minor can be solved efficiently.

Finally, consider the third case of the trinity lemma, where $G$ contains a large wall $W$ and we are given a flat layout of $W$, its extensions and the apices. This is the tricky bit. However, one can show that if $W$ is large enough, then it must contain a subwall $W'$, which is still large, does not contain any of the $s_i$'s or $t_i$'s and is "homogeneous" with respect to the apices. Informally, homogeneous means that every type of a small part of the wall with respect to the apices is realised sufficiently often all over the subwall $W'$. In [76], Robertson and Seymour show how such a homogeneous subwall can be constructed efficiently. To simplify the presentation, assume that $W'$ has actually no direct connection to the apices (other than those using vertices of $W \setminus W'$). Now suppose there are $k$ vertex-disjoint paths connecting $s_i$ and $t_i$, $1 \leq i \leq k$. Some of these paths may use parts of $W'$. As none of the endpoints $s_i, t_i$ is in $W'$, the paths merely cross $W'$, although they may do so in a rather irregular and complicated way. However, it can be shown that if $W'$ is homogeneous and large enough, then any such set of paths can be rerouted so as to avoid a central vertex $v$ of the wall (recall from above that the central vertices are those in the middle of the wall). This implies, that $k$ vertex-disjoint paths connecting $s_i, t_i$ exist in $G$ if, and only if, such paths exist in $G-v$. Hence, we can remove the central vertex $v$ and start the whole procedure again on the smaller graph.

It seems intuitively obvious that on a very large wall, everything that can be routed through the wall can be routed without using the central vertex. A formal proof of this is extremely complicated and uses a major part of the deep structure theory developed in the graph minor series.

As mentioned above, the solution to the disjoint paths problem was given by Robertson and Seymour in [76]. In fact, they solve the following more general problem. A *rooted graph* $(G, v_1, \ldots, v_k)$ is a graph $G$ together with vertices $v_i \in V(G)$. A rooted graph $(H, t_1, \ldots, t_k)$ is a minor of $(G, v_1, \ldots, v_k)$, if there is a minor map $\mu$ from $H$ to $G$ such that $v_i \in \mu(t_i)$ for all $1 \leq i \leq k$.

**Theorem 5.9** (*Robertson, Seymour* [76]) *The following problem is fixed-parameter tractable with a cubic fpt algorithm.*

> $p$-ROOTED-MINOR
>     *Input:* Rooted graphs $(G, v_1, \ldots, v_k)$, $(H, t_1, \ldots, t_k)$.
> *Parameter:* k.
>     *Problem:* Is $(H, t_1, \ldots, t_k)$ a minor of $(G, v_1, \ldots, v_k)$?

Clearly, this implies Theorem 5.8 and also Corollary 5.3. This is a truly remarkable consequence of the proof of the graph minor theorem. Note, however, that the statement is purely existential. For every minor ideal there is a finite set of excluded minors and for each member $H$ of the set we can decide in cubic time, whether a graph $G$ contains $H$ as a minor. The theory does not yield an algorithm to compute a set of excluded minors and hence it only states the existence of a polynomial time membership test but not an actual algorithm. We come back to this in Section 5.4 where we consider ways in which to overcome this non-constructive element in the theory.

### 5.3 The Structure of $H$-Minor Free Graphs

The proof of the graph minor theorem relies on a structure theory for graphs $G$ excluding a fixed graph $H$ as a minor. We have already seen some of the results developed in the proof. In this section we focus on describing the structure of graphs in terms of simple building blocks into which they can be decomposed.

The key to the decomposition theorem we are going to describe is once again the grid theorem, or in this case the trinity lemma as described in the previous section. Clearly, as $G$ excludes a fixed graph $H$ as a minor, it is obvious that, if we choose the values for $k$ and $h$ correctly, of the three cases of the trinity lemma, the second is impossible: if $G$ excludes $H$ it cannot contain a large clique minor. Further, if $G$ has small tree-width, then it can be decomposed into subgraphs of constant size. Hence, we primarily have to deal with the third case, where $G$ has large tree-width but does not contain a large clique minor.

Recall our exploration of the trinity lemma in the previous section. Let us assume that $G$ is highly connected. If not, we first decompose it into parts that are highly connected. We will come back to this later.

As $G$ has high tree-width it must contain a large wall as a subdivision. This wall may contain "crossings", in particular there may be a bounded number of apices and vortices. As explained before, apart from the vortices and the apices, the rest of the graph, the extensions, must fit nicely into the planar structure of the wall, i.e. they fit into the individual bricks. So far, however, we only have discussed the interior of the wall. There may be more to the graph, which is connected to the wall only through the perimeter. These connections cannot be too wild, though, as otherwise we would again find a large clique minor.

We can now subdivide the exterior cycle of the wall into a bounded number of regions and glue some of them together. In this way we obtain a graph that can be embedded into a surface of bounded genus: any such surface can be obtained from a convex polygon in the plane by gluing some edges together. Hence, after removing a bounded number of apices and vortices we obtain a graph that can be embedded into a surface of bounded genus. We say that $G$ has *almost bounded genus*. Recall that we assumed that $G$ is highly connected. If it is not, then we can decompose it into pieces with this property. This realisation is the main structural theorem in Robertson and Seymour's proof of the graph minor theorem: *if $\mathcal{C}$ is a class of graphs excluding a fixed minor $H$, then every graph $G \in \mathcal{C}$ can be decomposed into graphs that have almost bounded genus*.

We still have to make precise what we mean by "decomposing a graph". Intuitively, we recursively find a small separator in the graph and split the graph along the separator until the remaining graph is highly connected, and hence no such separators can be found. However, by doing so some information is lost. Let $G$ be a graph and $X$ be a small separator. We want to decompose the graph into subgraphs each containing $X$ and a component of $G - X$. Clearly, in a graph obtained from $X$ and a component $C$ of $G \setminus X$, we lose the connections between elements of $X$ through the other components of $G \setminus X$. In particular, elements of $X$ which are far apart in $X \cup C$ can be close together in other components and hence in $G$. This loss of information in the decomposition process needs to be avoided. A rather drastic approach, which we take here, is to add all possible edges between elements of the separator $X$, i.e. to turn $X$ into a clique.

Let $\mathcal{T} := (T, (B_t)_{t \in V(T)})$ be a tree-decomposition of a graph $G$ and let $t \in V(T)$ be a node with neighbours $t_1, \ldots, t_k$. The *torso* $[B_t]$ of the bag $B_t$ is defined as $G[B_t] \cup \bigcup_{i=1}^{k} K[B_{t_i}]$, where $K[B_{t_i}]$ is the complete graph on the vertex set $B_{t_i}$. The tree-decomposition $\mathcal{T}$ of $G$ is *over a class* $\mathcal{C}$ of graphs if the torsi of all bags in $\mathcal{T}$ belong to $\mathcal{C}$.      *torso*, $[B_t]$

*tree-decomposition over $\mathcal{C}$*

**Example 5.10** *Figure 5.10 shows a tree-decomposition of a graph over the class of triangles. Part b) shows the tree-decomposition and Part c) the corresponding torsi.*

A graph $G$ is called *decomposable* over a class $\mathcal{C}$ if it has a tree-decomposition over $\mathcal{C}$. For every class $\mathcal{C}$ we denote the class of graphs decomposable over $\mathcal{C}$ by $\mathfrak{D}(\mathcal{C})$. It is not hard to see that if $\mathcal{C}$ is minor closed then so is $\mathfrak{D}(\mathcal{C})$.      $\mathfrak{D}(\mathcal{C})$

**Example 5.11** *Let $\mathcal{C}_{k+1}$ be the class of graphs of order at most $k+1$ and let $\mathcal{T}_k$ be the class of graphs of tree-width at most $k$. Then $\mathcal{T}_k = \mathfrak{D}(\mathcal{C}_{k+1})$.*      ⊣

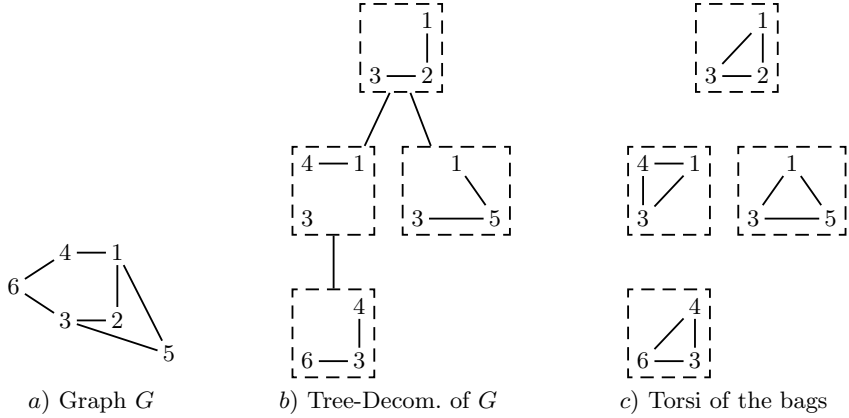

a) Graph $G$    b) Tree-Decom. of $G$    c) Torsi of the bags

**Fig. 16.** Tree-Decomposition over the class of triangles

Robertson and Seymour's structure theorem for classes of graphs excluding a minor can now be reformulated as follows.

**Theorem 5.12** (*Robertson, Seymour* [77]) *For every minor ideal $\mathcal{D}$ there is a class $\mathcal{C}$ of graphs of almost bounded genus such that $\mathcal{D} \subseteq \mathfrak{D}(\mathcal{C})$.*

We will not make the notion of "almost bounded genus" precise here and instead refer to [77] or to [30, Chapter 12] which contains a more elaborate introduction to the theory. For the applications we have in mind, we do not have to work with almost bounded genus graphs, vortices and apices directly but can use a simpler version of the structure theorem. This relies on the following lemma, proved by Grohe in [52].

*local tree-width*    The *local tree-width* is the function ltw : GRAPHS $\times \mathbb{N} \to \mathbb{N}$ defined as

$$\mathrm{ltw}(G, r) := \max\{\,\mathrm{tw}\bigl(G[N_r(v)]\bigr) : v \in V(G)\},$$

where $N_r(v)$ is the $r$ neighbourhood of $v$, i.e. the set of vertices of distance at most $r$ from $v$. That is, the local tree-width of a graph assigns to every radius $r \in \mathbb{N}$ the maximal tree-width of an $r$-neighbourhood in the graph $G$. See Section 7.3 for more on local tree-width.

**Lemma 5.13** (*Grohe* [52]) *Let $S$ be a surface. Then the class of all minors of graphs almost embeddable into $S$ has linear local tree-width.*

For all $\lambda, \mu \geq 1$ define

$$\mathcal{L}(\lambda) := \{G : \mathrm{ltw}(H, r) \leq \lambda \cdot r \text{ for all } H \preccurlyeq G\}$$

and

$$\mathcal{L}(\lambda, \mu) := \{G : \text{ there is } X \subseteq V(G), |X| \leq \mu \text{ s.th. } G \setminus X \in L(\lambda)\}.$$

Then, the previous lemma implies the following simpler structure theorem that will be used in later sections.

**Theorem 5.14** *For every minor ideal $\mathcal{D}$ there exist $\lambda, \mu \geq 1$ such that $\mathcal{D} \subseteq \mathfrak{D}(\mathcal{L}(\lambda, \mu))$.*

Furthermore, Grohe proves the existence of an algorithm for computing the decompositions over $\mathcal{L}(\lambda, \mu)$, based on the following lemma.

**Lemma 5.15** (*Grohe* [52]) *Let $\mathcal{C}$ be a minor closed class of graphs. Then there is a polynomial-time algorithm that, given a graph $G$, either computes a tree-decomposition of $G$ over $\mathcal{C}$ or rejects $G$, if no such decomposition exists.*

Taking $\mathcal{C}$ to be $\mathcal{L}(\lambda, \mu)$, the lemma implies the existence of an algorithm for computing tree-decompositions over $\mathcal{L}(\lambda, \mu)$. However, the algorithm outlined in [52] uses non-constructive elements of the graph minor theory and hence, while proving the existence of an algorithm, does not actually state one.

In [29], Demaine, Hajiaghayi and Kawarabayashi proved that the decompositions as guaranteed by Theorem 5.12 can be computed in polynomial time for every fixed class of graphs excluding at least one minor $H$.

**Theorem 5.16** (*Demaine, Hajiaghayi, Kawarabayashi* [29]) *For every fixed $H$, there is a polynomial-time algorithm for computing the decompositions of $H$-minor free graphs as stated in Theorem 5.12.*

From this, for each fixed $H$, a polynomial time algorithm which computes a tree-decomposition of an $H$-minor free graph $G$ over $\mathcal{L}(\lambda, \mu)$, for suitable values of $\lambda, \mu$, can easily be derived.

One may wonder why we only considered classes $\mathcal{L}(\lambda)$ of *linear* local tree-width instead of classes of graphs where the local tree-width is bounded by a polynomial $p(r)$ or even worse. In [24], Demaine and Hajiaghayi showed that minor closed classes of bounded local tree-width always have linear local tree-width. Hence, there is no need to consider non-linear local tree-width here, as all classes $\mathcal{L}(\lambda)$ are minor-closed.

### 5.4 Computing Excluded Minor Characterisations

Recall from Section 5.1 that every minor ideal can be characterised by a finite set of excluded minors (Theorem 5.2) and that for each fixed $H$ it is decidable in cubic time whether a graph $G$ contains $H$ as a minor (Theorem 5.9). As a consequence we obtain Corollary 5.3 stating that every minor ideal can be decided in cubic time. Note that the result contains a non-constructive element as it does not give a way to compute the excluded minors for a minor ideal. For instance, while we know that the class of knotlessly embeddable graphs can be decided in cubic time, no algorithm for doing so is actually known.

This naturally raises the question whether this non-constructive element can be removed from the proof, i.e. whether characterisations of minor ideals in terms of their excluded minors can be computed. Clearly, to state this precisely, we have to specify how we want to represent a minor ideal as an input to an algorithms and also what exactly we want to understand by a characterisation of a minor ideal in terms of excluded minors.

*obstruction*    Let $\mathcal{C}$ be a minor ideal. A graph $H$ is an *obstruction* for $\mathcal{C}$ if $H$ is an excluded minor of $\mathcal{C}$ but for all $H' \preccurlyeq H$ with $H' \neq H$ we have $H' \in \mathcal{C}$. Hence, obstructions
$\mathcal{O}(\mathcal{C})$    are minimal excluded minors. We denote the set of obstructions of $\mathcal{C}$ by $\mathcal{O}(\mathcal{C})$. It is easily seen that for all minor ideals $\mathcal{C}$, $\mathcal{O}(\mathcal{C})$ is unique up to isomorphism and it is finite by the Graph Minor Theorem. We will therefore take $\mathcal{O}(\mathcal{C})$ as the characterisation of minor ideals we want to compute.

This leaves us with the question how to specify a minor ideal as an input for algorithms. A natural choice is to provide a Turing-machine deciding the ideal and use this as input. However, Fellows and Langston [44] observed that there is no algorithm which, given a Turing-machine deciding a minor ideal $\mathcal{C}$, computes the set $\mathcal{O}(\mathcal{C})$. Later, Courcelle, Fellows and Langston [15] showed that there is no algorithm which, given an MSO-sentence defining a minor ideal $\mathcal{C}$, computes $\mathcal{O}(\mathcal{C})$.

On the other hand, it is known that obstructions can be computed for a number of natural minor ideals. For instance, for all $k \geq 1$ the obstructions can be computed for the class $\mathcal{T}_k$ of all graphs of tree-width $\leq k$ (see [60]), for the class $\mathcal{B}_k$ of all graphs of branch-width $\leq k$ (see [50]) and for the class $\mathcal{G}_k$ of graphs of genus $\leq k$ (this follows from [81] or a combination of [85] and [43]).

Fellows and Langston were the first to study algorithmic issues related to the graph minor theorem and ways to overcome its non-constructiveness. In [43], they propose a general method for computing obstruction sets based on a generalisation of the Myhill-Nerode theorem of formal language theory to "graph languages". Adler, Courcelle, Grohe and Kreutzer[8] present a similar method for computing obstruction sets based on definability in monadic second-order logic (see [1]). We will give a brief presentation of this method and illustrate it by an example. For all minor ideals $\mathcal{C}$ and $\mathcal{D}$, their union $\mathcal{C} \cup \mathcal{D}$ is minor closed and hence a minor ideal. We will show below that the set of obstructions for $\mathcal{C} \cup \mathcal{D}$ can be computed from $\mathcal{O}(\mathcal{C})$ and $\mathcal{O}(\mathcal{D})$.[9] The proof of this result also contains a nice application of the Trinity Lemma 5.7.

We first establish some lemmas which are all easily proved using well-known results from automata theory and the connection between monadic second-order logic on trees and tree-automata (see e.g. [84,10]).

**Lemma 5.17** *There is an algorithm which, given a formula $\varphi \in$ MSO defining a minor ideal $\mathcal{C}$, computes a formula $\psi \in$ MSO defining $\mathcal{O}(\mathcal{C})$.*

Proof. A graph $H$ is an obstruction for $\mathcal{C}$ if $H \notin \mathcal{C}$ but $H - v \in \mathcal{C}, H - e \in \mathcal{C}$ and $H/e \in \mathcal{C}$ for all $v \in V(H)$ and $e \in E(H)$. Given the formula $\varphi$ defining $\mathcal{C}$, this can be easily be formalised in MSO. □

The next lemma is based on a pumping lemma for tree-automata (see [10]).

---

[8] The proof presented here follows a suggestion by Bruno Courcelle simplifying the original proof of the result in [1].
[9] Note that the analogous problem for $\mathcal{C} \cap \mathcal{D}$ is trivial.

**Lemma 5.18** *There is an algorithm which, given a formula $\varphi \in$ MSO so that the class $\mathrm{Mod}(\varphi) := \{H : H \models \varphi\}$ is finite (up to isomorphism) and a $k \in \mathbb{N}$ such that $\mathrm{tw}(H) \leq k$ for all $H \in \mathrm{Mod}(\varphi)$, computes $\mathrm{Mod}(\varphi)$.*

*Proof (sketch).* Suppose $\varphi$ has only finitely many models each of tree-width $\leq k$. As we are given $k$ explicitly, we can use the interpretation defined in Section 3.3 to encode the models of $\varphi$ as coloured trees over a suitable alphabet and reduce the problem of computing the models of $\varphi$ to the problem of computing the corresponding tree-encodings. An upper bound for the size of these models can then be derived from a version of the pumping lemma of formal language theory for classes of trees definable by tree-automata. From this bound on the size, the actual models of $\varphi$ can easily be computed. $\square$

The previous lemmas together with the Graph Minor Theorem immediatly imply the following corollary which is the basis of the method for computing obstruction sets proposed in [1].

**Corollary 5.19** *There is an algorithm which, given a formula $\varphi \in$ MSO defining a minor ideal $\mathcal{C}$ and a $k \in \mathbb{N}$ such that $\mathrm{tw}(H) \leq k$ for all $H \in \mathcal{O}(\mathcal{C})$, computes the set $\mathcal{O}(\mathcal{C})$.*

As an application of the result we show that the obstructions for the union $\mathcal{C} \cup \mathcal{D}$ of minor ideals $\mathcal{C}, \mathcal{D}$ can be computed from the sets $\mathcal{O}(\mathcal{C})$ and $\mathcal{O}(\mathcal{D})$. For this, we have to show that $\mathcal{C} \cup \mathcal{D}$ is MSO-definable and to establish an upper bound on the tree-width of its obstructions.

It is easily seen that for any fixed graph $H$ there is an MSO-formula $\varphi_H$ which is true in a graph $G$ if, and only if, $H \preccurlyeq G$. This follows immediately from the definition of minors in terms of minor maps and images as presented in Section 5.1. To define $\mathcal{C} \cup \mathcal{D}$ in MSO note that $G \in \mathcal{C} \cup \mathcal{D}$ if, and only if, $G$ either excludes a minor from $\mathcal{O}(\mathcal{C})$ or a minor from $\mathcal{O}(\mathcal{D})$. As we have seen, this is MSO-definable and a corresponding formula can easily be computed. It remains to establish a bound on the tree-width of the obstructions.

**Lemma 5.20** *Let $\mathcal{C}$ and $\mathcal{D}$ be minor ideals and let $\mathcal{U} := \mathcal{C} \cup \mathcal{D}$. There is an algorithm which, given $\mathcal{O}(\mathcal{C})$ and $\mathcal{O}(\mathcal{D})$ as input, computes a number $k \in \mathbb{N}$ such that $\mathrm{tw}(H) \leq k$ for all $H \in \mathcal{O}(\mathcal{U})$.*

*Proof (sketch).* Suppose $G \in \mathcal{O}(\mathcal{U})$. Hence, $G \notin \mathcal{U}$ but $G - v \in \mathcal{U}$ for all $v \in V(G)$. It follows that there are $H \in \mathcal{O}(\mathcal{C})$ and $I \in \mathcal{O}(\mathcal{D})$ such that $H \preccurlyeq G$ and $I \preccurlyeq G$. Let $k := \max\{|H|, |I|\} + 1$ and choose $h$ "large enough", where the meaning of large enough will become clear later.

By the Trinity Lemma 5.7, either a) $\mathrm{tw}(G) \leq f(k, h)$ for some computable function $f$, or b) $K_k \preccurlyeq G$ or c) there is a subset $X \subseteq V(G)$ with $|X| < \binom{k}{2}$, a wall $W$ of height $h$ in $G \setminus X$, and a flat layout of the compass of $W$ in $G \setminus X$.

Suppose c) applies. It follows from a result by Robertson and Seymour in [76] that if $h$ is chosen large enough then there is a vertex $v$ in the wall $W$ (the middle vertex) such that $G - v$ still contains $H$ and $I$ as minors, contradicting the minimality of the obstruction $G$. Hence, case c) is impossible. The idea to choose the middle vertex is same as in the proof of Theorem 5.8 described in Section 5.2.

For b), if $G$ contains a $K_k$ minor then there is a strict subgraph $G' \subsetneq G$ containing a $K_{k-1}$ minor. Hence, by the choice of $k$, $G'$ contains $H$ and $I$ as minors, contradicting the minimality of $G$. Thus, case b) is impossible as well.

Finally, in a) the tree width of $G$ is bounded by a computable function in $h$ and $k$ and we have found a uniform upper bound for the tree-width of $G$ which concludes the proof. □

**Corollary 5.21** ([1]) *For all minor ideals $\mathcal{C}, \mathcal{D}$ the set $\mathcal{O}(\mathcal{C} \cup \mathcal{D})$ is computable from the sets $\mathcal{O}(\mathcal{C})$ and $\mathcal{O}(\mathcal{D})$.*

Using a similar approach it was shown in [1] that obstructions can be computed for other natural minor ideals. In particular, if $\mathcal{C}$ is a minor ideal whose obstructions are known, then the obstructions can be computed for the class $\mathcal{C}_{\text{apex}}$ of *apex graphs over $\mathcal{C}$*, defined as

$$\mathcal{C}_{\text{apex}} := \{G : \text{there is } v \in V(G) \text{ such that } G - v \in \mathcal{C}\}.$$

However, there remain interesting open problems.

**Open Problem 5.22** *1. Is there an algorithm which, given $\lambda \geq 0$, computes the obstructions $\mathcal{O}(\mathcal{L}(\lambda))$? See Section 5.3 for a definition of $\mathcal{L}(\lambda)$ and $\mathcal{L}(\lambda, \mu)$. Note that, by using the computability of $\mathcal{O}(\mathcal{C}_{\text{apex}})$ from $\mathcal{O}(\mathcal{C})$, the set $\mathcal{O}(\mathcal{L}(\lambda, \mu))$ can be computed from $\mathcal{O}(\mathcal{L}(\lambda))$, for all $\mu \geq 0$.*

*2. If $\mathcal{C}$ is a minor ideal whose obstructions are given, can we compute the obstructions of the class $\mathfrak{D}(\mathcal{C})$ of graphs tree-decomposable over $\mathcal{C}$?*

A solution for both open problems would be particularly interesting as every minor ideal is a subclass of a class $\mathfrak{D}(\mathcal{L}(\lambda, \mu))$ for some $\lambda, \mu \geq 0$.

## 6 Monadic Second-Order Logic Revisited

Recall from Section 3.5 that for each $k$, the $\text{MSO}_2$-theory of the class $\mathcal{T}_k$ of graphs of tree-width at most $k$ is decidable. The aim of this section is to prove a kind of converse, also due to Seese.

**Theorem 6.1** (*Seese* [79]) *If $\mathcal{C}$ is a class of graphs with decidable $\text{MSO}_2$-theory, then $\mathcal{C}$ has bounded tree-width.*

The proof of the theorem crucially relies on the excluded grid theorem (Theorem 5.4) and the fact that the MSO-theory of grids is undecidable. The latter can easily be established using tiling systems or by a direct encoding of the run of Turing-machines using MSO-formulas (see e.g. [8]).

Suppose $\mathcal{C}$ has a decidable $\text{MSO}_2$-theory but unbounded tree-width. Then, by the excluded grid theorem, for all $n \geq 1$, there is a graph $G_n \in \mathcal{C}$ containing $G_{n \times n}$ as a minor. The key to the theorem is to show that grid minors can be defined in MSO$_2$. Hence, the (undecidable) MSO-theory of grids can be reduced to the MSO-theory of $\mathcal{C}$ contradicting the assumption that the latter is decidable.

We start by showing how walls can be formalised in MSO$_2$. The extension to grids follows easily. Let $G$ be a graph and consider an MSO$_2$-formula formalising the following.

1. There are two sets $\mathcal{H}$ and $\mathcal{V}$ of edges, each of which induces a set of pairwise vertex disjoint paths (which we will think of as horizontal and vertical paths in a wall).
2. For all $P \in \mathcal{H}$ and $Q \in \mathcal{V}$, $P \cap Q$ is a subpath of both, $P$ and $Q$. Further, $V(P \cap Q) \cap V(H) = \varnothing$ for all $H \in (\mathcal{V} \cup \mathcal{H}) \setminus \{P, Q\}$.
3. There is a path $L \in \mathcal{V}$ such that the intersection of $L$ with each $Q \in \mathcal{H}$ contains an endpoint of $Q$ ($L$ is the left-most vertical path in the wall). Once we have $L$, we can give the horizontal paths $P \in \mathcal{H}$ a direction, where we say that $p \in V(P)$ is to *the left* of $p' \in V(P)$, if the subpath of $P$ containing $p'$ and a vertex in $L$ also contains $p$.
4. There is a path $T \in \mathcal{H}$ such that the intersection of $T$ with each $P \in \mathcal{V}$ contains an endpoint of $P$ ($T$ is the top-most horizontal path in the wall). As with horizontal paths, we can now use $T$ to give the vertical paths $P \in \mathcal{V}$ a direction and say that $p \in V(P)$ is *above* $p' \in V(P)$.
5. For each path $P \in \mathcal{V}$ except $L$ there is a path $P' \in \mathcal{V}$ (the path immediately to the left of $P$) such that for all $Q \in \mathcal{H}$: if $p \in V(P \cap Q)$ and $p' \in V(P' \cap Q)$ are vertices in the intersection of $Q$ and $P, P'$, then $p'$ is to the left of $p$ in $Q$ and there is no $S \in \mathcal{H}$ such that any $s \in V(S \cap Q)$ lies in the subpath of $Q$ between $p$ and $p'$.
6. The analogue condition for horizontal paths.

Clearly, the various conditions are MSO$_2$-definable. Now, if $\mathcal{V}$ and $\mathcal{H}$ satisfy the conditions above, then they generate a wall in $G$ and conversely, the disjoint horizontal and vertical paths in a wall satisfy the conditions. Finally, it is easily seen that the class of grids can be defined in the class of walls and hence grid minors are MSO$_2$-definable in graphs.

Note that here we crucially use the fact the we are working with MSO$_2$-formulas and hence can quantify over the edge sets of disjoint paths. In MSO$_1$ we could only try to quantify over the vertex set of disjoint paths. However, if there are sufficiently many edges between these vertices, there is no way we can give the paths an orientation, e.g. define paths being to the left of others. And clearly, we cannot expect clique-minors to be definable in MSO$_1$ as, by Theorem 4.20, the MSO$_1$-theory of graph classes of bounded clique-width is decidable and hence there are classes with decidable MSO$_1$-theory but unbounded tree-width.

## 7 First-Order Model-Checking

In Section 3.4 and 4.3 we showed that the model-checking problem for monadic second-order logic is solvable in linear time for any fixed formula on classes of graphs of bounded tree- or clique-width. There is not much hope for extending these results to other or larger classes of graphs such as planar graphs or graphs of bounded degree. This follows immediately from the following theorem by Garey, Johnson and Stockmeyer and the fact that 3-colourability is MSO-definable.

**Theorem 7.1** (*Garey, Johnson, Stockmeyer* [49]) *3-colourability is* NP-*complete on the class of planar graphs of degree at most* 4.

However, first-order logic is tractable on much larger classes of graphs. For instance, Seese [80] showed that first-order logic admits linear time model-checking (for a fixed formula) on any class of graphs of bounded degree. The same complexity bound was later obtained by Frick and Grohe [47] for planar graphs and classes of graphs of bounded *local tree-width*, a notion that properly extends both planarity and bounded degree (see below). The important property of first-order logic that makes these results possible is *locality*.

The section is structured as follows. In Section 7.1 we introduce the concept of locality and present Gaifman's theorem. In Section 7.2 we apply locality to obtain fixed-parameter algorithms for first-order model-checking on graph classes of bounded degree. The algorithms developed in this section can be applied in a much more general context using the concept of localisation of graph invariants. This will be formally defined in Section 7.3. In Section 7.4 we present fixed-parameter algorithms for first-order model-checking on $H$-minor free graphs. Finally, in Section 7.5 we discuss fixed-parameter tractability for first-order model-checking in a broader context and exhibit some limits to tractability.

### 7.1 Locality of First-Order Logic

Let $G$ be a graph. Recall that the *distance* $d^G(u, v)$ between two vertices $u, v \in V(G)$ is the length of the shortest path from $u$ to $v$ or $\infty$ if there is no such path. Further, for every $v \in V(G)$ and $r \in \mathbb{N}$ we define the *r-neighbourhood* of $v$ in $G$ as the set

$$N_r^G(v) := \{w \in V(G) : d^G(v, w) \leq r\}$$

of vertices of distance at most $r$ from $v$. For a set $W \subseteq V(G)$ we set $N_r^G(W) := \bigcup_{v \in W} N_r^G(v)$. We omit the index $\cdot^G$ whenever $G$ is clear from the context.

If $\sigma$ is a signature and $A$ is a $\sigma$-structure, we define the distance $d^A(a, b)$ and the $r$-neighbourhood $N_r^A(a)$ in terms of the Gaifman-graph $\mathcal{G}(A)$ of $A$,[10] i.e. $N_r^A(a)$ is the set of elements of distance at most $r$ from $a$ in the Gaifman-graph.

It is easily seen that for any fixed $r \in \mathbb{N}$ "distance at most $r$" is first-order definable, that is, for every $r \in \mathbb{N}$ there is a formula $\mathrm{dist}_{\leq r}(x, y)$ such that for all structures $A$ and all $u, v \in V(A)$

$$A \models \mathrm{dist}_{\leq r}(u, v) \qquad \text{iff} \qquad d^A(u, v) \leq r.$$

Similarly, there are formulas $\mathrm{dist}_{>r}(x, y)$ and $\mathrm{dist}_{<r}(x, y)$ defining distance $> r$ and $< r$ respectively. To improve readability we will write $\mathrm{dist}(x, y) \leq r$ instead of $\mathrm{dist}_{\leq r}(x, y)$ and likewise for the other formulas.

---
[10] See Section 2 for a definition of Gaifman-graphs.

A first-order formula $\varphi(x)$ is *r-local* if for every structure $A$ and all $a \in V(A)$

$$A \models \varphi(a) \quad \text{iff} \quad A\big[N_r^A(a)\big] \models \varphi,$$

where $A\big[N_r^A(a)\big]$ denotes the substructure of $A$ induced by $N_r^A(a)$. Hence, truth of an $r$-local formula at an element $a$ in a structure only depends on its $r$-neighbourhood. A formula $\varphi(x)$ is *local* if it is $r$-local for some $r \in \mathbb{N}$.

A *basic local sentence* is a first-order sentence of the form

$$\exists x_1 \ldots \exists x_k \Big( \bigwedge_{1 \leq i < j \leq k} \text{dist}(x_i, x_j) > 2r \wedge \bigwedge_{i=1}^{k} \vartheta(x_i) \Big)$$

where $\vartheta(x)$ is local. In 1981, Gaifman showed that every first-order sentence is equivalent to a Boolean combination of basic local sentences.

**Theorem 7.2 (***Gaifman* [48]**)** *Every first-order sentence is equivalent to a Boolean combination of basic local sentences. Furthermore, there is an algorithm that, given a first-order formula as input, computes an equivalent Boolean combination of basic local sentences.*

A first-order formula is in *Gaifman Normal Form* (GNF), if it is a Boolean combination of basic local sentences. Gaifman's original proof is by an explicit translation of first-order formulas into formulas in GNF. A proof sketch along this lines can also be found in the survey paper [53]. A different, model-theoretical proof can be found in [37, Section 2.5].

The translation of formulas into Gaifman normal form is effective. However, it has recently been shown [23] that this translation may involve a non-elementary blow-up in the size of the sentence.

**Theorem 7.3 (***Dawar, Grohe, Kreutzer, Schweikardt* [23]**)** *Let $\sigma := \{E\}$ be the signature of graphs. For every $h \geq 1$ there is an FO$[\sigma]$-sentence $\varphi_h$ of size $\mathcal{O}(h^4)$ such that every FO$[\sigma]$-sentence in Gaifman normal form that is equivalent to $\varphi_h$ on the class of finite trees has size at least tower$(h)$, where tower$(h)$ denotes a tower of 2s of height $h$.*

From a practical point of view, this renders algorithms using Gaifman's theorem useless, no matter what their theoretical complexity might be.

**Example 7.4** *Recall that a* dominating set $X$ *in a graph* $G$ *is a set* $X \subseteq V(G)$ *such that for all* $v \in V(G)$, $v \in X$ *or there is a* $u \in X$ *and* $\{u, v\} \in E(G)$. *For* $k \in \mathbb{N}$, *the formula*

$$\varphi_k := \exists x_1 \ldots \exists x_k \forall y \Big( \bigvee_{1 \leq i \leq k} (x_i = y \vee Eyx_i) \Big)$$

*is true in a graph $G$ if, and only if, $G$ has a dominating set of size at most $k$.*

*To convert this into an equivalent sentence in Gaifman normal form, we first observe that no connected graph of diameter at least $3k + 1$ can have a*

*dominating set of size at most k. Here, the diameter of a graph is the maximum of the distance between any two vertices.*

*Hence, on connected graphs, the formula $\varphi_k$ above is equivalent to the conjunction of the basic local sentence*

$$\psi := \neg \exists x_1 \exists x_2 \mathrm{dist}(x_1, x_2) > 3k+1,$$

*saying that the diameter of $G$ is greater than $3k+1$, and the basic local sentence $\exists x \chi(x)$, where $\chi(x)$ is the $3k+1$-local formula*

$$\exists y_1 \in N_{3k+1}(x) \ldots \exists y_k \in N_{3k+1}(x) \forall z \in N_{3k+1}(x) \bigvee_{1 \leq i \leq k} (y_i = z \vee E z y_i).$$

*Note that this formula correctly defines the existence of a dominating set of size $k$ only in connected graphs, as in graphs with more than one component there may exist a dominating set of size $k$ even though there are vertices $x_1, x_2$ of distance greater than $3k+1$. Adapting the formula to this case requires a little more effort.* ⊣

### 7.2 First-Order Logic on Graphs of Bounded Degree

As a first application of the use of Gaifman's locality theorem for algorithmic meta theorems we consider graph classes of bounded degree.

**Definition 7.5** *A class $\mathcal{C}$ of graphs has* bounded degree *if there is a $d \in \mathbb{N}$ such that $\Delta(G) \leq d$ for all $G \in \mathcal{C}$.*

In 1996, Seese [80] showed that model-checking for a fixed first-order sentence can be done in linear time on graph classes of bounded degree.

**Theorem 7.6** (*Seese* [80]) *For any class $\mathcal{C}$ of graphs of bounded degree and any fixed first-order sentence it can be decided in linear time whether $G \models \varphi$ for a graph $G \in \mathcal{C}$. In other words, first-order model-checking on $\mathcal{C}$ is fixed-parameter tractable by a linear fpt algorithm.*

*Proof.* The proof method we use here is essentially the method used by Frick and Grohe to show a similar result for planar graphs.

Let $\varphi$ and $G \in \mathcal{C}$ be given. We first convert $\varphi$ into Gaifman normal form, i.e. into a Boolean combination of basic local sentences. As Boolean combinations are easy to deal with, we only need to consider basic local sentences of the form

$$\psi := \exists x_1 \ldots \exists x_k \Big( \bigwedge_{1 \leq i < j \leq k} \mathrm{dist}(x_i, x_j) > 2r \wedge \bigwedge_{i=1}^{k} \vartheta(x_i) \Big)$$

where $\vartheta(x)$ is $r$-local for some $r \in \mathbb{N}$.

To check whether $\psi$ is true in $G$ we proceed in two steps. First, we test for all $v \in V(G)$ if $G[N_r^G(v)] \models \vartheta$. As $G$ has degree bounded by some constant $d$, the size of $N_r^G(v)$ is constant and hence this can be decided in constant time.

Colour all vertices $v$ *red* for which $G\bigl[N_r^G(v)\bigr] \models \vartheta$ and let $Q$ be the set of *red* vertices. Now, $G \models \psi$ if $Q$ contains $k$ vertices of pairwise distance $> 2r$.

In the second step we search for $k$ such vertices. For this, we use the greedy algorithm shown in Figure 17. The algorithm proceeds as follows. In lines 2–6 of

```
1: L := ∅
2: while Q ≠ ∅ do
3:     choose v ∈ Q arbitrarily
4:     L := L ∪ {v}
5:     Q := Q \ N_{2r}(v)
6: end while
7: if |L| ≥ k then
8:     accept G
9: else
10:    if G[N_{2r}(L)] ⊨ ∃x_1 ... x_k(⋀_{i≠j} dist(x_i, x_j) > 2r ∧ ⋀_i "x_i is red") then
11:        accept G
12:    else
13:        reject G
14:    end if
15: end if
```

**Fig. 17.** Algorithm to find $k$ vertices of pairwise distance $> 2r$

the algorithm, we try to choose $k$ *red* vertices of pairwise distance $> 2r$ greedily. If we succeed, i.e. if the set $L$ contains $k$ elements, then we are done and accept $G$. Otherwise, we know that $L$ contains fewer than $k$ vertices which are all *red* and of pairwise distance $> 2r$ and also that any other *red* vertex is within distance $\leq 2r$ of an element of $L$ (otherwise we could add the vertex to $L$). Hence, all *red* vertices of $G$ are contained in the $2r$-neighbourhood $N := N_{2r}[L]$ of $L$. Again, $N$ is of constant size and hence we can check in constant time whether $N$ contains $k$ *red* vertices of pairwise distance $> 2r$. This is done in line 12 by testing whether the graph induced by the neighbourhood satisfies the first-order formula stating that there are $k$ distinct *red* vertices of pairwise distance $> 2r$. □

The previous theorem gives a simple example how locality can be used to obtain efficient model-checking algorithms for first-order logic. As it turns out, a similar scheme can be employed in many cases.

**Theorem 7.7** *Let $\mathcal{C}$ be a class of graphs such that the following problem is fixed-parameter tractable:*

> *Input:* $\varphi \in \mathrm{FO}$, graph $G \in \mathcal{C}$, $v_1, \ldots, v_k \in V(G)$ and $r \in \mathbb{N}$.
> *Parameter:* $r + k + |\varphi|$.
> *Problem:* Decide $G\bigl[N_r^G(v_1, \ldots, v_k)\bigr] \models \varphi$.

*Then model-checking for first-order logic is fixed-parameter tractable on $\mathcal{C}$.*

*Proof.* We proceed as in the proof of Theorem 7.6. By Gaifman's theorem, we may assume that $\varphi$ is a basic local sentence $\exists x_1 \ldots \exists x_k \big(\bigwedge_{i \neq j} \operatorname{dist}(x_i, x_j) > 2r \wedge \bigwedge_i \vartheta(x_i)\big)$, where $\vartheta(x)$ is an $r$-local formula for some $r \in \mathbb{N}$.

In the first step, we compute the set $Q$ of vertices $v \in V(G)$ such that $G[N_r(v)] \models \vartheta(v)$. By assumption, for each $v \in V(G)$ this can be done in time $f(r + 1 + |\vartheta|) \cdot |G|^{\mathcal{O}(1)}$, for some computable function $f : \mathbb{N} \to \mathbb{N}$, and hence the total running time is $f(r + 1 + |\vartheta|) \cdot |G|^{\mathcal{O}(1)}$.

In the second step we aim to find $k$ vertices in $Q$ whose distance is pairwise $> 2r$. Using the algorithm of Figure 17 this can be done in time $f(2r \cdot k + \mathcal{O}(k)) \cdot |G|^{\mathcal{O}(1)}$. Hence, the total running time is $f(2r \cdot k + \mathcal{O}(k)) \cdot |G|^{\mathcal{O}(1)}$. □

While this theorem may appear somewhat artificial, we will see a number of interesting applications of it by considering localisations of graph invariants such as tree-width or rank-width.

### 7.3 Localisation of graph invariants

Let GRAPH denote the class of all finite graphs.

$loc_f(G, r)$    **Definition 7.8** *A* graph invariant *is a function $f$ : GRAPH $\to \mathbb{N}$. For every graph invariant $f$ we define its* localisation *$loc_f$ : GRAPH $\times \mathbb{N} \to \mathbb{N}$ as*

$$loc_f(G, r) := \max \Big\{ f\Big(G[N_r(v)]\Big) : v \in V(G) \Big\}.$$

*A class $\mathcal{C}$ of graphs has* bounded local $f$, *if there is a computable[11] function $h : \mathbb{N} \to \mathbb{N}$ such that $loc_f(G, r) \leq h(r)$ for all $G \in \mathcal{C}$ and $r \in \mathbb{N}$.*

That is, to compute $loc_f(G, r)$ we compute the $r$-neighbourhoods $N := N_r(v)$ of all vertices $v \in V(G)$ and for each such $N$ the value $f(N)$. $loc_f(G, r)$ is then the maximum of these values. In particular, if the problem: given $G$ and $k$, where $k$ is the parameter, to decide whether $f(G) \leq k$ is fixed-parameter tractable, then so is the problem: given $G, r, k$, where $r + k$ is the parameter, to decide if $loc_f(G, r) \leq k$.

**Example 7.9** *Of particular interest is the localisation of tree-width, called* local tree-width *(see also the discussion at the end of Section 5.3). There are a number of interesting examples for graph classes with bounded local tree-width.*

1. *Every graph class of bounded tree-width also has bounded local tree-width (bounded by a constant).*
2. *The class of planar graphs has bounded local tree-width. More precisely, Robertson and Seymour [75] showed that every planar graph of radius $r$ has tree-width $\leq 3r + 1$.*

---
[11] As we are asking for $h$ to be computable, we should call this *effectively bounded local $f$*. But this would make the notation even more clumsy and we therefore refrain from mentioning effectiveness in the sequel.

3. Any class of graphs of bounded degree. This is easily seen as the $r$-neighbourhoods of graphs of degree at most $d$ contain $< d^{r+1}$ vertices. ⊣

Similar to local tree-width we can define local rank-width or clique-width, where we take $f : \text{GRAPH} \to \mathbb{N}$ to be the function assigning to each graph its rank- or clique-width.

Another interesting example is the localisation of the following graph invariant. Let $mec : \text{GRAPH} \to \mathbb{N}$ (*minimal excluded clique*) be the function assigning to each graph $G$ the minimal order of a clique that is not a minor of $G$, i.e. $\quad mec(G)$

$$mec(G) := \min\{k : K_k \not\preceq G\}.$$

Graph classes with locally bounded *mec* are called graph classes with *locally excluded minors* and have been studied by Dawar, Grohe and Kreutzer in [21]. Clearly, every graph class $\mathcal{C}$ with an excluded minor $H$ also locally excludes $H$, i.e. has bounded local *mec*. The converse fails, though, as is witnessed by the following class of graphs. For $k \in \mathbb{N}$ let $S_k$ be the graph obtained from $K_k$ by replacing all edges by internally vertex disjoint paths of length $k$. Now take $\mathcal{C} := \{S_k : k \in \mathbb{N}\}$. Obviously, the minor closure of $\mathcal{C}$ is the class of all graphs, i.e. $\mathcal{C}$ does not exclude a minor. However, it locally excludes minors, as every $k$-neighbourhood of graphs $G \in \mathcal{C}$ excludes $K_k$. Hence, $f : \text{GRAPH} \times \mathbb{N} \to \mathbb{N}$ defined as $f(G, r) := r$ dominates the local *mec* of $\mathcal{C}$.

Note, that $\mathcal{C}$ has bounded local tree-width and hence also provides an example separating proper minor ideals and graph classes of bounded local tree-width. It is easily seen that every class of graphs of bounded local tree-width also locally excludes minors. The converse fails again, as not even every minor ideal has bounded local tree-width. This is witnessed by the class of apex graphs defined as

$$\mathcal{C}_{apex} := \{G : \text{ there is } v \in V(G) \text{ such that } G - v \text{ is planar}\}.$$

In particular, this class contains all grids with one additional vertex adjacent to every vertex in the grid. Hence, $\mathcal{C}_{apex}$ has unbounded local tree-width but clearly excludes $K_6$.

**Lemma 7.10** *The concept of locally excluded minors strictly generalises both excluded minors and bounded local tree-width. That is, every class of graphs that excludes a minor or has bounded local tree-width, also locally excludes minors. The converse fails in both cases.*

The aim of this section is to prove the following theorem.

**Theorem 7.11** *Let $f$ be a graph invariant such that the following is fixed-parameter tractable.*

> MC(FO, $f$)
>     *Input:* Graph $G$ and $\varphi \in$ FO.
> *Parameter:* $f(G) + |\varphi|$.
>     *Problem:* Decide whether $G \models \varphi$.

*Then for every class $\mathcal{C}$ of locally bounded $f$, the problem* $\mathrm{MC}(\mathrm{FO}, \mathcal{C})$ *is fixed-parameter tractable.*

*Proof.* Let $g : \mathbb{N} \to \mathbb{N}$ be a bound for $loc_f(G, \cdot)$ for all $G \in \mathcal{C}$. We first suppose that $f$ is *induced subgraph monotone*, i.e. $f(H) \leq f(G)$ for all $H, G$ such that $H$ is an induced subgraph of $G$, and further has the property that if $G_1, G_2$ are vertex disjoint graphs, then $f(G_1 \cup G_2) \leq \max\{f(G_i) : i = 1, 2\}$. Note that graph invariants such as tree-width, branch-width, clique-width and rank-width all have these properties.

Then the result follows from Theorem 7.7 as follows. Given $\varphi \in \mathrm{FO}$, $G \in \mathcal{C}$, $v_1, \ldots, v_k \in V(G)$ and $r \in \mathbb{N}$, we first compute $H := G[N_r^G(v_1, \ldots, v_k)]$ in polynomial time. Clearly, every component of $H$ has radius at most $k \cdot r$ and hence $f(H) \leq loc_f(G, k \cdot r) \leq g(k \cdot r)$. The assumptions of this lemma then imply that the assumptions of Theorem 7.7 are satisfied and thus we can decide $H \models \varphi$ by fpt-algorithms.

If $f$ does not have the properties above, we can no longer apply Theorem 7.7 directly. Instead, we have to repeat its proof. We leave the details to the reader. □

**Corollary 7.12** *First-order model-checking is fixed-parameter tractable on graph classes of*

– *bounded local tree-width*
– *bounded local rank- or clique-width.*

In the next section we will show that first-order model-checking is fixed-parameter tractable on graph classes excluding at least one minor. We will later consider localisation in this context and show an analogous result for graph classes locally excluding a minor.

### 7.4  First-Order Logic on $H$-Minor Free Graphs

The aim of this section is to show that first-order model-checking is fixed-parameter tractable on every class $\mathcal{C}$ of graphs excluding at least one minor $H$. If we take $|\varphi|$ as the parameter, this was first shown by Flum and Grohe [45] in 2001. That is, for every fixed $H$, the problem is tractable under the parametrization $|\varphi|$. However, the exponential of the polynomials occurring in the running time analysis can depend on $H$. As it turns out, this parametrization is not strong enough to apply our method of localisation to the problem. In [21], therefore, Dawar, Grohe and Kreutzer consider the problem under the parametrization $|\varphi| + |H|$ and show fixed-parameter tractability for this case.

Let us first consider the case where $H$ is fixed and $|\varphi|$ is the parameter. In the light of the previous sections, the proof of the theorem seems rather straightforward: given $G \in \mathcal{C}$, Theorem 5.14 tells us that there are $\lambda, \mu \geq 1$ such that $G$ has a tree-decomposition over $\mathcal{L}(\lambda, \mu)$, i.e. a tree-decomposition such that the torsi of its bags have bounded local tree-width after removal of a few elements, and Theorem 5.16 tells us how to compute the decomposition in polynomial time. Furthermore, we already know how to deal with graphs in $\mathcal{L}(\lambda)$ of bounded local tree-width and extending this to graphs in $\mathcal{L}(\lambda, \mu)$ poses no real problem. And indeed, this is the general idea to show that FO model-checking is FPT on $H$-minor free graphs, although formally implementing the idea requires some care and additional lemmas. To make this precise it is convenient to introduce further notation.

A graph $G$ is the *clique sum* of graphs $G_1$ and $G_2$, denoted $G = G_1 \oplus G_2$, *clique sum*, $\oplus$ if $G_1 \cap G_2$ is a complete graph and $G$ is obtained from $G_1 \cup G_2$ by possibly deleting some edges from $E(G_1 \cap G_2)$. Formally, $V(G) = V(G_1) \cup V(G_2)$, $G_1 \cap G_2$ is a clique and there is a (possibly empty) set $X \subseteq E(G_1 \cap G_2)$ such that $E(G) = E(G_1 \cup G_2) - X$. We write $G = G_1 \oplus_{\overline{v}} G_2$ to indicate that $G$ is the $\oplus_{\overline{v}}$ clique-sum of $G_1$ and $G_2$ and that $V(G_1 \cap G_2) = \overline{v}$.

Recall that a tree-decomposition of a graph $G$ is *over* a class $\mathcal{C}$ of graphs if the torsi $[B_t]$ of all its bags belong to $\mathcal{C}$, where the torso of a bag $B_t$ is obtained from $G[B_t]$ by turning the intersections of $B_t$ with neighbouring bags $B_s$ into cliques. Hence, the graph $G$ is obtained as the clique-sum of its bags, an observation that we will use in the following proofs.

We begin by proving an extension of Courcelle's theorem, this time not by a reduction to trees but by computing MSO-types directly. Recall the definition of MSO and FO $q$-types and the Feferman-Vaught theorem from Section 2.3.

**Lemma 7.13** *Let $\operatorname{tp}_q$ be one of $\operatorname{tp}_q^{\mathrm{FO}}$ and $\operatorname{tp}_q^{\mathrm{MSO}}$. The following problem is fixed-parameter tractable: given*

- *a labelled graph $G$ of tree-width $\leq k$,*
- *tuples $\overline{v}_i \in V(G)^{r_i}$, $0 \leq i \leq m$ for some $m$, such that $G[\overline{v}_i]$ is a clique, and*
- *$q$-types $\Theta_1, \ldots, \Theta_m$,*

*compute $\operatorname{tp}_q(G, \overline{v}_0)$ for all graphs $G' = G \oplus_{\overline{v}_1} H_1 \oplus_{\overline{v}_2} \cdots \oplus_{\overline{v}_m} H_m$ such that $\operatorname{tp}_q(H_i, \overline{v}_i) = \Theta_i$. The parameter is $q + k$.*

*Proof.* Given $G$, we first compute an ordered tree-decomposition $(T, (\overline{b}_t)_{t \in V(T)})$ of $G$ of width at most $k$ (see Definition 3.25). Note that, as the $\overline{v}_i$ induce cliques in $G$, for each $i$ there is at least one $t_i$ such that $\overline{v}_i \subseteq \overline{b}_{t_i}$. Hence, we can assume that for each $0 \leq i \leq m$ there is a leaf $t \in V(T)$ such that $\overline{v}_i = \overline{b}_t$ and that no other leaf contains a vertex from any of the $\overline{v}_i$ for $1 \leq i \leq m$.

For each $t \in V(T)$, let $T_t$ be the subtree of $T$ rooted at $t$ and let $\mathcal{B}_t$ be the set $\mathcal{B}_t := \bigcup_{s \in V(T_t)} \overline{b}_s$. Beginning from the leaves we inductively compute $\operatorname{tp}_q(G[\mathcal{B}_t], \overline{b}_t \overline{v}_0)$ for each node $t \in V(T)$. Here, the notation $\operatorname{tp}_q(G[\mathcal{B}_t], \overline{b}_t \overline{v}_0)$ indicates that in $G[\mathcal{B}_t]$ we compute the type of $\overline{b}_t$ and all vertices of $\overline{v}_0$ contained in $\mathcal{B}_t$. For leaves $t$ with $\overline{b}_t = \overline{v}_i$, for some $1 \leq i \leq m$, we can infer the type $\operatorname{tp}_q(G[\overline{b}_t], \overline{b}_t \overline{v}_0)$ from $\Theta_i$. For other leaves we can compute their types directly, as they only contain at most $k + 1$ elements. For inner nodes $t$ with children $t_1, t_2$ we apply Lemma 2.3. □

As the previous lemma applies to MSO-types, Courcelle's theorem is clearly a special case of it. Hence, the proof here provides an alternative way of establishing Courcelle's theorem. While the two approaches may seem to be somewhat different, the underlying principle is the same. Recall that in our original proof of Courcelle's theorem, we encoded graphs $G$ of tree-width $\leq k$ in labelled trees $T$ and then rewrote the formula $\varphi$ on $G$ to a new formula $\varphi'$ on $T$ such that $G \models \varphi$ if, and only, if $T \models \varphi'$. On the tree-encoding, we then applied results from automata theory which establish that MSO model-checking is fixed-parameter tractable on trees. More specifically, the MSO-formula $\varphi'$ is translated into an automaton $\mathcal{A}_\varphi$ which accepts $T$ if, and only if, $T \models \varphi'$. Although it is not usually proved this way, essentially the automaton has a state for each possible $q$-type and its transition relation combines types similar to what is done in Lemma 2.3.

But back to first-order model-checking on graph classes excluding a minor. Essentially the previous lemma allows us to deal with tree-decompositions over graphs of bounded tree-width, which clearly is not enough for our purposes.

**Lemma 7.14** *Let* $\mathrm{tp}_q$ *denote* $\mathrm{tp}_q^{\mathrm{FO}}$. *The following problem is fixed-parameter tractable for all* $\lambda, \mu$: *given*

- *a labelled graph* $G \in \mathcal{L}(\lambda, \mu)$,
- *tuples* $\overline{v}_i \in V(G)^{r_i}$, $0 \leq i \leq m$ *for some* $m$, *such that* $G[\overline{v}_i]$ *is a clique, and*
- $q$-*types* $\Theta_1, \ldots, \Theta_m$,

*compute* $\mathrm{tp}_q(G, \overline{v}_0)$ *for all graphs* $G' = G \oplus_{\overline{v}_1} H_1 \oplus_{\overline{v}_2} \cdots \oplus_{\overline{v}_m} H_m$ *such that* $\mathrm{tp}_q(H_i, \overline{v}_i) = \Theta_i$. *The parameter is* $q$.

*Proof.* The proof is by induction on $\mu$. For $\mu = 0$, we adapt the proof of Theorem 7.7 using Lemma 7.13 locally. Now let $\mu > 0$ and let $G \in \mathcal{L}(\lambda, \mu), \overline{v}_i, \Theta_i$ be an instance of the problem. By definition, $G$ contains a vertex $v \in V(G)$ such that $G \setminus v \in \mathcal{L}(\lambda, \mu - 1)$. Note that for all $\lambda', \mu', \mathcal{L}(\lambda', \mu')$ is a minor ideal and hence has a cubic time membership test by Corollary 5.3. Thus, in time $\mathcal{O}(|G|^4)$ we can find such a vertex $v$. Let $G_2$ be the coloured graph obtained from $G$ by introducing a new colour $C$ by which we label all neighbours of $v$ and then eliminating $v$ from $G$. By construction, $G_2 \in \mathcal{L}(\lambda, \mu - 1)$. Furthermore, it is an easy exercise to translate first-order formulas $\varphi$ over $G$ to formulas $\varphi'$ over $G_2$ such that $G \models \varphi$ if, and only if, $G_2 \models \varphi'$. Hence, the $q$-type of $G' = G \oplus_{\overline{v}_1} H_1 \oplus_{\overline{v}_2} \cdots \oplus_{\overline{v}_m} H_m$ can be recovered from the $q$-type of $G'_2 = G_2 \oplus_{\overline{v}_1} H_1 \oplus_{\overline{v}_2} \cdots \oplus_{\overline{v}_m} H_m$, and the latter is computable by the induction hypothesis. □

The previous two lemmas are the main ingredients for the proof of the following theorem.

**Theorem 7.15** (*Flum, Grohe* [45]) *Let* $\mathcal{C}$ *be a class of graphs excluding at least one minor. Then the following problem is fixed-parameter tractable.*

> MC(FO, $\mathcal{C}$)
>     *Input:* $G \in \mathcal{C}$, $\varphi \in \mathrm{FO}$.
> *Parameter:* $|\varphi|$.
>     *Problem:* Decide $G \models \varphi$.

*Proof.* Let $G$ and $\varphi$ be given and let $q$ be the quantifier-rank of $\varphi$. Using Theorem 5.16, we first compute a tree-decomposition $(T, \gamma)$ of $G$ over $\mathcal{L}(\lambda, \mu)$, for some $\lambda, \mu$. We view $T$ as a directed tree with root $r$.

For each $t \in V(T)$, $t \neq r$, with parent $s \in V(T)$, let $\overline{v}_t := B_t \cap B_s$. Recall that in the torsi of $B_t$ and $B_s$, $\overline{v}_t$ induces a clique. For the root $r$ we define $\overline{v}_r$ as the empty tuple. Furthermore, for each $t \in V(T)$ let $T_t$ be the subtree of $T$ rooted at $t$ and let $\mathcal{B}_t := \bigcup_{s \in V(T_t)} B_s$. Finally, for $t \in V(T)$ let $G_t := G[\mathcal{B}_t] \cup K[\overline{v}_t]$. Note that for all $t \in V(T)$, $\overline{v}_t \leq k$, where $k := \lambda + \mu$, as $\overline{v}_t$ induces a clique in the torso $[B_t]$ of $B_t$. As $[B_t] \in \mathcal{L}(\lambda, \mu)$ and graphs in $\mathcal{L}(\lambda, \mu)$ cannot contain a clique of order $> \lambda + \mu$ we obtain $|\overline{v}_t| \leq k$. Hence, as $\lambda, \mu$ only depend on the excluded minor of $\mathcal{C}$ and therefore are fixed, we obtain a fixed upper bound for the size of $\overline{v}_t$, $t \in V(T)$.

To decide $G \models \varphi$, we aim at computing the type $\text{tp}_q(G, \overline{v}_r)$. We can then simply check whether $\varphi \in \text{tp}_q(G, \overline{v}_r)$. Towards this aim, starting at the leaves and proceeding bottom-up, we apply Lemma 7.14 at each node to compute the type $\text{tp}_q(G_t, \overline{v}_t)$. □

The previous theorem shows that for every fixed graph $H$, first-order model-checking is fixed-parameter tractable, with parameter $|\varphi|$, on every class of graphs excluding $H$. However, the algorithm as described above is not fixed-parameter tractable in the parameter $|H| + |\varphi|$ as we use a non-constructive approach in Lemma 7.14 and also the algorithm described in [29] seems to use the minor $H$ in an inappropriate way for parameterized complexity.

We therefore turn to a different parametrization of the problem, where we take the parameter to be $|\varphi| + |H|$. This problem was studied by Dawar, Grohe and Kreutzer in [21]. The approach taken there is similar to the method outlined above. However, instead of using tree-decompositions over $\mathcal{L}(\lambda, \mu)$, [21] uses a slightly weaker form of decompositions, called *weak* decompositions over $\mathcal{L}(\lambda, \mu)$. The main result in [21] is that for every $H$, every graph excluding $H$ has a weak decomposition over some $\mathcal{L}(\lambda, \mu)$ (which is relatively straightforward to show) and that these decompositions can be computed by an fpt-algorithm with parameter $H$ (which requires considerably more work). Once this is shown, the proof method outlined above can be adapted to weak decompositions yielding the following result.

**Theorem 7.16** (*Dawar, Grohe, Kreutzer* [21]) *The following problem is fixed-parameter tractable.*

> $p$-MC(FO)
>    *Input:* $G, H$ such that $H \not\preceq G$, $\varphi \in$ FO.
> *Parameter:* $|\varphi| + |H|$.
>    *Problem:* Decide $G \models \varphi$.

An immediate consequence of the theorem is the following. Recall from Section 7.3 the definition of the *minimum excluded clique* number $mec(G)$ of a graph $G$ and of locally excluded minors. For any function $f : \mathbb{N} \to \mathbb{N}$ let $\mathcal{C}_f$ be the class of graphs $G$ such that $mec(G) \leq f(|G|)$.

**Corollary 7.17** *There is an unbounded function $f : \mathbb{N} \to \mathbb{N}$ such that $\mathrm{MC}(\mathrm{FO}, \mathcal{C}_f)$ is fixed-parameter tractable.*

Another consequence of the theorem is that it allows us to apply the framework of localisation as developed in Section 7.3 to obtain the following result.

**Corollary 7.18** *Let $\mathcal{C}$ be a class of graphs locally excluding a minor. Then the problem*

> $\mathrm{MC}(\mathrm{FO}, \mathcal{C})$
> *Input:* $G \in \mathcal{C}$, $\varphi \in \mathrm{FO}$.
> *Parameter:* $|\varphi|$.
> *Problem:* Decide $G \models \varphi$.

*is fixed-parameter tractable.*

The previous result has a number of algorithmic applications.

**Corollary 7.19** *1. The following problem is fixed-parameter tractable.*

> $p$-DOMINATING SET
> *Input:* Given graphs $G, H$ such that $H \not\preccurlyeq G$ and $k \in \mathbb{N}$.
> *Parameter:* $k + |H|$.
> *Problem:* Decide whether $G$ contains a dominating set of size $\leq k$.

*Analogous results hold for all other first-order definable parameterized problems, such as* INDEPENDENT SET *and* CLIQUE *and also for problems such as deciding for a fixed graph $G'$ whether $G'$ has a homomorphism into $G$, or $G'$ is an (induced) subgraph of $G$, where here the parameter can be taken to be $|H| + |G'|$.*

*2. Let $\mathcal{C}$ be a class of graphs locally excluding a minor. Then problem such as* DOMINATING SET, INDEPENDENT SET *etc. are fixed-parameter tractable on $\mathcal{C}$. Furthermore, the problem, given graphs $H$ and $G$ such that $G \in \mathcal{C}$, to decide whether $H$ is homomorphic to $G$ or $H$ is an (induced) subgraph of $G$ can be decided by fpt algorithms with parameter $|H|$.*

### 7.5 Limits to First-Order Model-Checking

In the previous sections we have seen various examples for classes of graphs or structures that allow efficient model-checking for first- or monadic second-order logic, where efficient means fixed-parameter tractable. The picture described there (and illustrated in Figure 18 below) is as yet far from being complete with a number of incomparable graph invariants which allow parameterized algorithms for first-order model-checking. It is an interesting and important challenge to explore the boundary of parametrized tractability for first-order model-checking. In an ideal situation, we would be able to completely classify the classes $\mathcal{C}$ of structures into those where $\mathrm{MC}(\mathrm{FO}, \mathcal{C})$ is FPT and where it is not. So far, work in this direction has mainly focussed on establishing parameterized model-checking algorithms for various classes of graphs defined in terms of graph properties. It is unlikely that in this way, or at least with the properties studied so far, we can fully explore tractability for first-order model-checking. For, FO- or MSO model-checking is preserved under interpretations.

**Lemma 7.20** *If $\mathcal{C}$ is a class of graphs such that $\mathrm{MC}(\mathrm{FO}, \mathcal{C})$ is fixed-parameter tractable on $\mathcal{C}$ and $\mathcal{D}$ is a class of graphs first-order interpretable in $\mathcal{C}$ as described in Section 2.3, then first-order model-checking is fixed-parameter tractable on $\mathcal{D}$.*

**Corollary 7.21** *If $\mathrm{MC}(\mathrm{FO}, \mathcal{C})$ is fixed-parameter tractable then so is $\mathrm{MC}(\mathrm{FO}, \mathcal{D})$ for the class $\mathcal{D} := \{G := (V, V^2 \setminus E) : (V, E) \in \mathcal{C}\}$ of graphs whose complements are in $\mathcal{C}$.*

Hence, if there is a graph property that precisely describes when FO model-checking is tractable, it has to be closed under edge-complementation or more generally under first-order interpretations. Clearly, none of the graph structure classes considered in this paper are closed under edge-complementation.

In addition to studying further classes of graphs obtained from graph invariants it may therefore be beneficial to consider constructions that allow us to construct new classes $\mathcal{C}$ of graphs with tractable model-checking from other, known classes of graphs. For instance, one could try to generalise the constructions using tree-decompositions over classes of graphs. It is easily seen that if $\mathcal{C}$ is a class of graphs for which the appropriate version of Lemma 7.14 holds, then first-order model-checking is also tractable on the class of graphs that can efficiently be tree-decomposed over $\mathcal{C}$. We refrain from giving a formal definition of this as, so far, its only application seems to be Theorem 7.15. Tree-decompositions are a special case where Feferman-Vaught style theorems can be applied. It may be worthwhile to consider further constructions that allow us to define new tractable model-checking intances from the classes we already know.

The previous lemma also has interesting consequences in its negative form, that is, it can be used to show intractability results as demonstrated in the next lemma.

**Lemma 7.22** *For $k \in \mathbb{N}$ let $\mathcal{AD}_k$ be the class of graphs of maximum average degree at most $k$, where the maximum average degree of a graph $G$ is the maximum of the average degrees of all subgraphs of $G$. For $k \geq 4$, $\mathrm{MC}(\mathrm{FO}, \mathcal{AD}_k)$ is $\mathrm{AW}[*]$-hard, i.e. fixed-parameter intractable.*

*Proof.* Recall from Section 2.4 that $\mathrm{MC}(\mathrm{FO}, \mathrm{GRAPH})$, the model-checking problem for FO on the class of all finite graphs, is $\mathrm{AW}[*]$-complete. Further, FO model-checking on the class of all graphs $G$ can easily be reduced to FO model-checking on the class of incidence graphs $I(G)$. As incidence graphs have maximum average degree at most 4, the result follows immediately. □

Hence, graph classes of bounded maximum average degree provide a first non-trivial upper bound for parameterized tractability of FO model-checking. Complementing the efforts to establish fixed-parameter tractability results for first-order model-checking on various classes of graphs and structures, it would be interesting to prove stronger intractability results. For MSO there is hope that a clean characterisation of tractable cases might be achievable essentially using existing machinery. For first-order logic we are still very far from a good understanding of its tractable and intractable cases and much more work is needed approaching the tractabilitiy frontier from both sides.

Towards another graph property that may yield fixed-parameter algorithms for first-order logic, consider again the proof of the previous lemma. Essentially, given a graph $G$ we subdivide every edge once to obtain the incidence graph. For first-order logic, this does not pose much of a problem as we can easily rewrite the formula to deal with the subdivision. Similarly, if we replace every edge by a path of length $k$, i.e. subdivide a bounded number of times, then again we obtain small maximum average degree but we can easily rewrite first-order formulas to deal with these paths of fixed length.

Note that this essentially means that we replace every vertex by a graph of fixed radius, e.g. in the case of $k=3$ we replace every vertex by a star. Hence, if we are interested in paramaterized tractability, then we should require our graphs to have bounded maximum average degree even after we contract neighbourhoods of fixed radius. This idea is formalised in the notion of bounded expansion introduced by Nešetřil and Ossona de Mendez in [64,65,66].

Let $H, G$ be graphs such that $H \preccurlyeq G$ and let $\mu$ be a minor map from $H$ to $G$. The *radius* of $\mu$ is defined as the maximal radius of any of the subgraphs $\mu(v)$, for $v \in V(H)$. We write $H \preccurlyeq_r G$ if there is a minor map $\mu$ of radius $\leq r$ from $H$ to $G$. The *greatest reduced average density (grad)* $\nabla_r(G)$ *of rank $r$ of $G$* is defined as

$$\nabla_r(G) := \max\left\{ \frac{|E(H)|}{|V(H)|} : H \preccurlyeq_r G \right\}.$$

A class $\mathcal{C}$ of graphs has *bounded expansion* if there is some function $f : \mathbb{N} \to \mathbb{N}$ such that $\nabla_r(G) \leq f(r)$ for all $r \in \mathbb{N}$ and $G \in \mathcal{C}$. As shown by Nešetřil and Ossona de Mendez [63], every class excluding a minor also has bounded expansion. On the other hand, there are classes of graphs which locally exclude a minor, in fact even classes of graphs of bounded local tree-width, which do not have bounded expansion. (See [53, Example 4.7] for an example of a class of bounded local tree-width but unbounded expansion.) The converse is true as well, i.e. there are classes of graphs of bounded expansion that do not locally exclude a minor.

Classes of bounded expansion overcome the problem with first-order model-checking outlined above: the graphs in the class must have small average degree even after we contract neighbourhoods of fixed radius. Hence our interpretation from the class of all graphs into the class of graphs of maximum average degree at most 4 does not yield a class with bounded expansion.

**Open Problem 7.23** *Is* $\mathrm{MC}(\mathrm{FO}, \mathcal{C})$ *fixed-parameter tractable on every class $\mathcal{C}$ of bounded expansion?*

## 8   Conclusion

This paper gives an overview of algorithmic meta-theorems developed in recent years. See Figure 18 for a diagrammatic summary of the results presented in this paper.

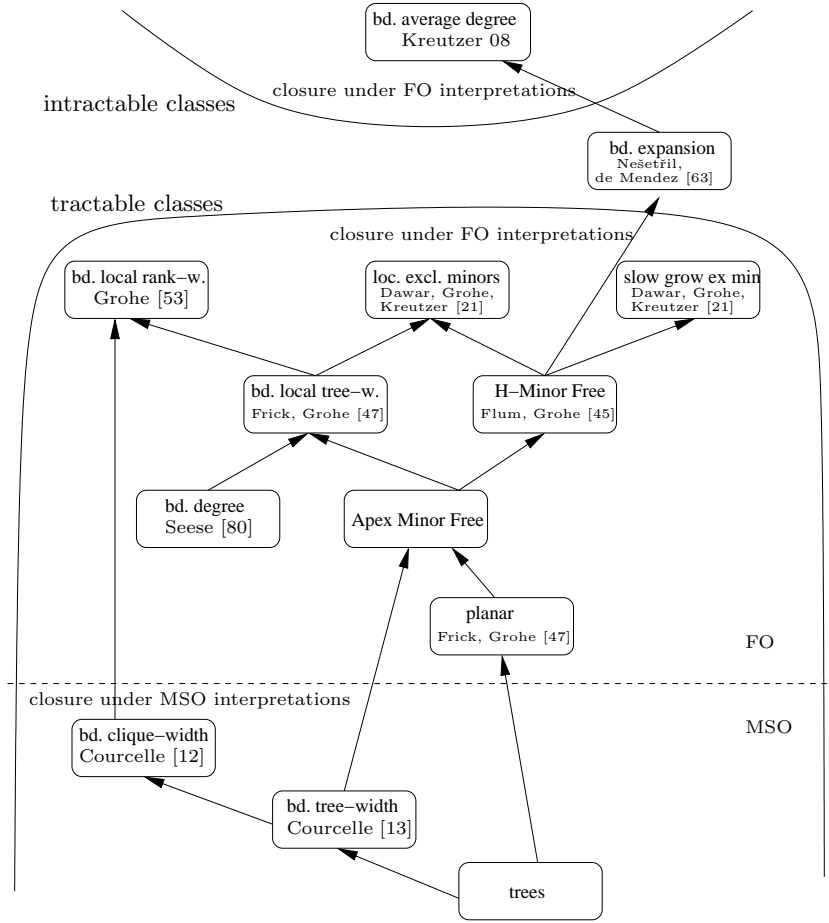

**Fig. 18.** Summary of results

As we have seen, first-order model-checking is fixed-parameter tractable on a wide range of graph classes defined by standard graph invariants such as tree-width or excluded minors. By localising these invariants we obtained even further tractable classes. However, we are still very far from a clear picture of where first-order model-checking is tractable and where it is not. Further research, in particular into intractability results is needed before we can hope for a clean and smooth theory.

**Acknowledgements**

I want to take this opportunity to express my gratitude towards Martin Grohe for the long discussions we had on graph minors and for his advise, patience and inspiration during the time I worked with him at the Humboldt University in Berlin.

Many thanks to Javier Esparza for proofreading this manuscript and for his patience as an editor. Also, many thanks for proofreading this manuscript to Martin Grohe, Paul Hunter, Michael Kreikenbaum, Andras Salamon, Marc Thurley, Mark Weyer.